\documentclass[amssym,useAMS,usenatbib]{mnras}
\usepackage{array}
\usepackage{graphicx}
\usepackage{amsmath}
\usepackage{amssymb}

\def\spose#1{\hbox to 0pt{#1\hss}}
\def\simlt{\mathrel{\spose{\lower 3pt\hbox{$\mathchar"218$}}
     \raise 2.0pt\hbox{$\mathchar"13C$}}}
\def\simgt{\mathrel{\spose{\lower 3pt\hbox{$\mathchar"218$}}
     \raise 2.0pt\hbox{$\mathchar"13E$}}}
\def\be{\begin{equation}}
\def\ee{\end{equation}}
\def\bce{\begin{center}}
\def\ece{\end{center}}
\def\bea{\begin{eqnarray}}
\def\eea{\end{eqnarray}}
\def\ben{\begin{enumerate}}
\def\een{\end{enumerate}}

\def\brr{\begin{array}}
\def\err{\end{array}}
\def \simlt {\mathrel{\spose{\lower 3pt\hbox{$\mathchar"218$}}
     \raise 2.0pt\hbox{$\mathchar"13C$}}}
\def \simgt {\mathrel{\spose{\lower 3pt\hbox{$\mathchar"218$}}
     \raise 2.0pt\hbox{$\mathchar"13E$}}}
\DeclareMathOperator{\Tr}{Tr}

\def\xpk{\xi_{\rm pk}}
\def\gtrsim{\lower.5ex\hbox{$\; \buildrel > \over \sim \;$}}
\def\bnpk{\bar{n}_{\rm pk}}

\newmuskip\pFqskip
\mathchardef\pFcomma=\mathcode`, 

\newcommand*\pFq[6]{%
  \begingroup
  \begingroup\lccode`~=`,
    \lowercase{\endgroup\def~}{\pFcomma\mkern\pFqskip}%
  \mathcode`,=\string"8000
  {}_{#1}F_{#2}\left(#3,#4;#5;#6\right)%
  \endgroup
}

\def \prd {{\it Phys. Rev. D}}

\def \apj {{\it ApJ}}
\def \apjl {{\it ApJ Let.}}
\def \apjs {{\it ApJ Sup.}}

\def \aap {{\it A\&A}}
\def \mnras {{\it MNRAS}}
\def \nat {{\it Nature}}
\usepackage{hyperref} 

\begin{document}
\title[On the connectivity of the cosmic web: theory and implications]{On the connectivity of the cosmic web: theory and implications for cosmology and galaxy formation }
% =============================================================================

\author[Sandrine Codis, Dmitri Pogosyan  and Christophe Pichon]{\parbox{\textwidth}{
Sandrine Codis$^{1,2}$\thanks{codis@iap.fr}, Dmitri Pogosyan$^{3}$,  Christophe Pichon$^{1,4,5}$}
\vspace*{6pt}\\
\noindent
$^{1}$ Institut d'Astrophysique de Paris, CNRS \& Sorbonne Universit\'e, UMR 7095,  98 bis boulevard Arago, 75014 Paris, France, \\
$^{2}$ Canadian Institute for Theoretical Astrophysics, University of Toronto, 60 St. George Street, Toronto, ON M5S 3H8, Canada,\\
$^{3}$ Department of Physics, University of Alberta, 412 Avadh Bhatia Physics Laboratory, Edmonton, Alberta, T6G 2J1, Canada,  \\
$^{4}$ Korea Institute of Advanced Studies (KIAS) 85 Hoegiro, Dongdaemun-gu, Seoul, 02455, Republic of Korea,\\
$^{5}$ Institute for Astronomy, University of Edinburgh, Royal Observatory, Blackford Hill, Edinburgh, EH9 3HJ, United Kingdom.
}

% =============================================================================
\maketitle

% =====================================================================
\begin{abstract}
Cosmic connectivity and multiplicity, i.e. the number of filaments globally or locally connected to a given cluster is a natural probe of the growth of structure and in particular of the nature of dark energy. It is also a critical ingredient driving the assembly history of galaxies as it controls mass and angular momentum accretion. \\
The connectivity of the cosmic web is investigated here via the persistent skeleton. This tool identifies topologically the ridges of the cosmic landscape which allows us to investigate how the nodes of the cosmic web are connected together. When applied to Gaussian random fields corresponding to the high redshift universe, it is found that on average the nodes are connected to exactly $\kappa=4$ neighbours in two dimensions and $\sim 6.1$ in three dimensions. Investigating spatial dimensions up to $d=11$, typical departures from a cubic lattice $\kappa=2d$ are shown to scale like the power 7/4 of the dimension. These numbers strongly depend on the height of the peaks: the higher the peak the larger the connectivity. \\
Predictions from first principles based on peak theory are shown to reproduce well the connectivity and multiplicity of Gaussian random fields and cosmological simulations. As an illustration, connectivity is quantified in galaxy lensing convergence maps and large dark haloes catalogues. As a function of redshift and scale the mean connectivity decreases in a cosmology-dependent way. As a function of halo mass it scales like 10/3 times the log of the mass. Implications on galactic scales are discussed.

\end{abstract}
% =====================


%=====================
\section{Introduction}
%=====================

Over the course of the last decades,
our understanding of the extragalactic  universe 
has significantly evolved:  the description of its components  
 has evolved
 from being (essentially) isolated to being multiply connected both on large scales, cluster scales and galactic scales.
  This interplay between large and small scales is driven in part 
 by gravity which tends to couple dynamically different scales in the framework of the so-called concordant cosmological model \citep{prunet}.
 This model  
 predicts a certain shape for the initial conditions, leading to a hierarchical formation scenario,
 which produces  the large scale structure, the most striking feature in the distribution of matter
on megaparsecs scale. This distribution was observed more than thirty
years ago by the first CfA catalog \citep{Lapparent} followed by many others
such as the SDSS \citep{sdss}, 2dF \citep{2df} or more recently DES \citep{2016PhRvD..94b2001A} galaxy redshift surveys. 

The  ``Cosmic Web'' picture  \citep{ks93,1996Natur.380..603B}  was developed to  explain the  origin of this network:
it
relates
the observed clusters of galaxies, and filaments that link them, 
to the geometrical
properties of the initial density field that are enhanced but not yet
destroyed by the still mildly non-linear evolution on those scales. 
It builds from  the ellipsoidal collapse  model   studied by \cite{lb64,lin65}, 
followed by Zel'dovich's work \citep{zeldo70}  which 
related the anisotropic nature of the gravitational collapse  to the formation of elongated and flattened structures. 
The concept of cosmic web emerged from those ideas and was extended in the peak-patch formalism \citep{1996ApJS..103....1B}: the origin of filaments and nodes lies in the asymmetries of the initial Gaussian random field (GRF hereafter) 
describing the primordial universe which is later amplified by gravitational collapse. 
These investigations stressed  the key role of the theory of random fields in cosmology and the importance of non-local tidal effects in weaving the cosmic web. The high-density peaks define the nodes of the evolving cosmic web and completely determine the filamentary pattern in between. 
 Building upon the cosmological peak theory described in \cite{BBKS}, local properties of the filamentary cosmic web can be predicted such as the length of filaments, the surface of walls or their curvatures \citep{pogo09}, while its topology can be fully characterised  \citep{Gott86,Mecke94,Matsubara0,GPP2012}, all of these observables carrying complementary cosmological information \citep{Zunckel11,Codis2013}.

While traditionally the emphasis is  placed on the statistical descriptors of the underlying random field via the hierarchy of N point correlation functions
\citep{Scoccimarro98,2017MNRAS.468.1070S},
in this paper we focus on the connectivity of this cosmic network as a mean to understand  its morphology and geometry. 
 Indeed, recent  ridge extractors \citep[such as the skeleton,][]{skel2D} allow for the definition 
of the filamentary cosmic web as a {\em  connected} network that continuously
link maxima and saddle-points of a scalar field together. Hence it is
of interest to try and understand the topological and
geometrical properties of the underlying density field through  the connectivity
and hierarchical relationship that the ridges introduce between the critical
points.  This can  be used to establish, in particular,
the percolation properties of the Web \citep{Colombi2000}.  Applied to cosmology, these ridges provide a formal definition of
the concept of individual filaments. Considering matter distribution
on large scales in the Universe, a natural definition of a {\em
  single} filament is the subset of the cosmic web that directly
links two haloes together. The transposition of such a definition to
the skeleton  allows the introduction of useful concepts such as
 neighbouring relationship between haloes in the cosmic web sense.
 This has implication on both large and small scales.
 
 From the point of view of constraining cosmological parameters, the redshift evolution 
 of the connectivity of the cosmic web on large scales can be used as a probe of fundamental physics. Indeed it can robustly estimate the growth of structure and therefore  the equation 
 of state of dark energy,  as the rate of acceleration of the universe  disconnects  haloes, 
 and gravitational  non-linear evolution induces filament coalescence.

 From a smaller scale perspective, 
the importance of the cosmic web's connectivity
is  sustained  by pan chromatic observations of the environment of galaxies which illustrate sometimes  
spectacular merging processes, 
 following the pioneer work of e.g. \cite{Schweizer} (motivated by theoretical investigations such as  \cite{Toomre}).
The importance of anisotropic accretion on cluster and dark matter halo scales  
\citep{aubert,Steinmetz,2015ApJ...813....6K,aubertpichon} down to central galaxies \citep{2011arXiv1106.0538K,2015arXiv151200400W} is now believed to play a significant role in regulating 
the shape and spectroscopic properties 
of galaxies.
Indeed it has 
been claimed \citep[see e.g][]{ocvirk,dekel}
that the geometry of the cosmic inflow on a galaxy is strongly correlated to its history and nature, as can be seen through the observed \citep{2016MNRAS.457.2287A,2017A&A...597A..86P,Malavasi2016b,2017MNRAS.466.1880C,kraljic18,laigle+17}, virtually measured \citep[among many others]{codisetal12,2015MNRAS.446.1458M,dubois14} and predicted \citep[see for instance][]{Kaiser1984,codis2015,2015MNRAS.447.2683A,biaspaper} correlations between cosmic web and galactic properties such as its mass, spin, shape,
temperature or entropy distribution. 
One of the puzzles of galaxy formation involves understanding how galactic disks reform
after minor and intermediate mergers, a process which is undoubtedly controlled by the drifting of filaments through cosmic time and anisotropic gas inflow therein which carries coherent angular momentum \citep{pichonetal11,2015MNRAS.452..784P}.
At high redshifts, those streams -- the so-called cold flows -- can penetrate haloes deeply into their core and feed the central galaxy.
The coplanarity of those flows is predicted by numerical simulations \citep{2012MNRAS.422.1732D}, evidenced by the observation of planes of satellite galaxies around their central \citep[e.g.][]{1969ArA.....5..305H,2013Natur.493...62I} and could be explained by predicting cosmic connectivity as a function of the rareness of the nodes and prominence of filaments, which is one of the motivations of this paper.

 Section~\ref{sec:ext} first summarizes our understanding of the statistics of extrema, their spatial correlations and how Morse theory link them to the actual cosmic web.
Section~\ref{sec:GRF} then investigates the connectivity of $d$ dimensional  GRF both locally, 
and globally within the context of peak theory and numerical simulations of Gaussian random fields.
Section~\ref{sec:cosmic} focuses next on the cosmic connectivity: first
  for convergence maps 
then for
 the three dimensional dark matter distribution and dark halo catalogues in concordant $\Lambda$CDM simulation.
 Finally, Section~\ref{sec:conclusion} wraps up.

%=====================
\section{Critical sets of GRF}\label{sec:ext}
%=====================

\begin{figure}
\centering
\includegraphics[width=1\columnwidth]{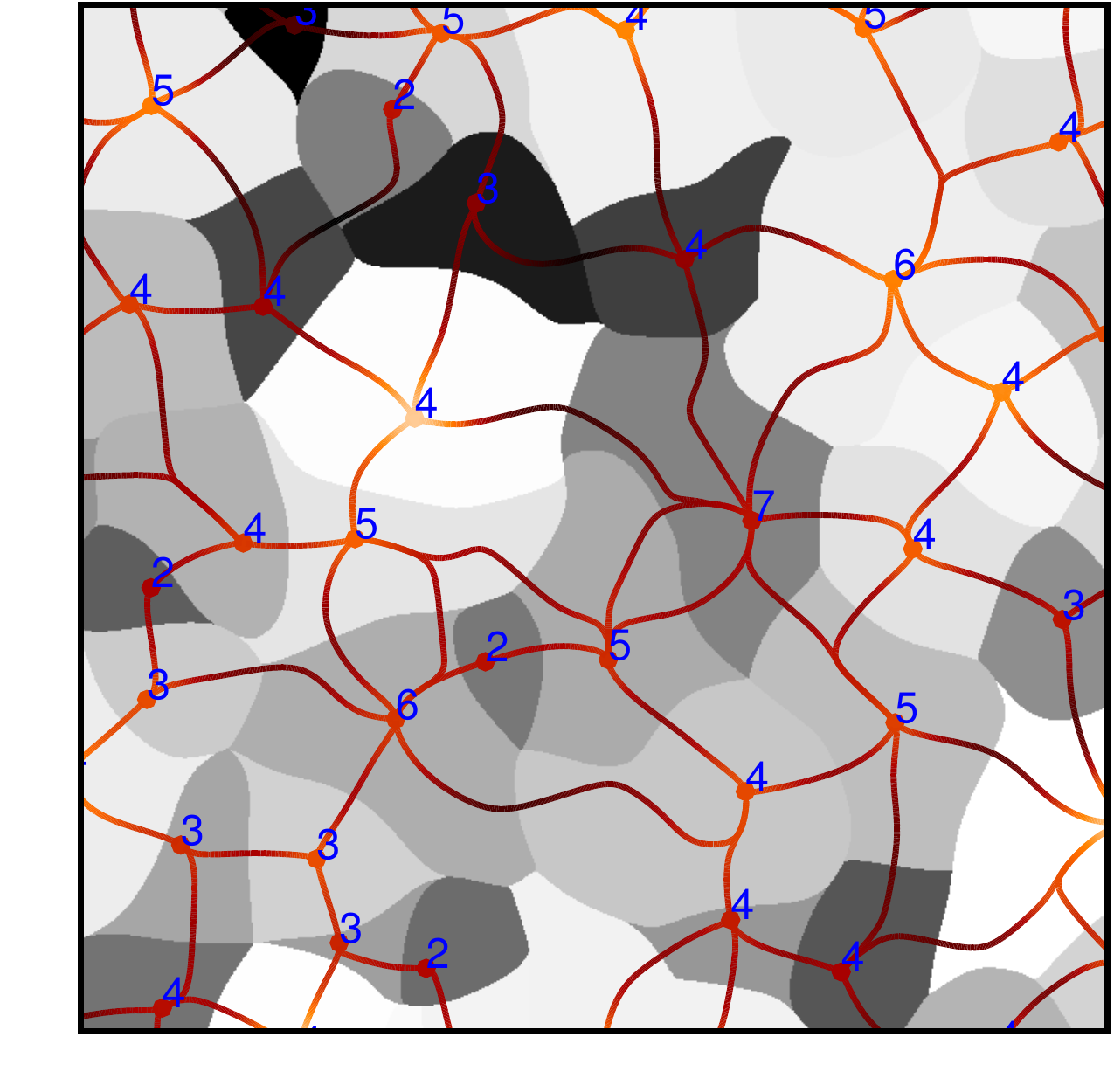}
\caption{
The connectivity, $\kappa$, of maxima vertices on top of the peak patches for a two-dimensional GRF field. The ({\sl blue}) numbers
represents $\kappa$ for the corresponding node. The colour coding of the skeleton ({\sl in shades of red}) reflects the underlying density of the field.
} \label{fig:PP-skel}
\end{figure}

The concept of random fields is central to cosmology.  Random fields
provide initial conditions for the  evolution of the matter distribution in the
Universe. They also describe the distribution of the cosmic microwave background on 
the celestial sphere and are key to understand the geometrical properties of the large-scale structure as was highlighted in this introduction both in the context of cosmology and galaxy formation. 
An example of a two dimensional GRF is displayed in Figure~\ref{fig:PP-skel}. Each peak (dots) is a node of the skeleton (in shades of red) and belongs to $\kappa$ valleys whose boundaries form the filaments. This number $\kappa$ is defined as the connectivity of the node, namely the number of peaks one peak is connected to. In this section, we will first recap some properties of extrema\footnote{Note that in this paper we will use the word ``extrema'' interchangeably with ``critical points'' to refer to all points satisfying the condition of zero gradient meaning both maxima, minima {\em and saddle points}.} in GRF before turning to the theory of the connectivity of GRF in Section~\ref{sec:GRF}.

\subsection{Extrema of GRF}\label{sec:extGRF}

Let us start by reviewing basic facts on the distribution of  {critical points}
in GRF. This field will be generically denoted $\rho$ (with a slight abuse of notation since in the cosmological context it will refer to the density contrast instead of the density field itself) and assumed to have zero mean.
The one-point distribution of extrema in 2D has been studied in
\cite{LonguetHiggins}, while an extensive study of 3D extrema in cosmological
settings goes back to \cite{BBKS}. In 2D, the set of extrema is composed of maxima,
saddle points and minima, identified by the signs of the eigenvalues $\lambda_{i}$, $i\in \{1,2\}$,
of the (Hessian) matrix of the second derivatives of the field denoted $\rho_{ij}$, $i,j\in \{1,2\}$, 
as $(--),(+-),(++)$, while in 3D there are two types of saddle points so that extrema come with signatures
$(---),(+--),(++-),(+++)$.
In this work, we are primarily
interested in the distribution of maxima (which we also often call peaks),
that are identified with
the nodes of the skeleton, and saddle points of 'filamentary' type, 
$(+-)$ in 2D or $(+--)$ in 3D through which the skeleton bridges connecting 
two maxima necessarily pass as described in Section~\ref{sec:Morse} below.

\subsubsection{Number density of extrema}

Let us first recall the well-known results for the 
total number density of maxima and 'filamentary' saddle points in GRF 
\begin{align}
{\rm 2D}:~ & \overline{n}_\mathrm{max} =  \frac{1}{8 \sqrt{3} \pi R_*^2}\,,
& & \overline{n}_\mathrm{sad} =  \frac{1}{4 \sqrt{3} \pi R_*^2} \label{eq:extremacounts-2D}\,,\\
{\rm 3D}:~ &\overline{n}_\mathrm{max}\! =\!\frac{29 \sqrt{15} \!-\! 18 \sqrt{10}}{1800 \pi^2 R_*^3} ,
& & \overline{n}_\mathrm{sad} \!=\! \frac{29 \sqrt{15} \!+\! 18 \sqrt{10}}{1800 \pi^2 R_*^3} ,
\label{eq:extremacounts-3D}
\end{align}
where the characteristic length $R_* \equiv \sigma_1/\sigma_2$ is defined
by the ratio of the variances of the field first,
$\sigma_1^2=\langle \nabla \rho \cdot \nabla \rho \rangle$,
and second, $\sigma_2^2=\langle (\Delta \rho)^2 \rangle$, derivatives.
A characteristic measure of peak separation is given by the radius of a sphere
that on average contains exactly one peak, 
$ {R}_p = ( \pi \overline{n}_\mathrm{max})^{-1/2} \approx 3.7 \; R_*$ and
$ {R}_p = (\frac{4 \pi}{3} \overline{n}_\mathrm{max})^{-1/3} \approx 4.2\; R_*$,
in 2D and 3D respectively. 
The universal relations
\begin{equation}
{\rm 2D}:  \overline{n}_\mathrm{sad}/\overline{n}_\mathrm{max}\!=\!2, \quad  {\rm 3D}: 
\overline{n}_\mathrm{sad}/\overline{n}_\mathrm{max}\!\approx\!3.055,
\end{equation}
are very important for connectivity discussion of GRFs.
Hereafter, all extrema heights will be measured  in units of
the rms $\sigma_0=\sqrt{\langle \rho^{2} \rangle}$ of the field by means of the rareness $\nu=\rho/\sigma_0$, 
all distances in  units of $  R_p$ and all
number densities in units of $\bar n_\mathrm{max}$. The latter effectively
means that instead of number densities we will be quoting the actual number
of extrema inside a sphere of radius $  R_p$. We shall use $d$ to designate
the dimensionality of space wherever general expressions can be used.

We are often interested in the number density of extrema that satisfy some
additional properties (e.g height).
Computing it requires the knowledge of the joint probability distribution function of
these properties, the gradient and the second derivatives of the field, all
evaluated at zero gradient.
The number density of extrema that satisfy these properties
is then obtained by averaging the determinant of
the Hessian $\det(\rho_{ij})$ with this distribution 
over the region of $\rho_{ij}$ with appropriate 
signs of eigenvalues for the given type of extrema,
\begin{equation}
\label{eq:next}
n_{\rm ext}(\mathrm{prop})\! =\!\frac{\displaystyle \! \int\!\! {\rm d} \rho_{ij}
P(\mathrm{prop}, \!\nabla\rho\!=\!0, \rho_{ij}) \!
\left| \det(\rho_{ij}) \!\right|\! \theta(\!\{\lambda_i\}\!)\!}{\overline{n}_\mathrm{max}}\!,
\end{equation}
the last condition being formally codified in the Heaviside $\Theta$-like
function $\theta(\!\{\lambda_i\}\!)$ equal, in particular to 
$\theta_p(\!\{\lambda_i\}\!) \equiv
\prod_{i=1}^{D}\Theta(-\lambda_{i})$ for peaks and
$\theta_s(\!\{\lambda_i\}\!) \equiv
\Theta(\lambda_1)\prod_{i=2}^{D}\Theta(-\lambda_{i})$ for filamentary saddles

\begin{figure*}
\begin{center}
\includegraphics[width= 0.95\columnwidth]{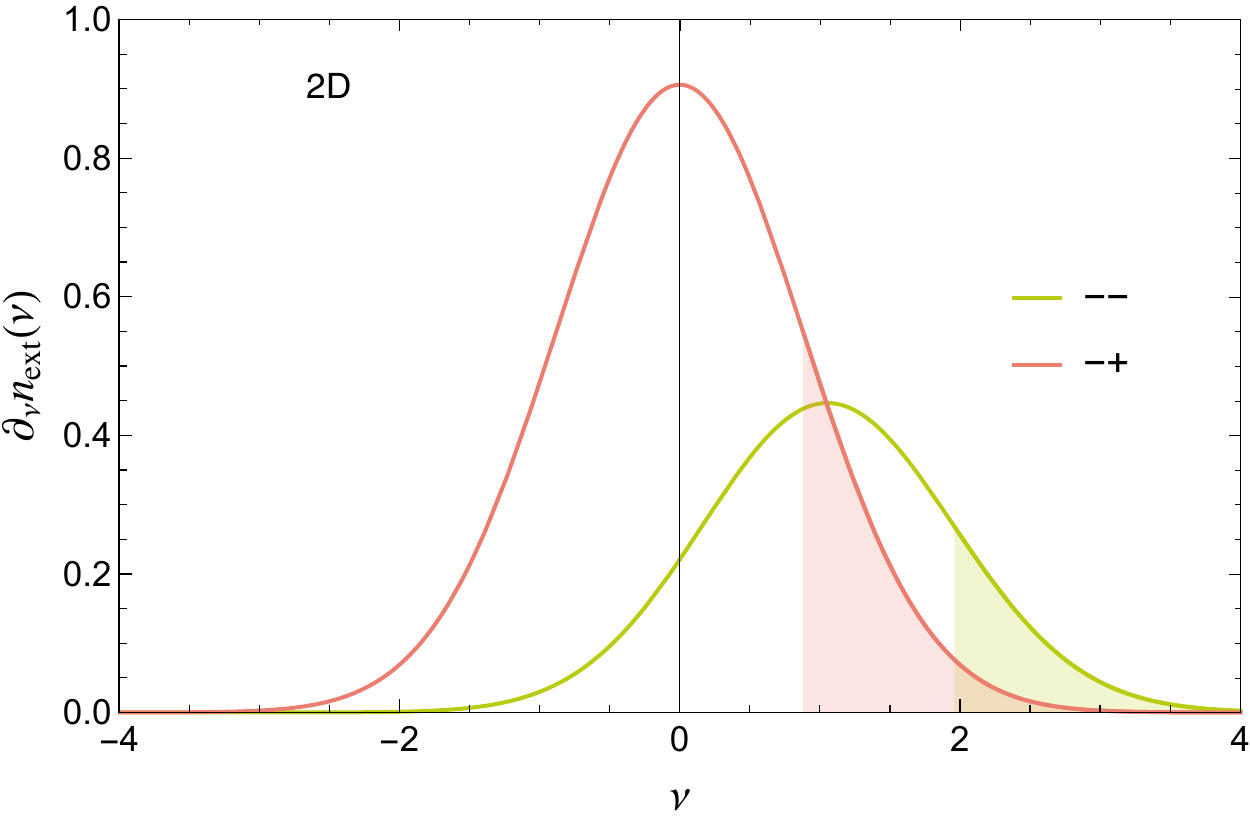} \hskip 1cm
\includegraphics[width= 0.95\columnwidth]{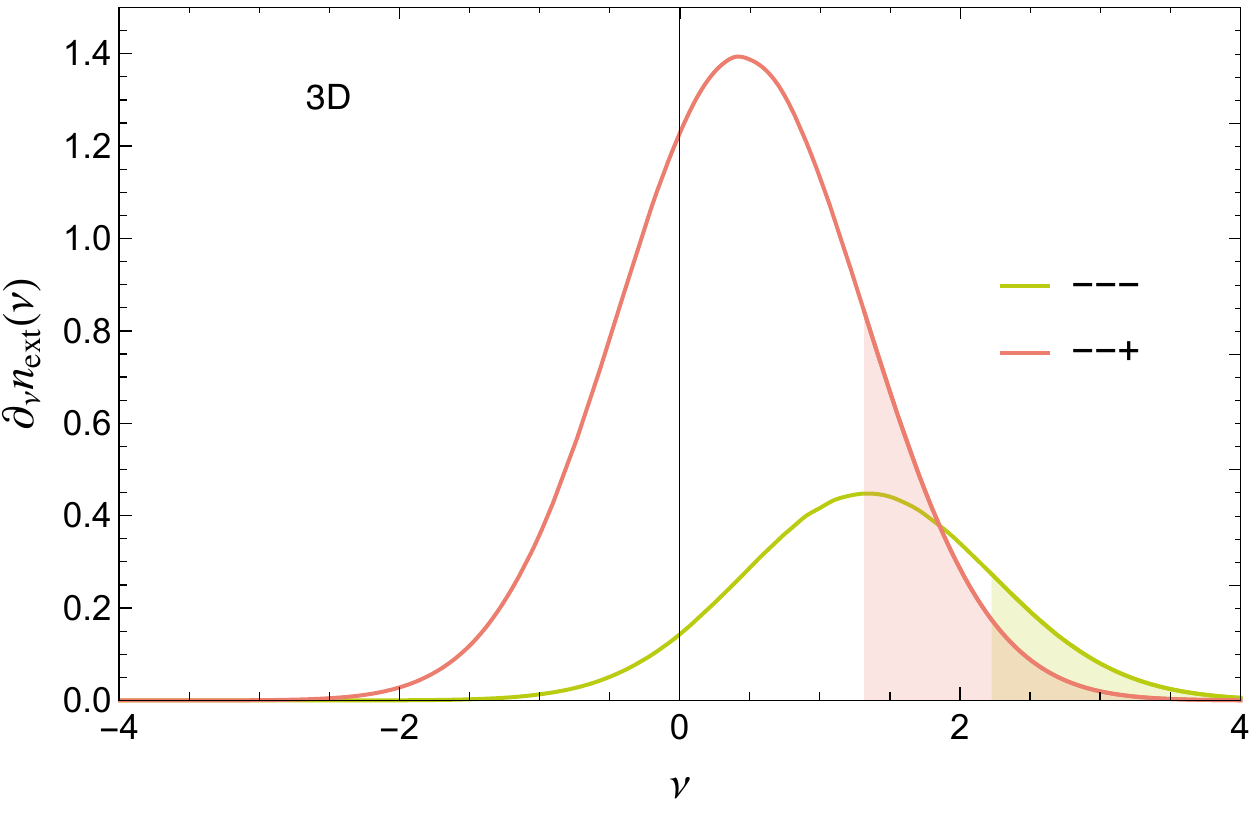} 
\caption{
Differential number of maxima and filamentary saddle points with height
$\nu$ in the spherical volume of radius $  R_p$ for 2D ({\sl left})
and 3D ({\sl right})
Gaussian random fields with $\gamma=0.58$. Shaded areas mark the regions of
high 'rare' maxima and filamentary saddles (above one sigma from the mean).
}
\label{fig:extrema}
\end{center}
\end{figure*}

Perhaps the most 
important property of an extrema is its height $\nu$. 
We shall denote as $n_\mathrm{ext}(\nu)$ the number of extrema
(in the spherical volume of radius $  R_p$)
exceeding the height $\nu$.
Figure~\ref{fig:extrema}
shows the correspondent differential numbers of maxima,
$\partial_{\nu} n_\mathrm{max}$ and saddle points $\partial_{\nu} n_\mathrm{sad}$ for GRF, which in 2D are given by analytical expressions
 \citep[ e.g ][]{BBKS,GPP2012}.
\begin{multline}
\partial_{\nu} n_\mathrm{max}\equiv\frac{\mathrm{d} n_\mathrm{max}}{\mathrm{d}\nu}=\frac{\sqrt{3(1-\gamma^{2})}}{\pi}\gamma\nu\exp\left({-\frac{\nu ^2}{2(1- \gamma ^2)}}\right)\\
+\sqrt{\!\frac{3}{\pi(6\!-\!4\gamma^{2})}} \exp\!\left({\!-\!\frac{3 \nu ^2}{6\!-\!4 \gamma ^2}}\right)\!\!\left[\!1\!+\!\textrm{Erf}\!\left(\!\!\frac{\gamma\nu}{\sqrt{(1\!-\!\gamma^{2})(6\!-\!4\gamma^{2})}}\!\!\right)\!\right]\\
+\sqrt{\!\frac{3}{2\pi}}\gamma^{2}(\nu^{2}-1) \exp\!\left({\!-\!\frac{ \nu ^2}{2}}\right)\!\!\left[\!1\!+\!\textrm{Erf}\!\left(\!\!\frac{\gamma\nu}{\sqrt{2(1\!-\!\gamma^{2})}}\!\!\right)\!\right],
\end{multline}
\begin{equation}
\partial_{\nu} n_\mathrm{sad}\equiv\frac{\mathrm{d} n_\mathrm{sad}}{\mathrm{d}\nu}=\sqrt{\frac{6}{\pi(3-2\gamma^{2})}} \exp\left({-\frac{3 \nu ^2}{6-4 \gamma ^2}}\right),
\end{equation}
while in 3D are evaluated by numerical integrations. 
The distributions
are governed by the spectral parameter 
$\gamma \equiv \sigma_1^2/(\sigma_0 \sigma_2)$ which reflects the correlation
between the normalised\footnote{In this paper, ``normalised'' means rescaled by its variance so that fields of zero mean and unit variance are considered.} field and trace of its Hessian at the same point,
$\gamma = -\langle \rho \Delta \rho \rangle/\sigma_{0}\sigma_{2}$.
We shall consider
the extrema as (relatively) high and rare if their height exceeds one
standard deviation from the mean. This threshold is $\gamma$-dependent\footnote{e.g for 2D saddles, the standard deviation from the mean is given by $\sqrt{1-2\gamma^{2}/3}$}
but as a guidance in future discussions we can use the values from the
$\gamma=0.58$\footnote{$\gamma=0.58$ corresponds to power-law power spectra
with spectral index $n_{s}=-1$ (resp. -2) in 2D (resp. 3D) and to a redshift
zero $\Lambda$CDM Universe on scales about $1$Mpc.
 }
case in Figure~\ref{fig:extrema}, namely $\nu \gtrsim 2$ for maxima and
$\nu \gtrsim 1$ for filamentary saddles. 

\subsubsection{Spatial correlations of GRF extrema}
\label{sec:corr_func}

A first information about the relative spatial distribution of extrema is given
by their two-point correlation function
\begin{equation}
\xi_{ab}(r)\equiv \frac{
\left\langle {\cal C}_\mathrm{a}(\mathbf{x})
{\cal C}_\mathrm{b}(\mathbf{x}+\mathbf{r}) \right \rangle}
{\left\langle{\cal C}_\mathrm{a}(\mathbf{x})\right\rangle
\left\langle {\cal C}_\mathrm{b}(\mathbf{x})\right\rangle}-1\,,
\end{equation}
where $a,b$ designate either peak or saddle and
${\cal C}_\mathrm{a}(\mathbf{x})=\left| \det(\rho_{ij}) \right|
\delta_{D}(\nabla\rho)\theta_a(\!\{\lambda_i\}\!)$
is the localised number density of extrema as used in equation~\eqref{eq:next}.
Additionally, one can add extra (and possibly different) constraints on the two extrema, for instance specific heights.
Here we briefly describe the features of 
peak-peak and peak-saddle correlations 
that are of main importance for connectivity discussion and refer the reader to Appendix~\ref{sec:appA} for further mathematical and computational details.

\begin{figure*}
\begin{center}
\includegraphics[width= 0.95\columnwidth]{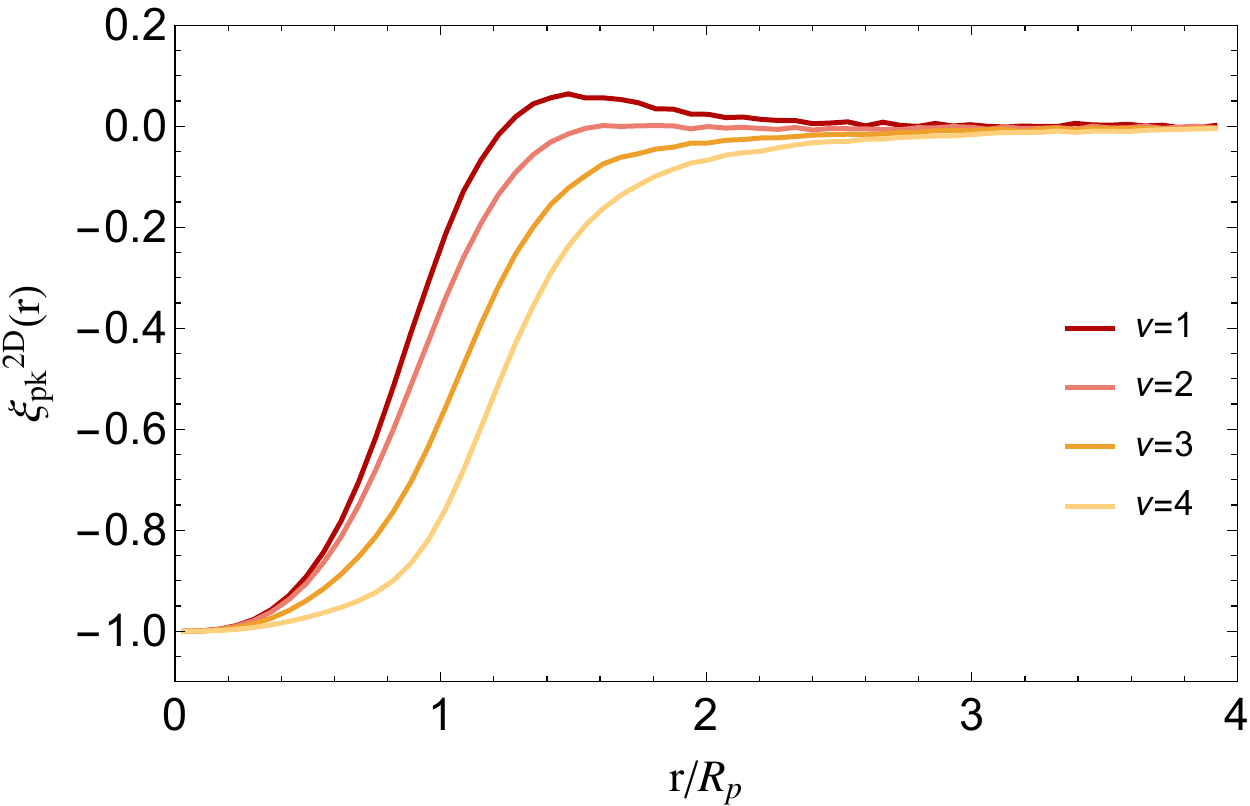} \hskip 1cm
\includegraphics[width= 0.95\columnwidth]{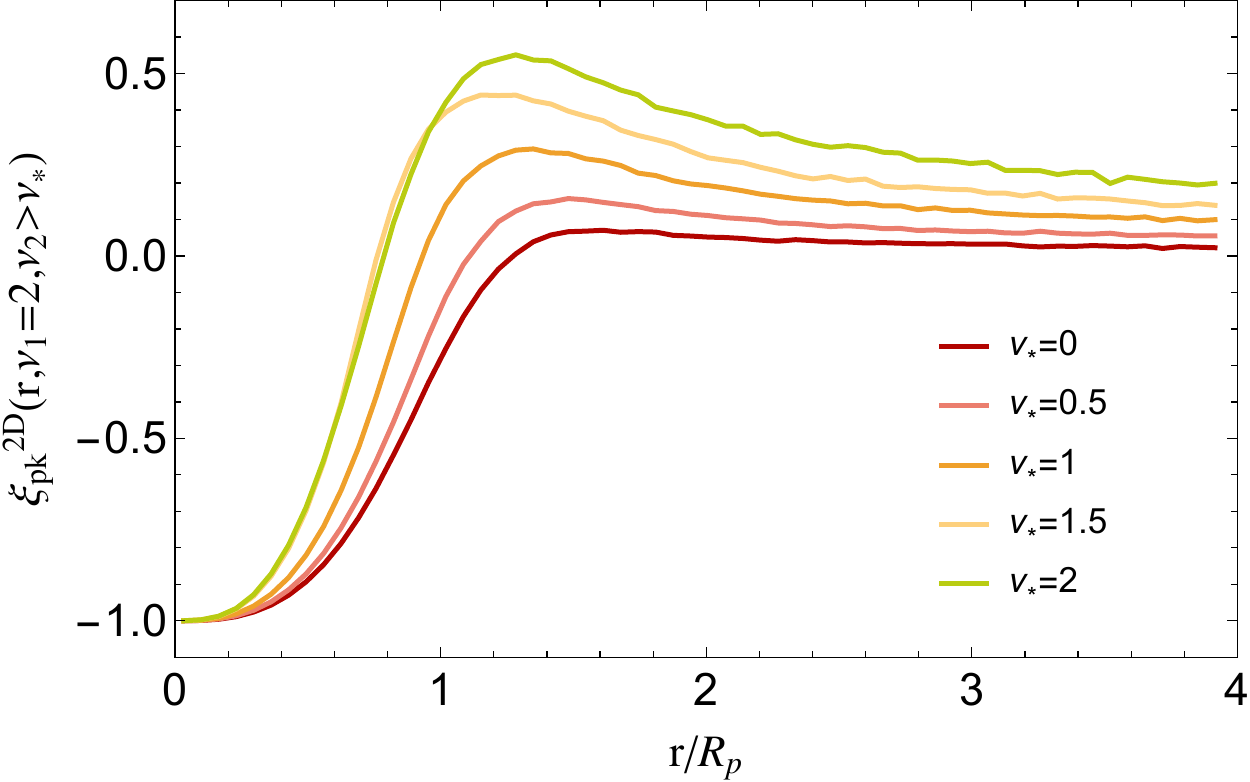}\\
\includegraphics[width= 0.95\columnwidth]{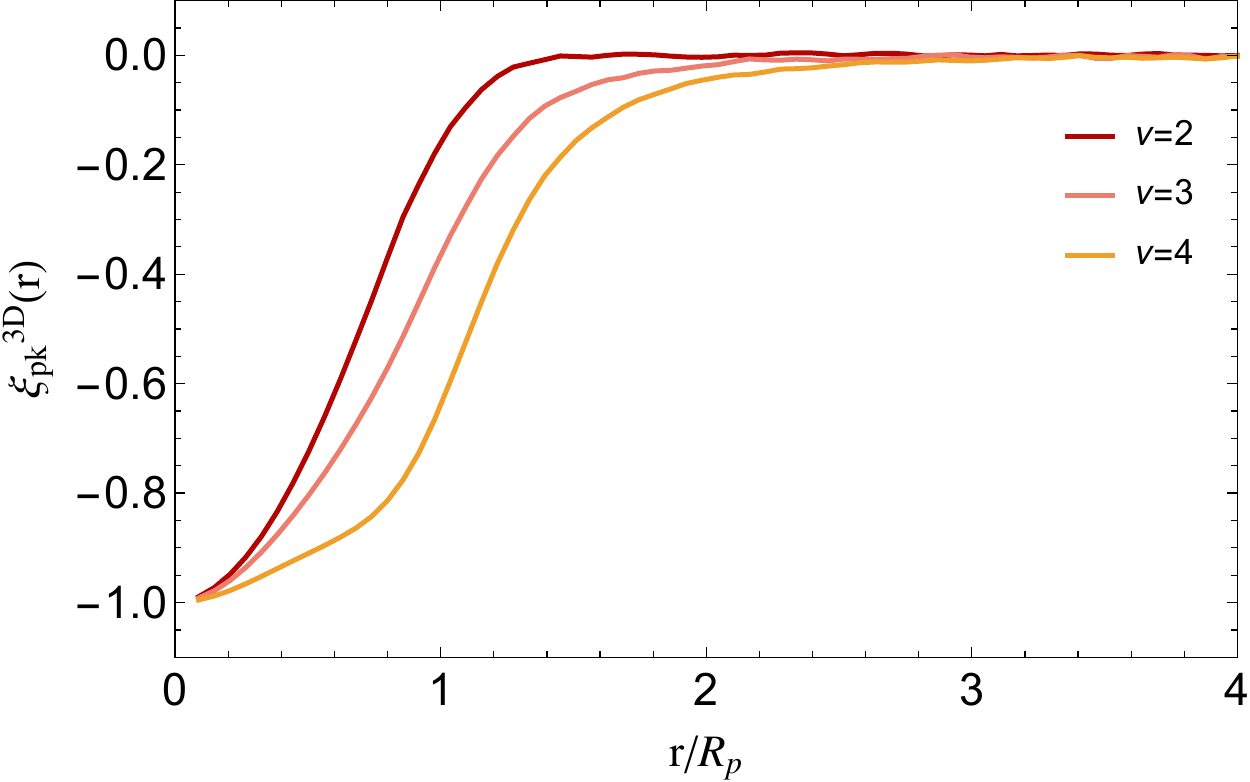} \hskip 1cm
\includegraphics[width= 0.95\columnwidth]{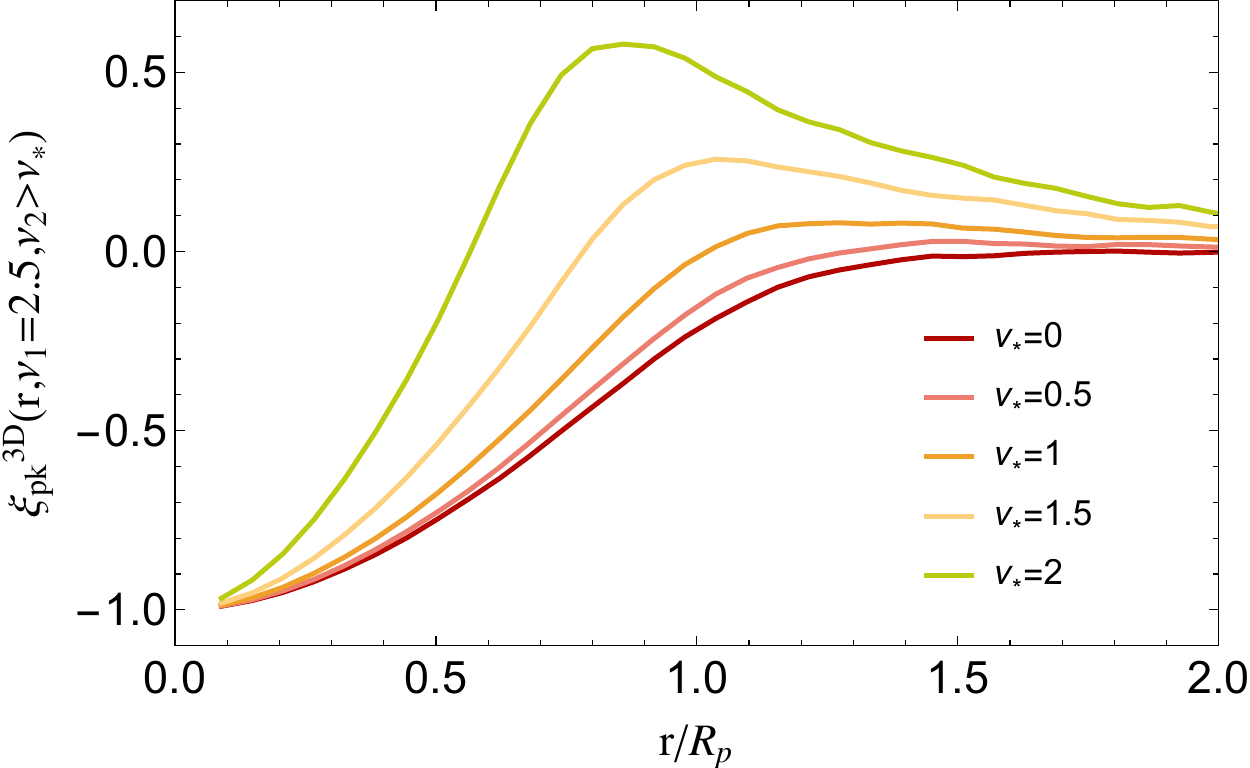} 
\caption{
{\sl Top row}: 2D GRF with $n_{s}=-1$, {\sl Bottom row}: 3D LCDM GRF smoothed on 5 Mpc$/h$.
{\sl Right-hand panels}: peak-peak correlation function when
the central peak has a specified height (from 2 to 4 as labeled) while
the height of its neighbour is unrestricted.
The separation is in units of the mean distance between peaks $   R_p$.
{\sl Left-hand panels}: correlation function of a peak with fixed height
$\nu_{c}=2.5$ (rare peak) and the second peak with
the height exceeding $\nu_{\star}$, for various thresholds
$\nu_{\star}$ between 0 and 2.
}
\label{fig:maxcorr}
\end{center}
\end{figure*}

For peak-peak correlations, 
the fundamental result of 
\citet{Kaiser1984} showed that high-density peaks at large separations
are positively correlated in excess of the mean correlation of the field (i.e positively biased).
A more detailed study of \cite{Baldauf16} %and Baldauf et al. (in prep.) in 1 and 3 dimensions 
demonstrated
that the peaks are anticorrelated (i.e avoid each other) at small separations
$R \simlt   R_p$, while an enhancement of the correlation between
pairs of peaks at
large separations is only pronounced if both peaks are of similar, and
high, height $\nu$. A representation of these results is given in
Figure~\ref{fig:maxcorr}.

We shall focus on the configurations where the central peak is rare (i.e
has high $\nu_c$)
while its neighbours have a distribution of heights $\nu_n$
above some threshold $\nu_*$, $\nu_n > \nu_*$. 
Figure~\ref{fig:maxcorr} shows that such a peak of GRF has a 
statistically well-defined exclusion neighbourhood where the probability
of finding another peak is suppressed.
In this paper, we use the extent of the exclusion region to define
the size of a peak-patch around the peak, $R_{\rm pp}$. 
When we consider only high neighbours above
the threshold (as displayed on the right panels of Figure~\ref{fig:maxcorr}),
we find that the high peaks demonstrate a sense of a 'first layer' of
neighbours, the correlation function having a pronounced
maximum therefore showing enhanced probability for neighbours to be at a particular
distance. It is natural to identify the peak patch radius $R_{\rm pp}(\nu_c,\nu_*)$
as the position where the peak-peak correlation function $\xi_{\rm pk}$ shows a maximum i.e when $ \xi_{\rm pk}'(r=R_{\rm pp}) = 0$.
From the point of view of the connectivity, it is these first neighbours that
the filament bridges will connect to, and the length of such filaments will
typically be of the order of $ R_{\rm pp}(\nu_c,\nu_*)$. 
When we study $R_{\rm pp}(\nu_c,\nu_*)$ as
function of $\nu_*$ at fixed $\nu_c$, we find that the distance to the next
peak decreases as $\nu_*$ increases to reach $\nu_c$, 
which is the essence of Kaiser bias\footnote{This trend reverses
with further increase of $\nu_* > \nu_c$ when
$R_{\rm pp}$ starts to increase, since the distant, now higher, peak begins to
dictate how far its neighbours can be.}.

The left-hand panels in Figure~\ref{fig:maxcorr} describe the case when
we consider neighbouring peaks of all heights.
The correlation function demonstrates that the extend
of the exclusion region $R_{\rm pp}(\nu_c)$ increases with the height
$\nu_c$ of the central peak. The definition for the exact boundary of
the patch in this case is somewhat ambiguous since the correlation
function does not exhibit a well-defined maximum and no clear
preferred position of the first neighbours is present.
Still, it is clear from our numerical results that 
the height dependence of exclusion radius for rare peaks, $\nu_c > 2$,
is roughly linear, with slope  $\sim 0.2$, i.e., 
$R_{\rm pp} \approx R_0 + \nu/5$ (in units of $R_p$),
both in 2D and in 3D (in particular one can track the width of the exclusion
zone at half-min level of the correlation $\xi_{\rm pk}=-0.5$). If we
calibrate this relation at $\nu=2$ by taking as the boundary the radius
where the correlation turns over to zero (i.e when its curvature is maximal), we have $R_{\rm pp} \approx 1.1 + \nu/5$
in 2D 
and $R_{\rm pp} \approx 0.9 + \nu/5$ in 3D. The accuracy of our correlation
computations does not warrant higher precision than $\sim 0.03$ in these relations.

\begin{figure*}
\begin{center}
\includegraphics[width= 0.95\columnwidth]{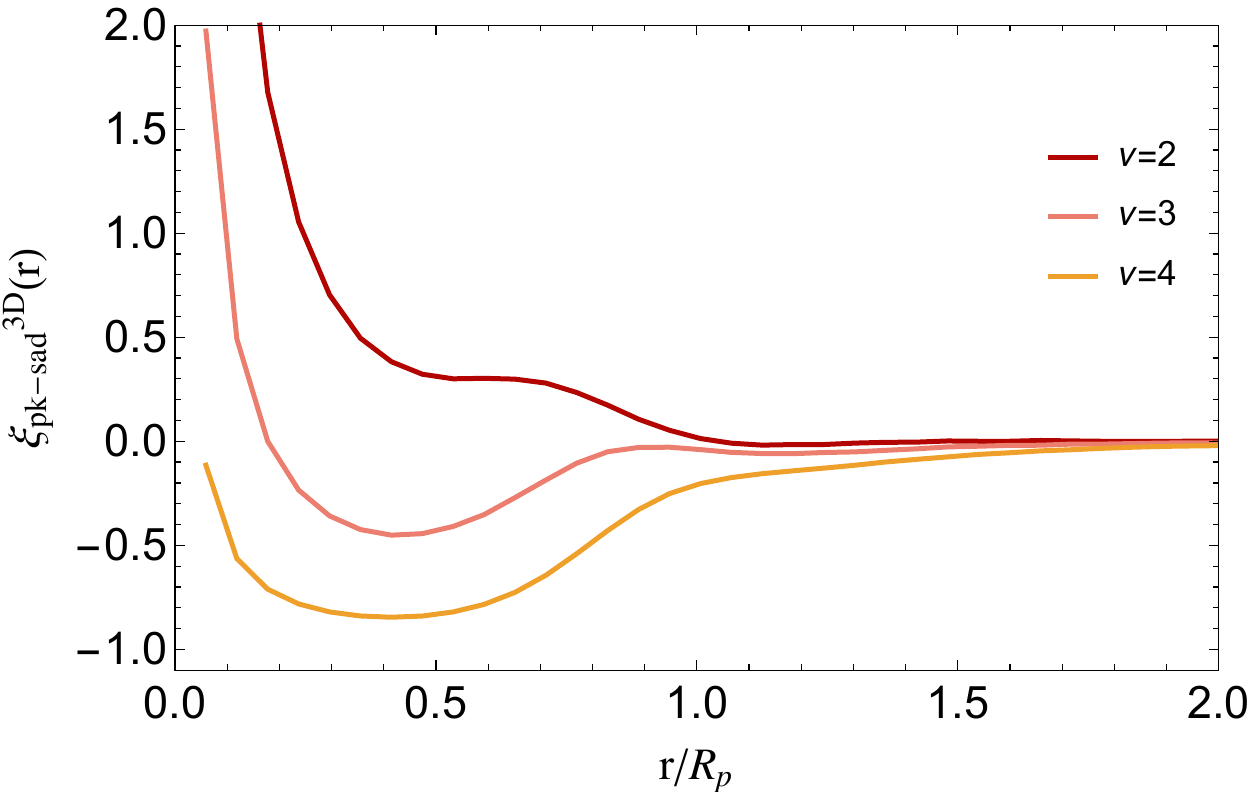} 
\hskip 1cm
\includegraphics[width= 0.95\columnwidth]{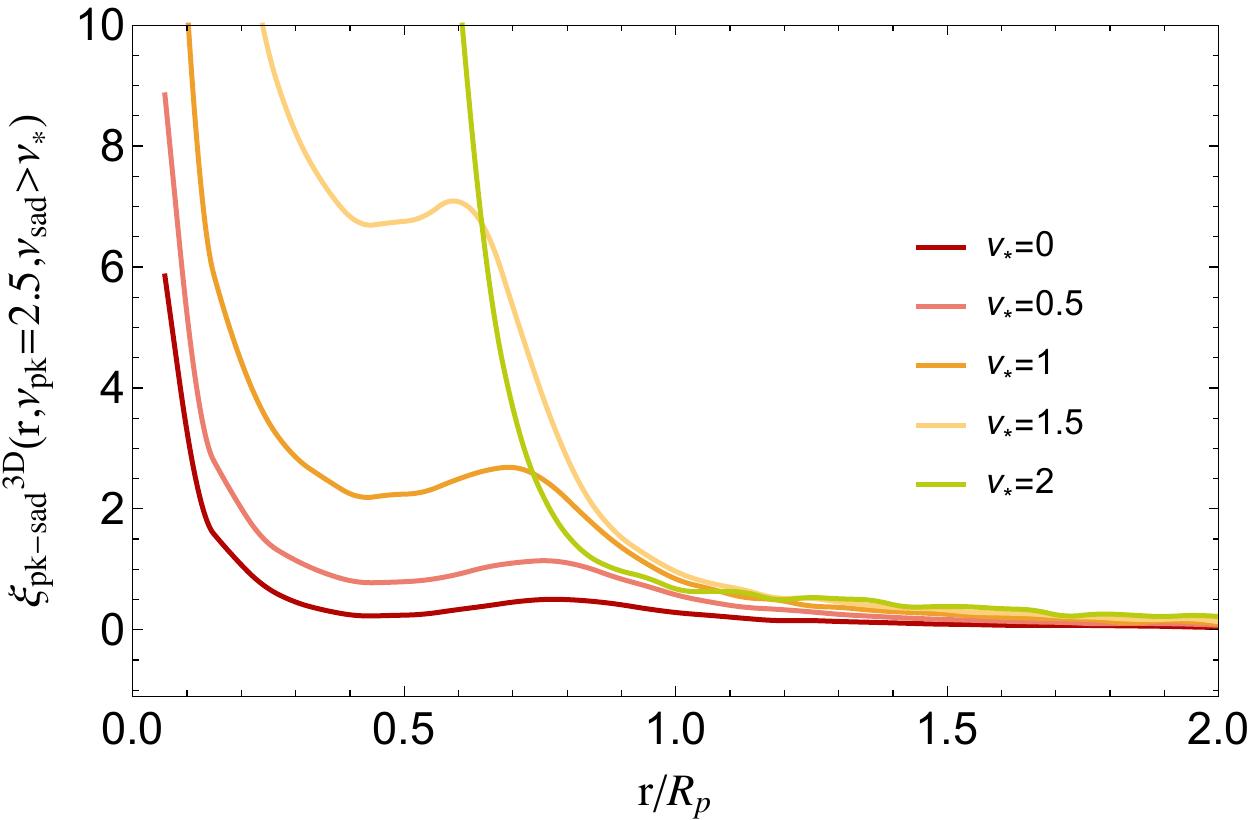}
\caption{
Same as Figure~\ref{fig:maxcorr} when the distant critical point is a filament
type saddle point.}
\label{fig:xisad}
\end{center}
\end{figure*}

Figure~\ref{fig:xisad} shows the correlations between a central peak and
its neighbouring saddle points, $\xi_{\rm pk\!-\!sad}$, computed in the same way
as the peak-peak correlations (only 3D plots are presented). 
We will need these results later in the 
paper when we will count the filamentary saddle points in the vicinity
of a peak as a handle on peak connectivity. For now, we note that
the presence of a high peak affects the density of saddles only at
$r < R_p$. Within $R_p$ a high peak attracts
saddles which are also high (even more so when the heights of peak and saddle get closer) but repels the lower ones.  This means that when all saddles are counted (left panel),
a very high peak ($\nu > 2.5-3$) 
predominantly repels the saddles (since most saddles will be
notably lower than the peak in this case) from their vicinity,
but less prominent peaks
($\nu \sim 2$) have actually an enhanced density of saddles near them.
For $r > R_p$ the density of saddles becomes close to the mean value.

\subsection{Morse theory and the skeleton picture}
 \label{sec:Morse}
 
 In oder to compute  connectivity of the cosmic web we must properly define and extract it as a set of contiguous filamentary branches. 

 \subsubsection{Ridge extractor algorithms}
Over the years, several methods have been developed for this purpose  \citep[see for instance][for a comparison of some of those cosmic web classification schemes]{2018MNRAS.473.1195L}. 
Two distinct hypotheses can be broadly identified at the heart of these approaches. 
One can either use the geometrical information contained in the {\it local} gradient and the Hessian
of the density or potential field \citep[e.g.][]{skel2D, AragonCalvo2007a,AragonCalvo2007b,Hahn2007a,Hahn2007b,Sousbie2008a,Sousbie2008b,Forero-Romero2009,Bond2010a,Bond2010b}, or the topology and connectivity of the density field 
 using the watershed transform \citep[]{AragonCalvo2010b} and/or  Morse theory \citep[e.g.][]{Colombi2000, Sousbie2008a,Sousbie2011a}. 
In the watershed  category, 
\cite{SCP} presented for instance
a method to compute
 the full hierarchy of the critical subsets of  a given
density field. 

\subsubsection{The skeleton of smoothed fields}
Formally, the skeleton of a {\sl continuous} field $\rho$ can be defined   in the context of Morse theory \citep{jost}
as  the set of critical lines (i.e. field lines which go through critical points) connecting the saddle points and the local maxima of that field while departing from the saddle along the first eigenvector of the curvature tensor
\citep{skel2D}.  
Peak (resp. void) patches of the density field can then be defined  \citep[see e.g.][]{SCP}  as the set of points converging to a specific local maximum (resp. minimum) while following  field lines in the direction (resp. the opposite direction) of the gradient. 
 One expects  filaments to  lie at  the intersection of  such void patches. 
   In particular, note that with this definition, one filament linking two maxima together necessarily passes through one and only one critical point which has to be a filament-type saddle point. 
Morse theory formalises this intuitive construct  by segmenting space  in
 respectively the void patches, the walls,
 the filaments and the peaks of the cosmic web. 
   The main advantage of  watershedding algorithms is to provide a fully connected set of critical lines, but two shortcomings remain: 
i)  The Morse  formalism only truly applies to Morse functions  that are smooth and non-degenerate (the Hessian is assumed not to be zero at   critical points). When dealing with discrete fields, one needs to smooth the density field before computing its skeleton and therefore  introduce a smoothing length.
ii) More dramatically, all watershed algorithms implemented on discretised meshes will over-produce filaments, because the segmentation is carried at finite resolution, where the underlying Morse theory is not satisfied.

\begin{figure}
\begin{center}
\includegraphics[width=0.9 \columnwidth]{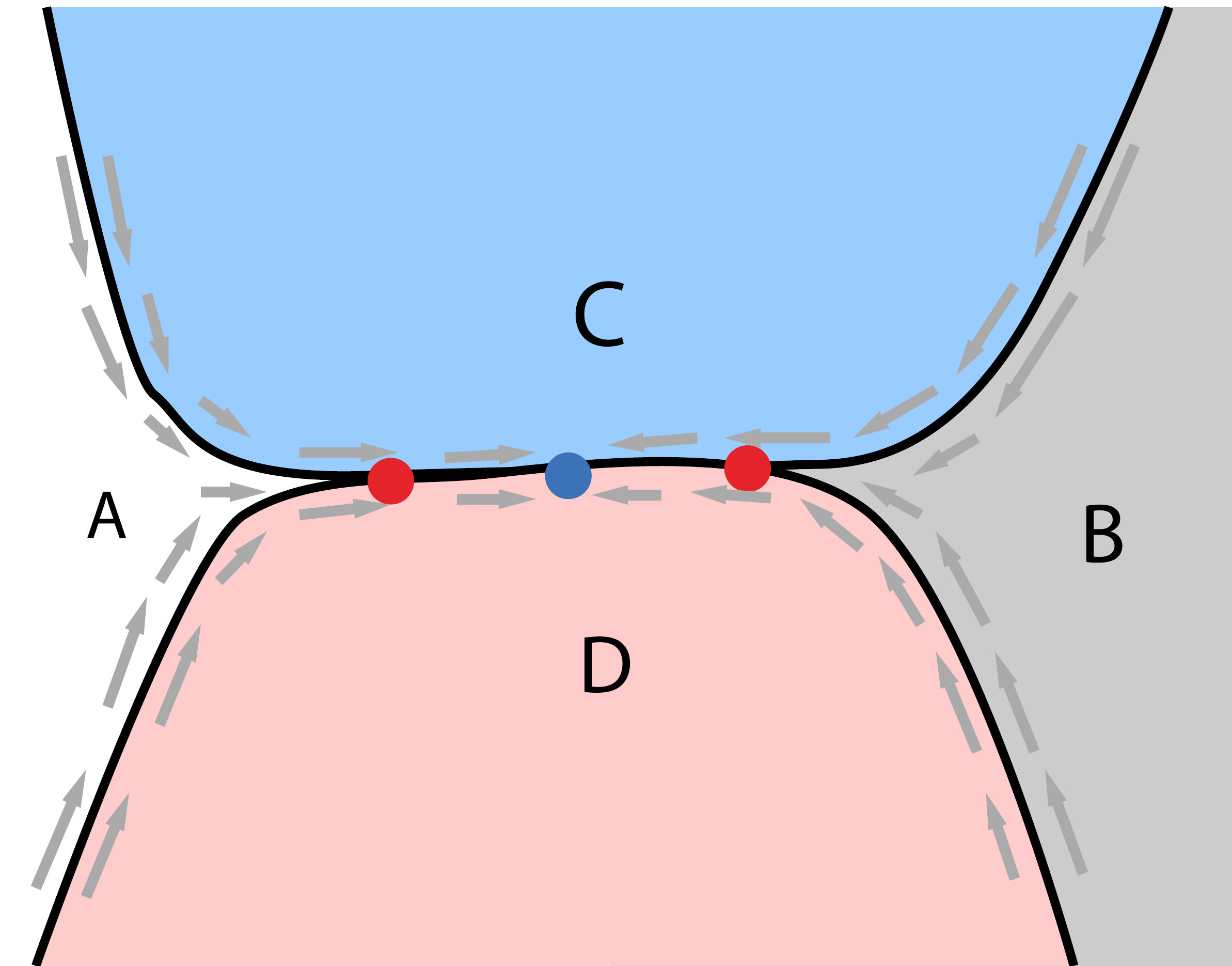} 
\caption{
Illustration of void patches  A-D whose intersections define the filaments. This figure can be interpreted in 2D, or in 3D
as a cross section perpendicular to the axis of the blue filament.  In the former case,
 the bifurcation points in red where two critical lines seem to merge away from the critical point in blue is  
 a resolution driven artifact \citep{pogo09}. In the latter case,
bifurcation lines in red are also artefact due to the resolution and would wrongly be identified as filaments
\citep[resp. at the intersection of voids A-C-D and B-C-D, see][for details]{Sousbie2011a}.  
}
\label{fig:bifurcation}
\end{center}
\end{figure}

These unexpected shortcomings are of prime importance when measuring cosmic connectivity as they  correspond to the appearance  of bifurcation points that need to be identified and properly accounted for. 
Indeed, when the field is constant over an extended region -- a generic situation when cloud-in-cell sampling is applied  on a grid -- even after smoothing(!), the Morse condition is not satisfied anymore and there can be an ambiguity on the exact location of the filamentary segments. This situation is ubiquitous as void patches cannot all be convex: typically some of them will be squashed by their stronger neighbours, hence their boundaries will be almost tangent until reaching the filament. At finite resolution,  the measured intersection line  occurs before the real location of the filament.
This  can be seen on the cross section shown in Figure~\ref{fig:bifurcation}. In 2D, it will occur on  locations different from minima, saddles and maxima, where an under-resolved critical line  splits in two (whereas Morse theory states that this can happen only at critical points). In 3D, it leads to pairs of spurious one-dimensional lines, wrongly identified as filaments by the watershedding algorithm. Those  were named bifurcation points and lines resp. by  \cite{pogo09}. 

\subsubsection{Discrete tracers and topological persistence}

In order to resolve the   above mentioned issues  the Discrete Persistent Structure Extractor  algorithm \cite[{\tt DISPERSE}][]{Sousbie2011a}\footnote{ The  code {\tt DISPERSE}  is publicly available \url{http://www.iap.fr/users/sousbie/disperse.html}} implementing {\sl discrete} Morse theory  \citep{forman} was developed.
This  geometric three-dimensional ridge finder  allows for a scale- and parameter-free topologically consistent extraction of all different components of  the fields. It can be applied either directly to 
the underlying DTFE density of points\footnote{This paper will restrict itself to regular cubic meshes, but all results apply directly to discrete point-like sample via DTFE \citep[see, e.g. ][for examples of implementation on observed surveys]{kraljic18,laigle+17,Malavasi2016b}. }  
or to a  regular  mesh,    from which the discrete Morse-Smale complex of the density function is computed 
   (by assigning a height to all simplices of the complex: vertices, edges, faces, volumes, etc). 
The grid  is  partitioned according to the discrete gradient flow of density into an ensemble of critical sets -- volumes, surfaces, curves and points corresponding to the voids, walls, filaments and clusters within the cosmic web,  the so-called ascending 3- 2- 1- and 0- manifolds of the discrete flow, respectively. 
The code uses  persistence ratio (the relative height of connected critical points as a measure of the significance of their topological connections) to filter out filaments dominated by the noise.  

Since the construct satisfies discrete Morse theory \citep{robins},  all
 critical lines only split at critical points as they should, but
multiple critical lines may overlap one another up to some bifurcation point
 (whose position depends on the sampling and smoothing). Filaments are not counted twice, only two branches 
 emanate  from each saddle point. 
The one minor drawback of the algorithm is that there is as-of-today no complete theory relating persistence 
to smoothing, so one must rely on calibration when comparing measurements   to 
predictions for  random fields smoothed on a given scale (see Appendix~\ref{sec:persistence-cuts}).
From a practical perspective, {\tt DISPERSE} keeps track via a structure stored
 at each critical points of how many saddle points are connected to it.
 It also stores the segments where bifurcation occurs.

%=====================
\section{Connectivity of GRF}\label{sec:GRF}
%=====================

Let us now turn to the theory of the connectivity of GRF as measured in GRF realisations or predicted from first principles.

%%%%%%%%%%%%%%%%%%%%%%%%%%%
\subsection{Connectivity of peakpatches}\label{connex}
%%%%%%%%%%%%%%%%%%%%%%%%%%%

\begin{figure*}
\includegraphics[width= 0.95\columnwidth]{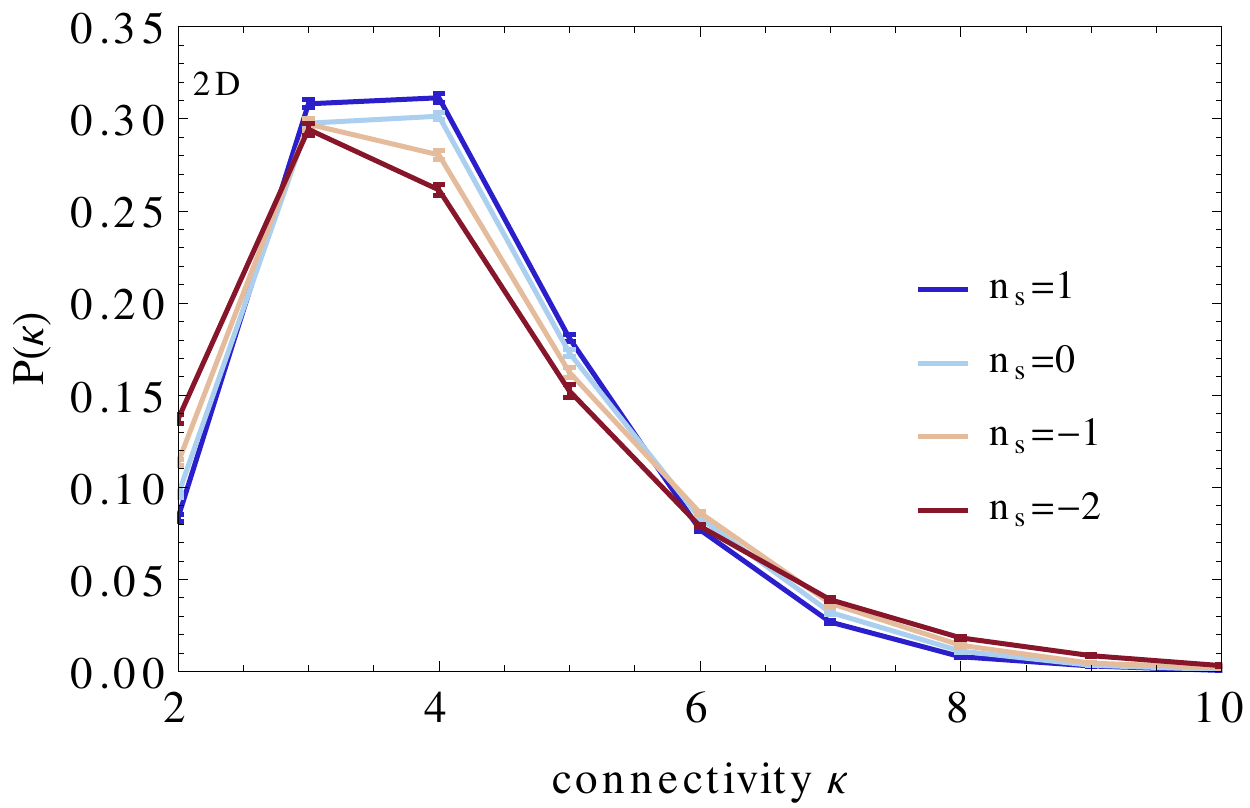} 
\includegraphics[width= 0.95\columnwidth]{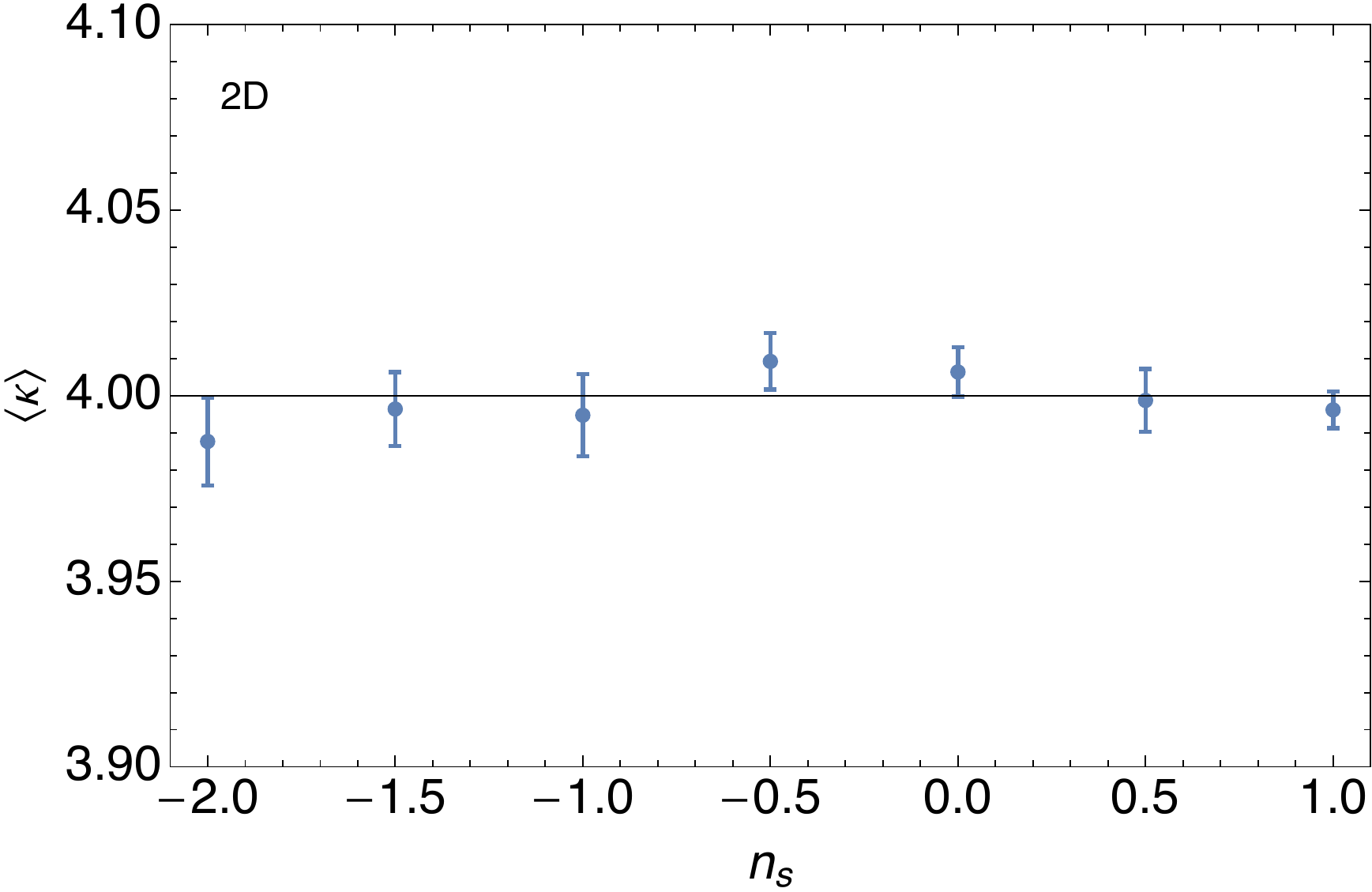} 
\caption{{\sl Left-hand panel:} PDF of the number of connectors in 2D evaluated from 10 GRF maps with spectral index from +1 (blue) to -2 (red). Error bars represent the error on the mean. {\sl Right-hand panel: } corresponding mean connectivity as a function of the spectral index. The mean number of connectors seems to be independent on the power spectrum and equal to four.}
\label{fig:connect2D}
\end{figure*}

\subsubsection{Numerical simulations}
%------------------------------------------------------------------------------------
To get two dimensional maps, we first generate 100 realisations of a GRF with power-law power spectrum $P(k)\propto k^{n_{s}}$ on a $2048^{2}$ grid. We then smooth those maps with a Gaussian kernel on 8 pixels. The skeleton of each map is extracted using {\tt DISPERSE} with a cut in persistence that depends on the power spectrum and is set so that the number of peaks in the maps matches the expected number of peaks $4096(n_{s}+4)/\sqrt{3\pi^{2}}$ with better than 0.5\% accuracy.
Adopted persistence cuts are given in Table~\ref{tab:persistencecuts}.
The spurious low-persistence peaks are due to the sampling of the Gaussian
field as illustrated in Figure~\ref{fig:cut}.

The skeleton obtained with {\tt DISPERSE} is further smoothed following a prescription 
described in \cite{SCP} which ensures that the number and positions of extrema are
fixed under the smoothing operation. Thus, smoothing only straightens
the skeleton segments and shifts bifurcation points 
but preserves topology and therefore connectivity.

A similar procedure is adopted in the three dimensional case for which we generate 20 realisations of a GRF on a $256^{3}$ grid with power-law power 
spectrum $P(k)\propto k^{n_{s}}$ and further smoothed on 4 pixels. We also use the total number of 3D peaks to choose the persistence cut. 
In this case, the  persistence cuts are found to be $p_{\rm min}=\{4.5,6,9,12\}\sigma_{0}/1000$ for $n_{s}=\{-3,-2,-1,0\}$. 
Note that those runs are generated for only one reference spectral index (-1 in 2D, -2 in 3D). For other spectral indices we will use only 10 runs because we restrict ourselves to marginals in those cases and statistics is therefore sufficient.
Once the skeletons are computed,   the number of saddle points connected to each peak are counted and  an histogram is computed  in order to get the PDF of the connectivity for different power spectra. We also attach to each node the value of its height so that the joint statistics of nodes' connectivity and height can be investigated.

\subsubsection{Peak Connectivity of 2D GRF}

The statistics of connectivity in 2D is shown on Figure~\ref{fig:connect2D}.
We first notice on the right-hand panel that the mean number of connectors found is four with no detectable dependence on the slope of the power spectrum. On average, cosmic web's nodes therefore follow a square lattice. This result is in agreement with expectations from extrema counts as will be described below. The left-hand panel displays the full PDF of 2D connectivity. Most peaks have a connectivity between 2 and 6 but a long tail pervades with (rare) peaks having a connectivity as large as 10. The shape of the power spectrum also matters with more negative spectrum being more skewed and positive spectrum being more symmetric around the mean.

\begin{figure*}
\includegraphics[width= 0.95\columnwidth]{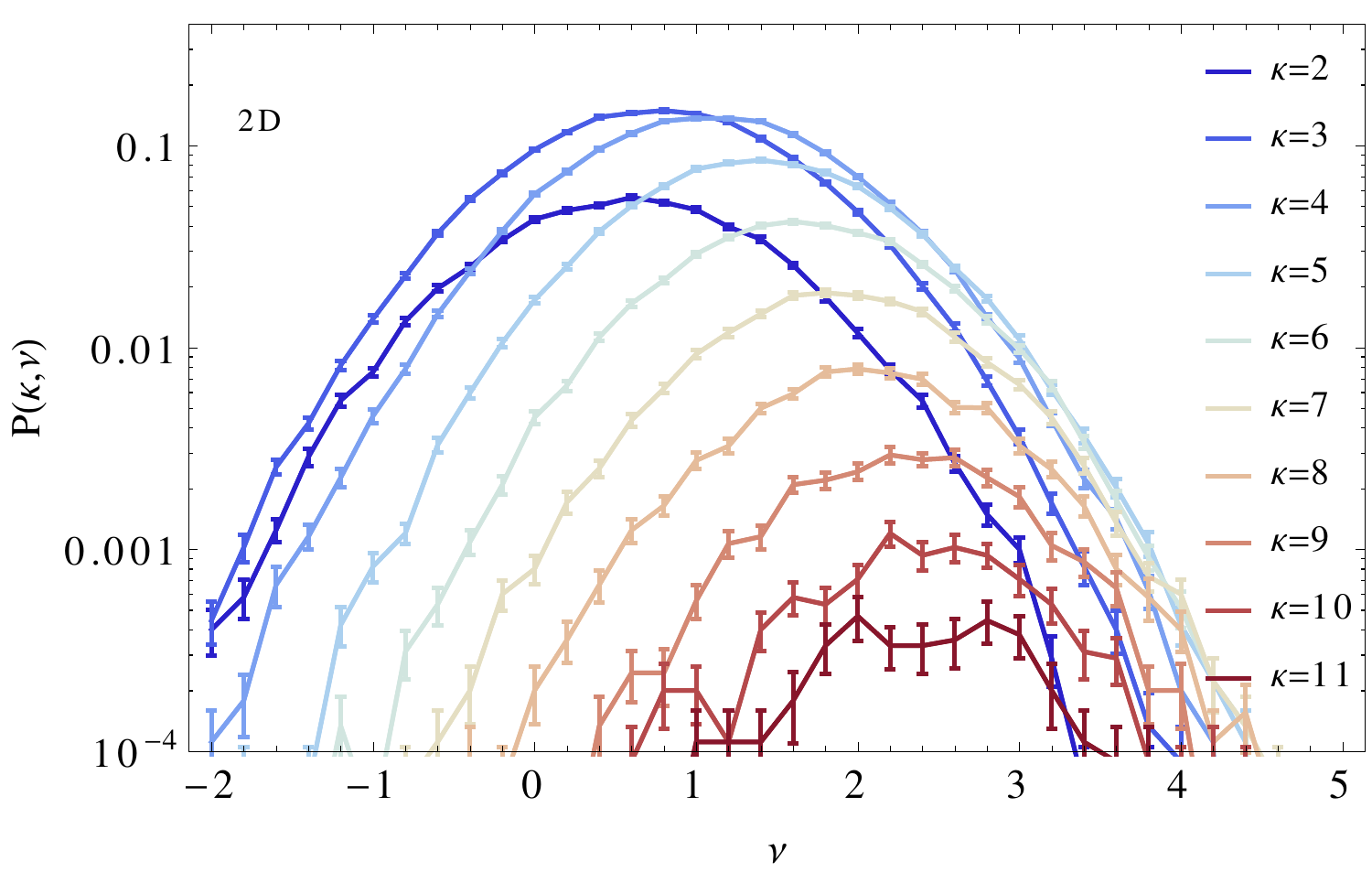} \hskip 1cm
\includegraphics[width= 0.97\columnwidth]{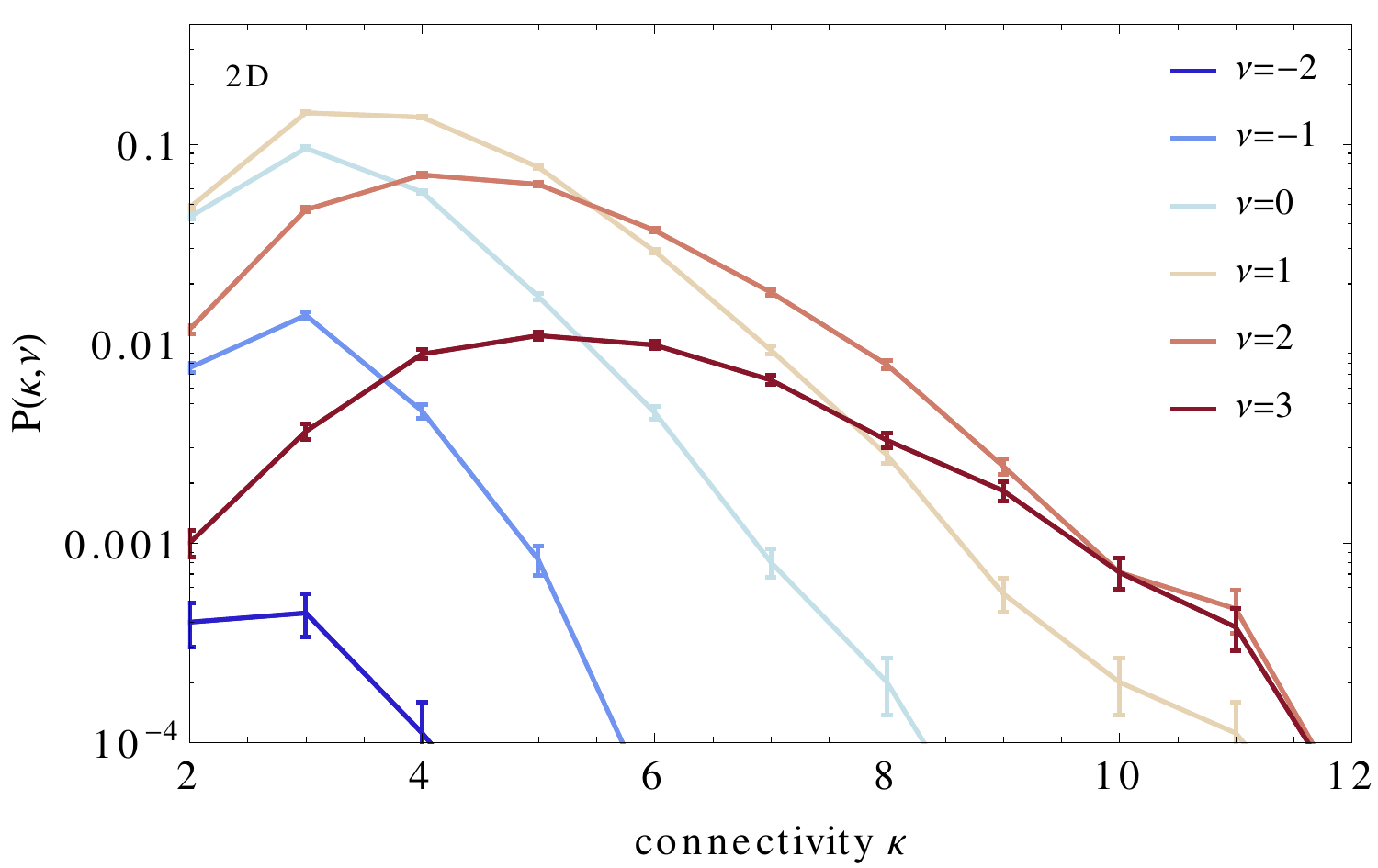} 
\caption{f
Joint PDF of the number of connectors and the peak height estimated from 100 realisations of a 2D GRF with spectral index $n_{s}=-1$.
Slices of the PDF at fixed connectivity and fixed height are respectively displayed on the {\sl left and right-hand panels}.}
\label{fig:JPDF2D}
\end{figure*}

\begin{figure*}
\includegraphics[width= 0.95\columnwidth]{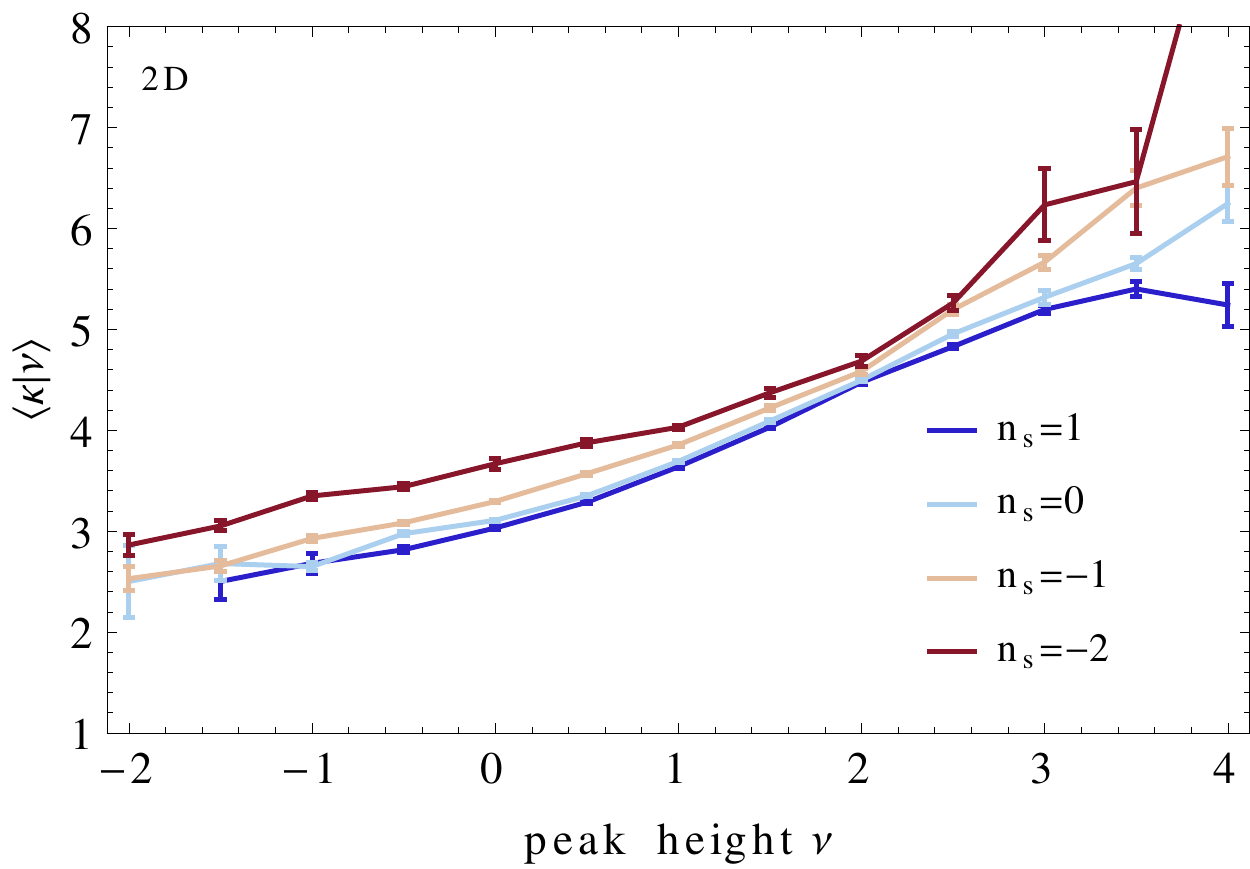} \hskip 1cm
\includegraphics[width= 0.97\columnwidth]{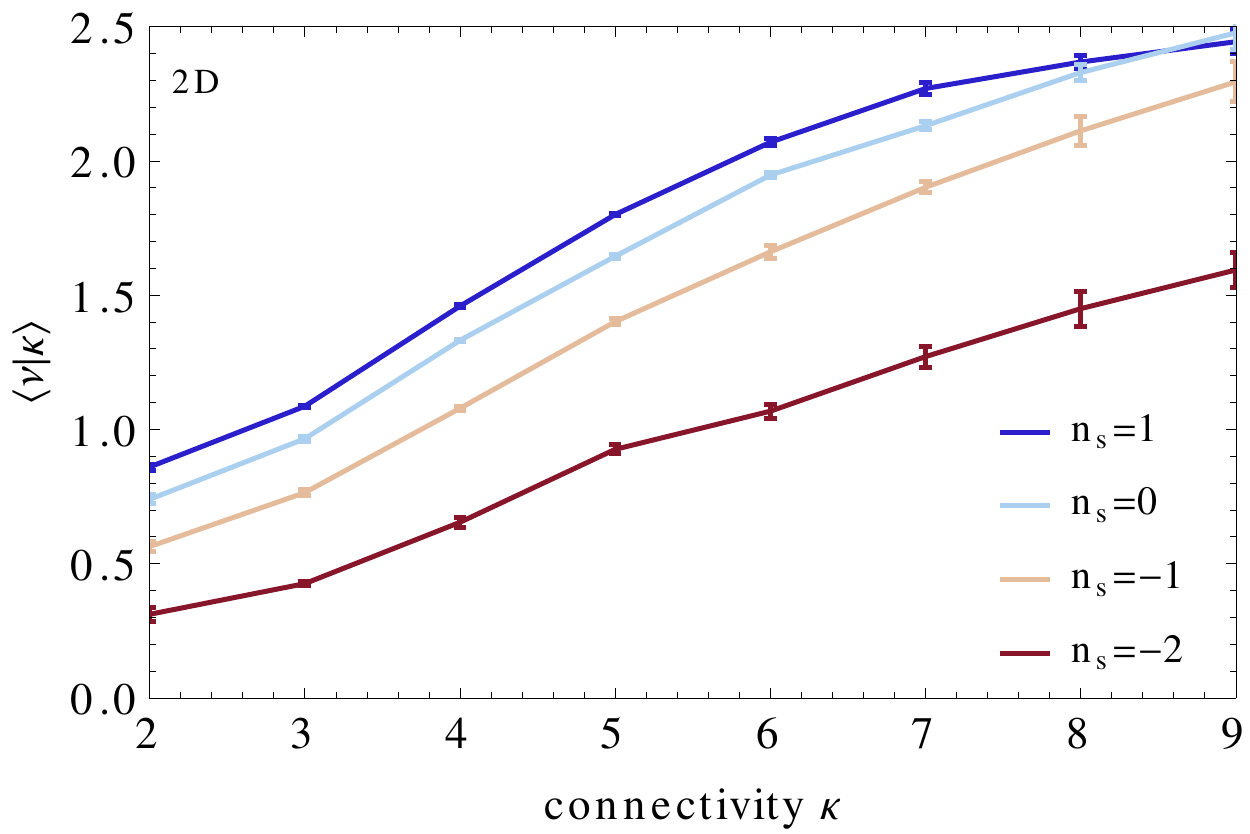} 
\caption{
{\sl Left-hand panel:} Mean number of connectors as a function of the peak height in 2D evaluated from 10 GRF maps with different spectral indices as labeled. {\sl Right-hand panel:} same as the left-hand panel for the mean peak height as a function of the number of connectors.}
\label{fig:connect2D-nu}
\end{figure*}

We anticipate that the number of connectors should also depend on peak height.
Hence, the joint PDF, ${\cal P}(\kappa,\nu)$, of the peak's number of
connections $\kappa$ and height $\nu$ is also measured. This is the central
result of this paper. Figure~\ref{fig:JPDF2D} displays this joint PDF as sets
of $\nu=const$ and $\kappa=const$ slices for a fixed spectral index $n_s=-1$.
As a complement, Figure~\ref{fig:connect2D-nu} describes the relation between
connectivity and peak height by plotting the conditional average number of
connectors at fixed height, $\langle \kappa | \nu \rangle $, and conversely,
the mean height for a given number of connectors,
$\langle \nu | \kappa \rangle$, this time for several spectral indices.
We find that the rarer the peak, the higher the connectivity.
This was expected since near a high contrast peak all eigenvalues tend to
become equal \citep{PB}. Therefore all incoming
directions become possible.
For low-density peaks, the number of connectors is found to be close to $\kappa=3$ 
rising almost linearly for positive contrasts, with very high peaks 
($\nu\approx 4$) reaching a mean connectivity of about six. Note that $\nu$ is an 
absolute density threshold, so that low-density maxima are predominantly peaks inside larger underdense (i.e void) regions. Thus we find that peaks in voids
have a reduced number of webbing connectors with surrounding structures. 
We note that the correlations between peak's height and connectivity is spectrum dependent.

\subsubsection{Peak Connectivity of 3D GRF}

\begin{figure*}
\includegraphics[width= 0.95\columnwidth]{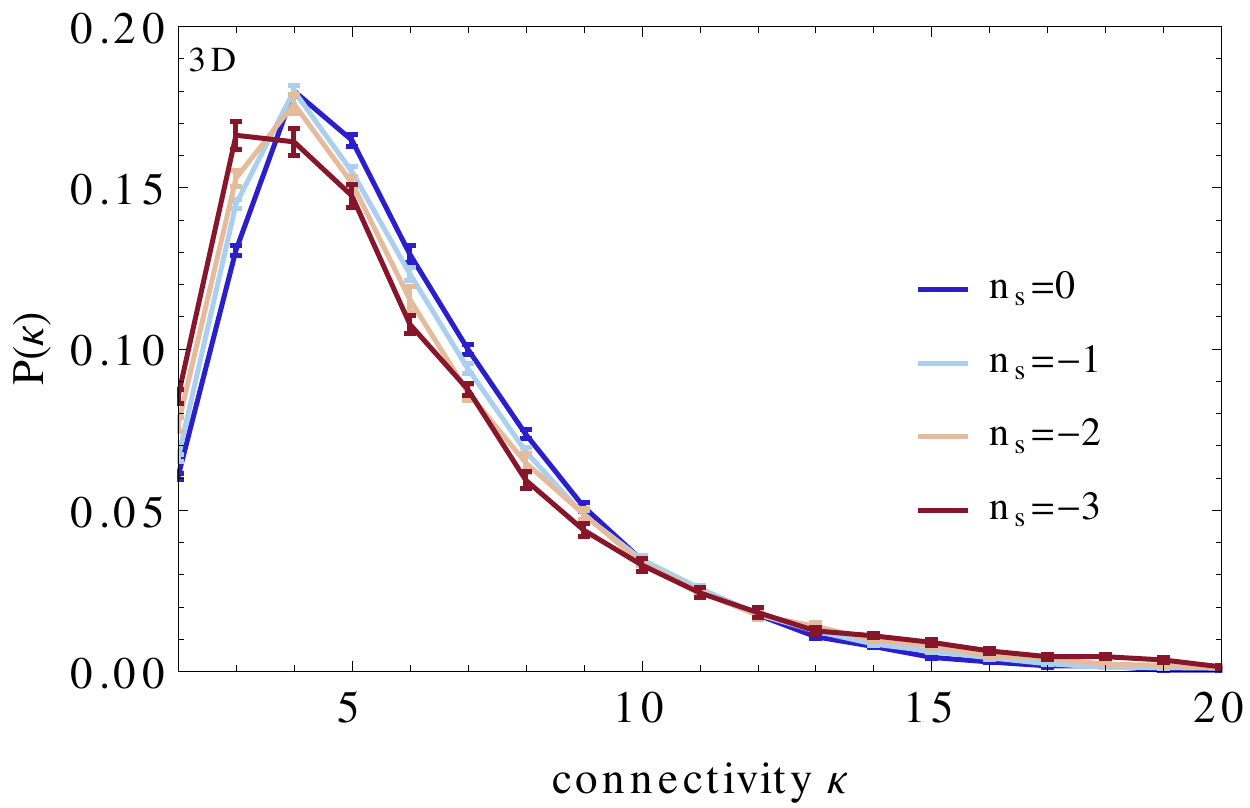} 
\includegraphics[width= 0.95\columnwidth]{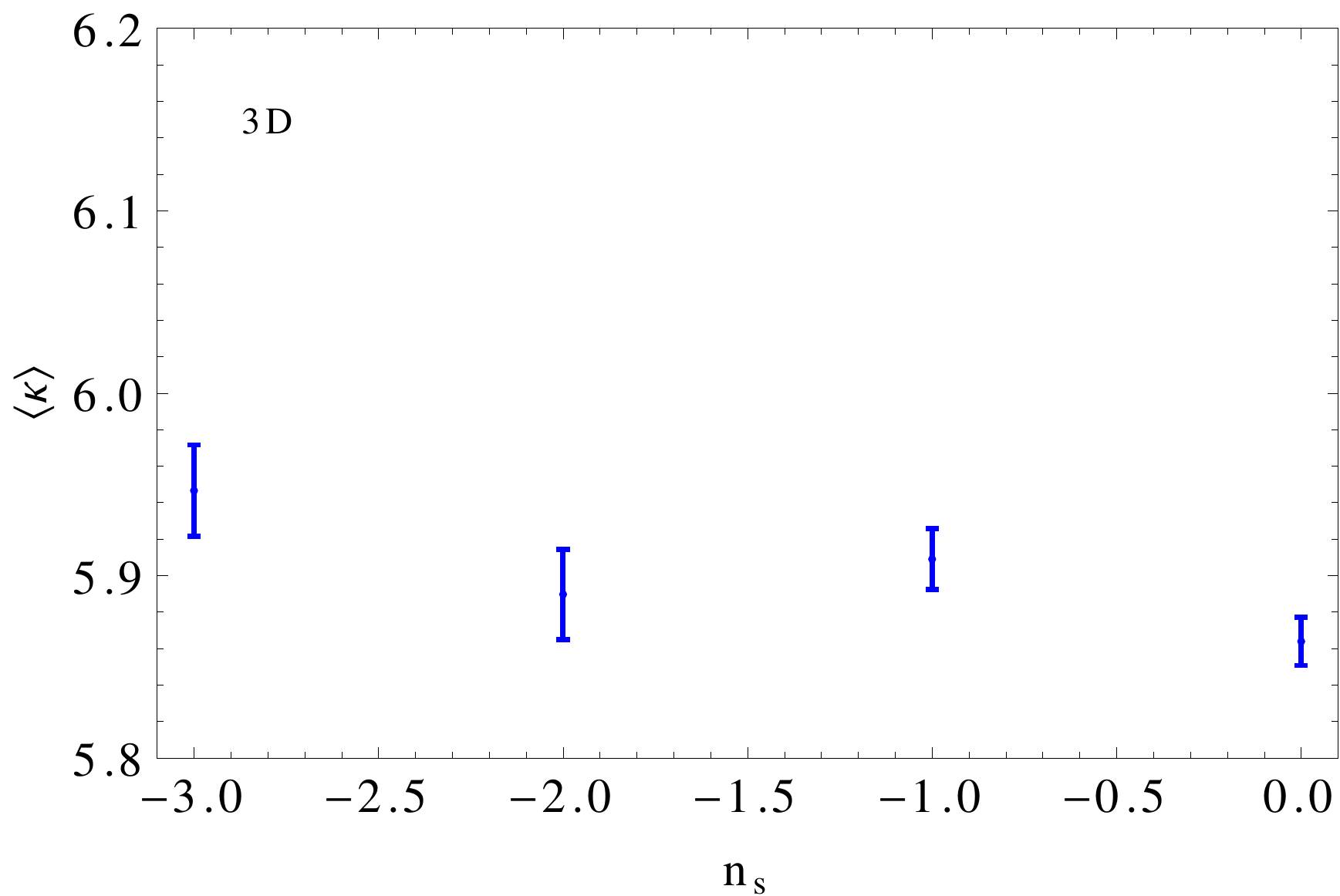} 
\caption{{\sl Left-hand panel: }PDF of the number of connectors in 3D evaluated from 10 GRF maps with spectral index from 0 (blue) to -3 (red). Error bars represent the error on the mean. {\sl Right-hand panel: }corresponding mean connectivity as a function of the spectral index. The mean number of connectors seems to be independent of the power spectrum and equal to four.}
\label{fig:connect3D}
\end{figure*}
\begin{figure*}
\begin{center}
\includegraphics[width= 0.95\columnwidth]{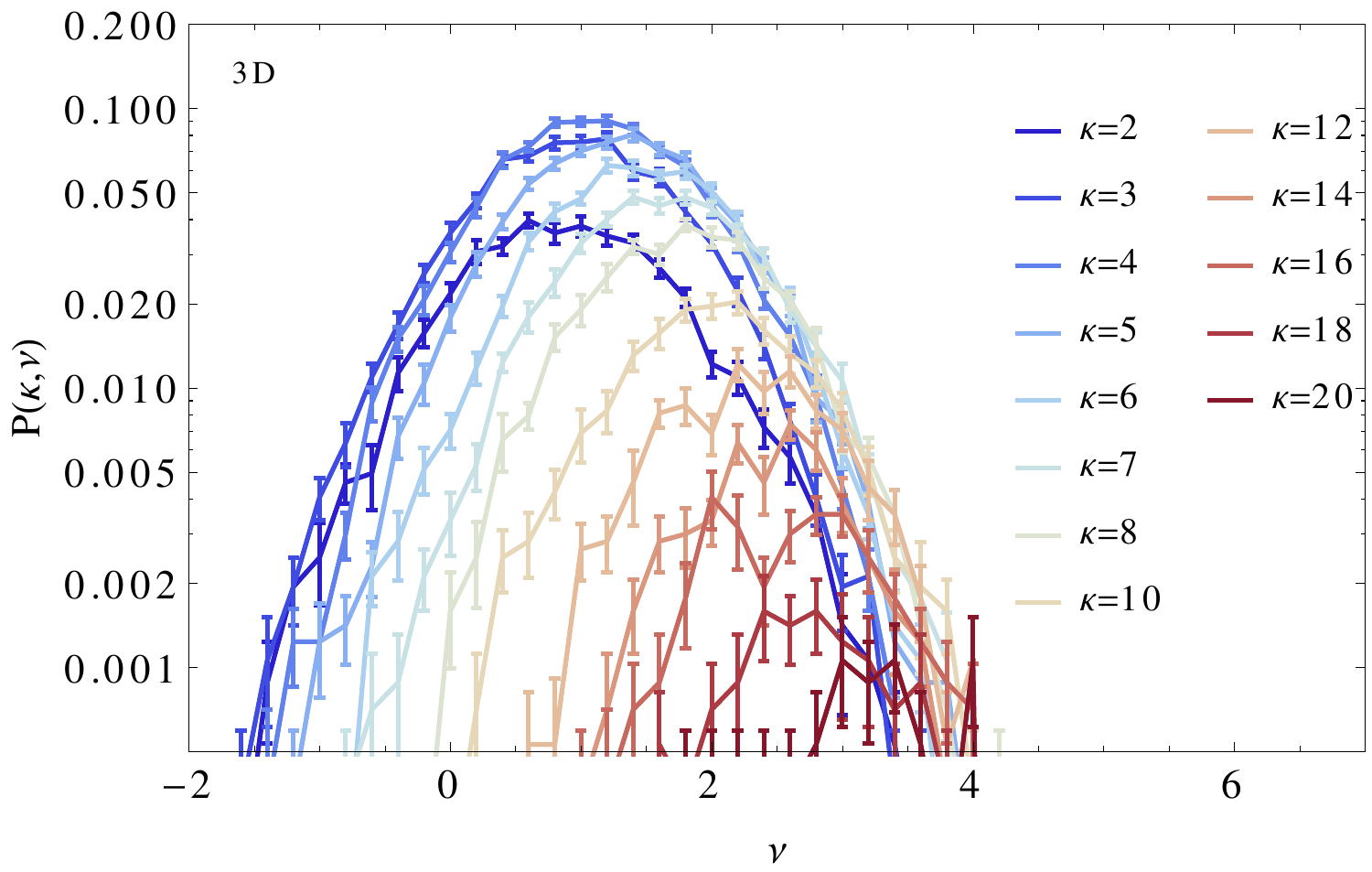} \hskip 1cm
\includegraphics[width= 0.95\columnwidth]{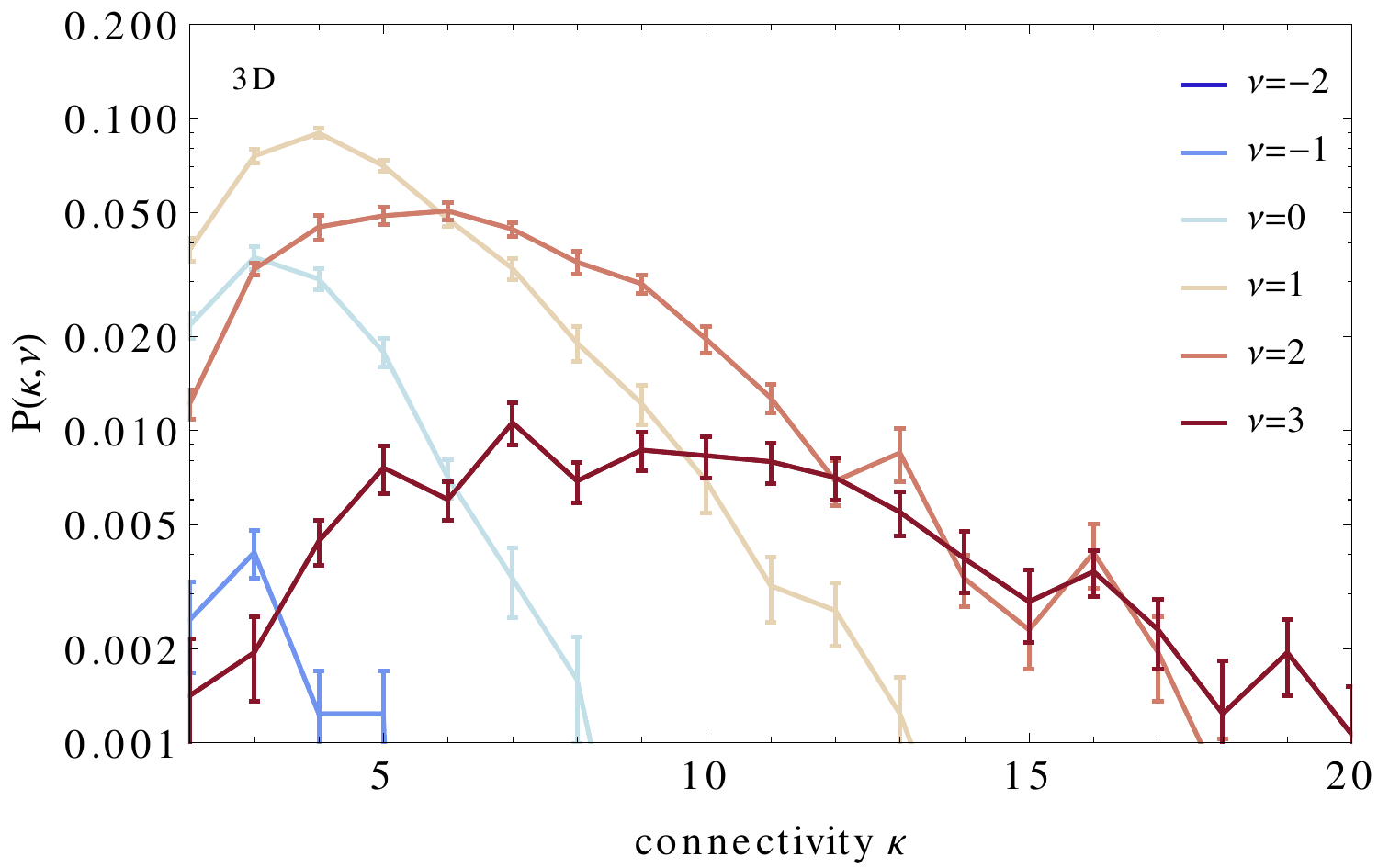}
\caption{
{\sl Left:}  The $\kappa=const$ slices of $P(\kappa,\nu)$ from 20 realizations
of 3D  $\gamma=0.72$ GRF$^{256}_{30}$  field, as functions of contrast $\nu$;
{\sl right:}  The $\nu=const$ slices of $P(\kappa,\nu)$ for the same
ensemble.  
}
\label{fig:GRF-3D-kappa-other}
\end{center}
\end{figure*}
\begin{figure*}
\includegraphics[width= 0.95\columnwidth]{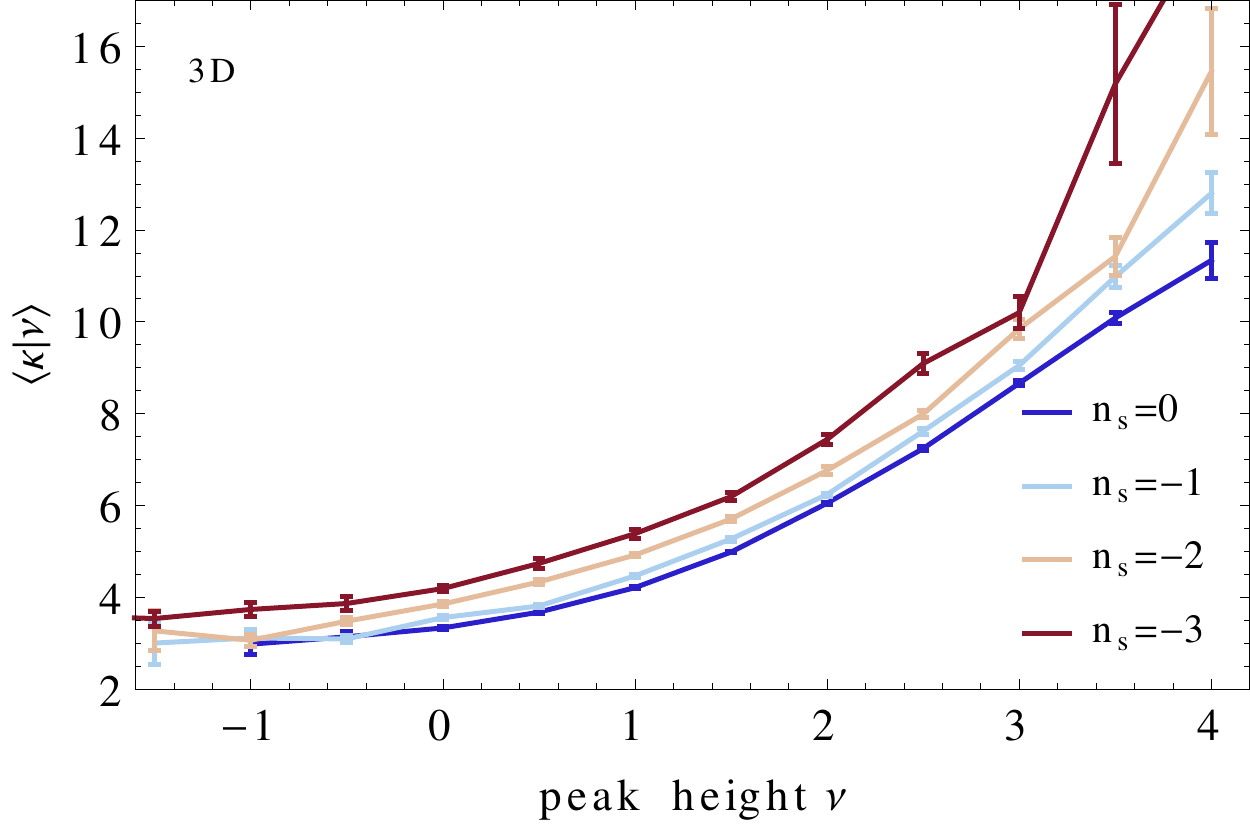} \hskip 1cm
\includegraphics[width= 0.95\columnwidth]{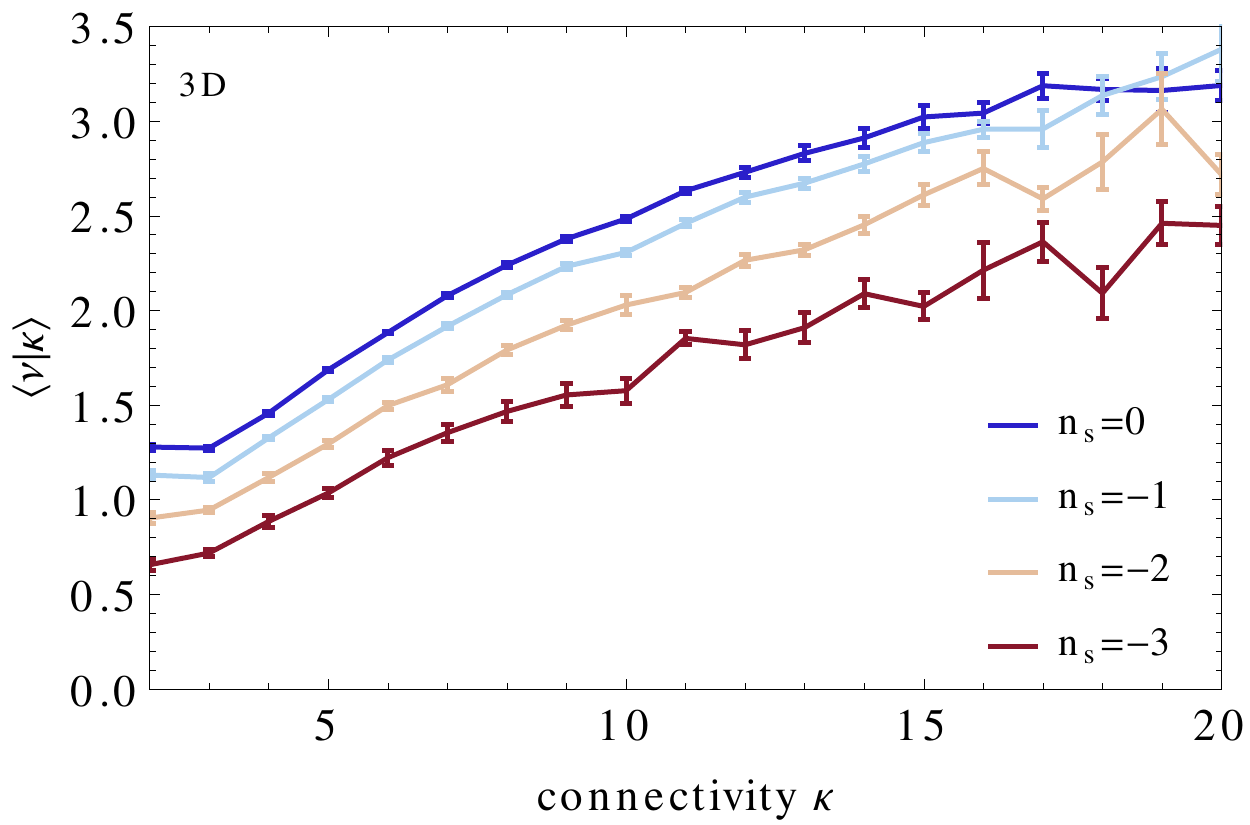} 
\caption{{\sl Left-hand panel: }Mean number of connectors as a function of the peak height in 3D evaluated from 10 GRF maps with different spectral indices as labeled. {\sl Right-hand panel:}  same as the left-hand panel for the mean peak height as a function of the number of connectors.}
\label{fig:connect3D-nu}
\end{figure*}

In three dimensions, we follow the same path as in the 2D case.
The mean and full PDF of 3D connectivity is shown in Figure~\ref{fig:connect3D}. On average, a 3D peak is connected to $\left\langle \kappa\right\rangle\approx 6$ saddle points with no detectable dependance on the spectral index. 
The mean 3D cosmic web is therefore close to a cubic lattice on average.
Neglecting the special case of $n_{s}=-3$, the full statistics of the number of connectors depends only marginally on the shape of the power spectrum, peaking at a connectivity of five and extending to quite high values of the order of twenty for the rarer objects. 
Note however that
%we have not detected the departure from 
%$\left\langle \kappa \right\rangle =6$ in our 3D measurements as
the statistics in our 3D measurements is not high enough 
and we may have some (small but non vanishing) boundary
effects given the small volume of the 3D maps we use in this paper.
It is however computationally expensive to significantly increase the volume
of each map because of the scalability limitations of $\tt DISPERSE$.

Figure~\ref{fig:GRF-3D-kappa-other} then displays the joint PDF of peak's height and connectivity $P(\kappa,\nu)$ and Figure~\ref{fig:connect3D-nu} shows the corresponding marginals namely the mean connectivity given height $\left\langle \kappa|\nu\right\rangle$ and mean height given connectivity $\left\langle \kappa|\nu\right\rangle$.
Similar to the 2D case, it is shown that the mean connectivity increases with peak height from three in underdense regions to more than ten connectors on average for the most massive peaks ($\nu\gtrsim 4$). The interplay between peak's connectivity and height does depend on the slope of the power spectrum: a peak of a given connectivity tends to be systematically higher for high (positive) values of the spectral index and smaller for more red-tilted spectra.

\subsection{Multiplicity of a peak, branches, bifurcations}\label{conmax}
%%%%%%%%%%%%%%%%%%%%%%%%%%%

Up to now we have focussed on the connectivity of a peak patch, defined as the
number of saddle points at the patch boundary through which 
the skeleton in the patch is connected to its neighbours. This definition
is topological and the connecting skeletons lines never formally intersect 
away of the critical points.  However, from the physical perspective
such lines may, and often do, pass parallel to each other over significant
distance, thus representing the vicinity of the same dense ridge.
Indeed, infinitesimally close
to the peak, the field can in general be described as a quadratic surface, thus
(unless degenerate) has one leading eigen-direction of the curvature tensor,
which corresponds to exactly two locally defined ridges,
vicinity of which is tracked
by every skeleton line emanating from the peak. Examples of such situation
can be seen in Figure~\ref{fig:PP-skel}.

To model physical overdense
filaments, one would like to join close 
skeleton lines into a single object, until the 
position that will now be identified with a bifurcation point.
Such procedure, and, thus, the position of bifurcation points
is smoothing scale dependent but can be made robust since the 
divergence of the 
two nearly parallel skeleton lines is usually exponentially quick.
Importantly, we do not count bifurcation points within 
one smoothing radius from the peak, treating the skeleton branches that
diverge each other so immediately as always distinct.
It is exactly them that are counted in the multiplicity of the peak.
This procedure is incorporated into {\tt DISPERSE}, where the implemented smoothing of the
skeleton preserves the positions of extrema (and thus 
global connectivity properties), but merges nearby skeleton segments and 
defines the bifurcation points of such ``physical'' skeleton.

Let us now count the actual number of 
%dense 
filaments incident onto a given
maximum. We call this measure multiplicity of the peak, $\mu$.
Clearly, this local (intra patch) measure
is in fact equal to the number of connecting saddle points minus
the number of bifurcation points within the peak patch
\begin{equation}
\mu=\kappa -n_{\rm bifurcation} ~.
\end{equation}

\begin{figure*}
\begin{center}
\includegraphics[width= 0.95\columnwidth]{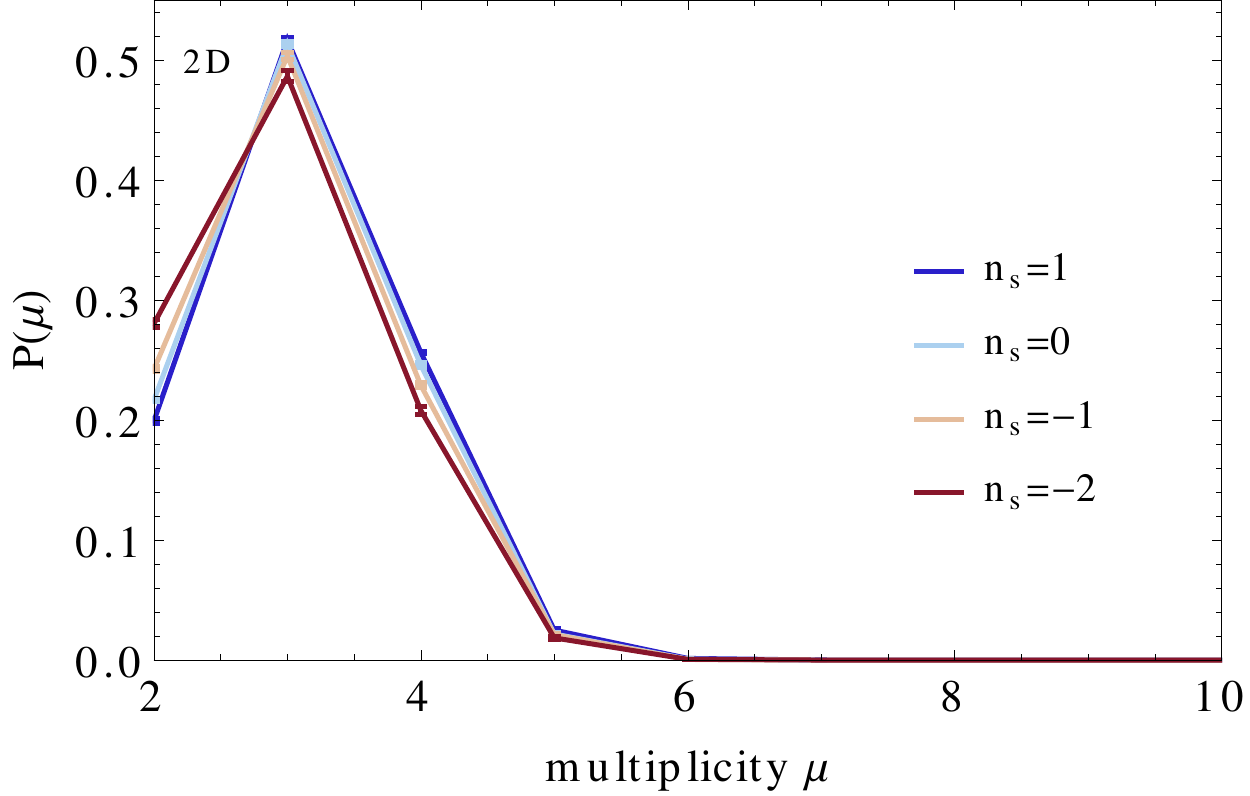} \hskip 1cm
\includegraphics[width= 0.95\columnwidth]{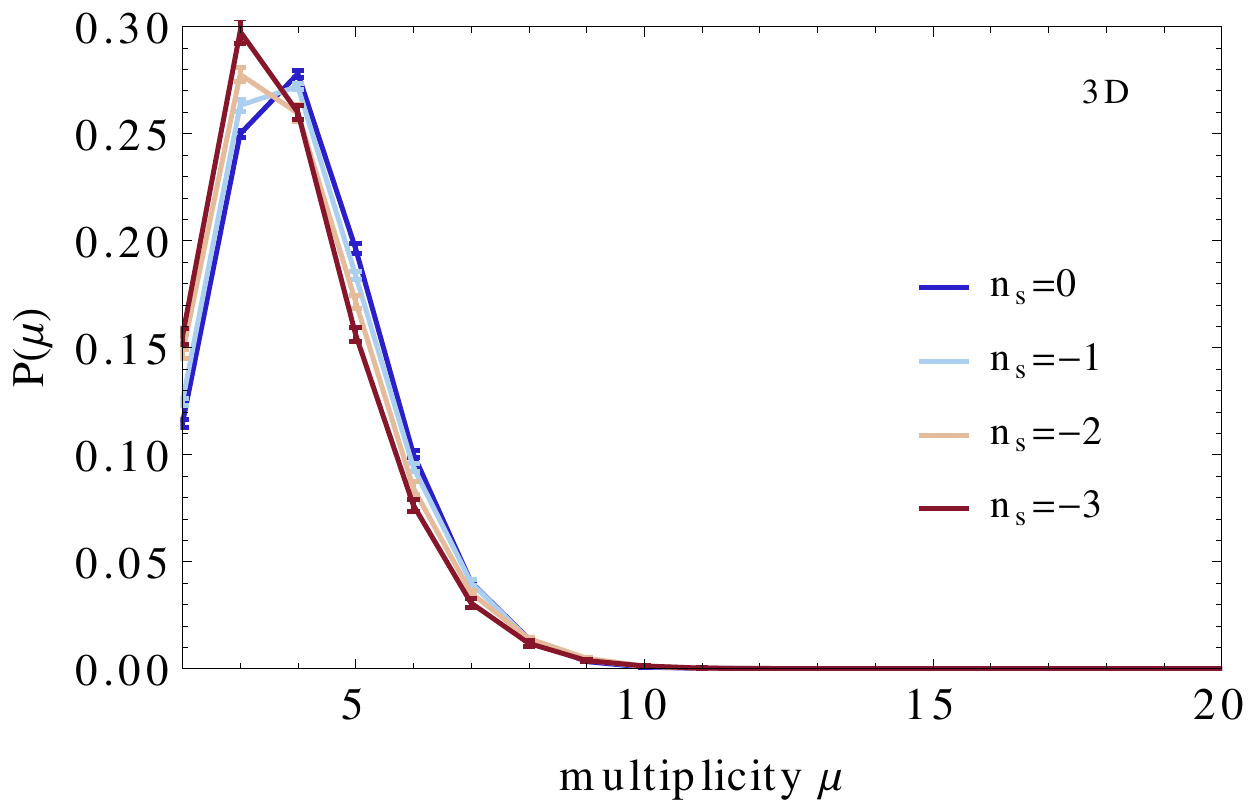} \hskip 1cm
\caption{
The PDF of the multiplicity $\mu$ of peaks in 10 GRF with various spectral indices.
This quantity reflects 
the number of incident skeleton filaments at the node.
{\sl Left-hand panel:} 2D estimate from 10 $2048^{2}$ GRF smoothed on 8 pixels, 
$\langle\mu\rangle \approx 3$ for all spectral slopes.
{\sl Right-hand panel:} Same quantity in 3D for 10  $256^{3}$ smoothed on 4 pixels, here $\langle \mu \rangle \approx 4$.
 }
\label{fig:GRF-2D-mu}
\end{center}
\end{figure*}

Figure~\ref{fig:GRF-2D-mu}
displays the corresponding PDF ${\cal P}(\mu)$ obtained
in our simulations, to be contrasted to Figure~\ref{fig:connect2D}
which represents the same distribution for $\kappa$.
The distribution of multiplicity is almost symmetric, centred  at 
$\langle \mu \rangle\approx3$ in 2D and $\langle \mu \rangle\approx4$ in 3D, and does not extend to numbers as large as for the $\kappa$ distribution.  Indeed, we have observed no peaks with multiplicity
exceeding 6 in 2D and 11 in 3D. A small dependence with the spectral index is found in this case ($\left\langle\mu\right\rangle$ varying from 3.85 to 4.1 for $n_{s}$ between -3 and 0). Hence, typically, four filaments (three in 2D) locally branch out from a peak and later bifurcate in order to connect to eventually 6 neighbouring nodes. This mean picture obviously varies from one node of the cosmic web to the other.

\begin{figure*}
\begin{center}
\includegraphics[width= 0.95\columnwidth]{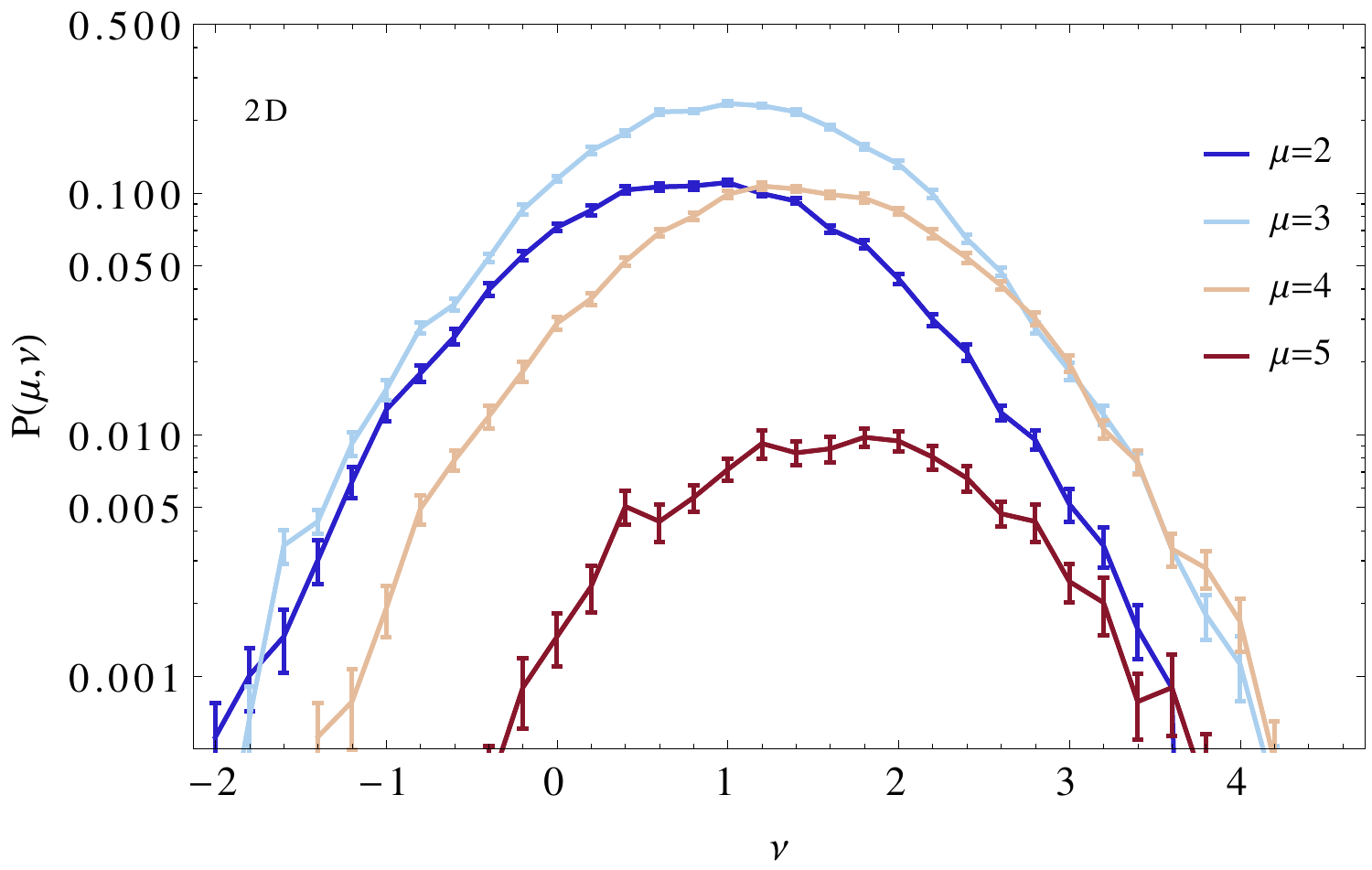} \hskip 1cm
\includegraphics[width= 0.95\columnwidth]{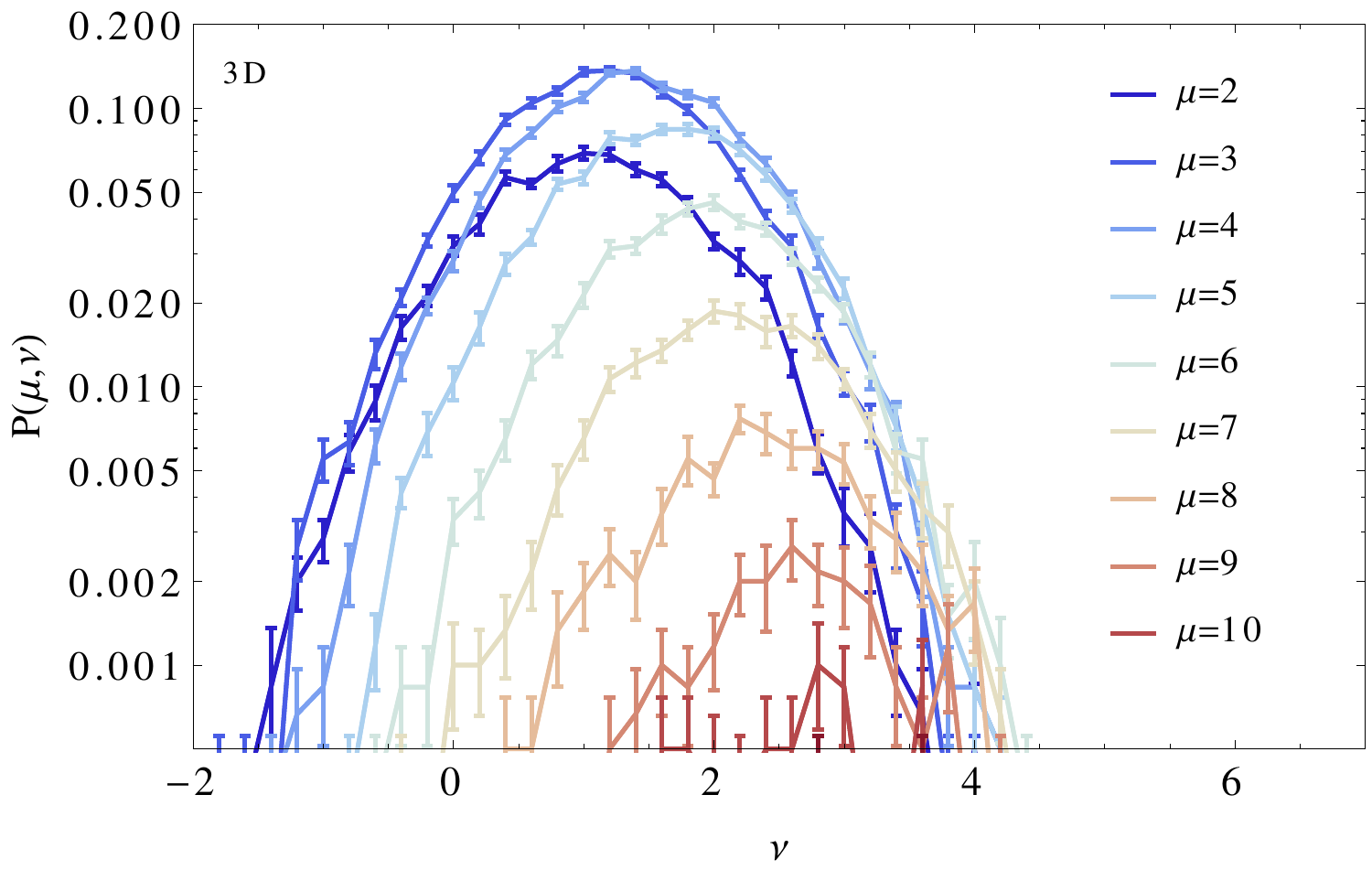} \hskip 1cm
\caption{
PDF of the multiplicity $\mu$ of peaks i.e their connectivity  {\sl corrected } for the number of bifurcations within the peakpatch.  
{\sl Left-hand panel:} Results obtained from twenty 2D ${2048}^{2}$GRF with spectral index $n_{s}=-1$;
the corresponding mean is $\langle \mu \rangle=$ 3 for all values of $n_{s}$.
{\sl Right-hand panel:} same quantity in 3D for 20  $256^{3}$ GRF. Here $\langle \mu \rangle=$ 4.
 }
\label{fig:GRF-2D-kappa-max}
\end{center}
\end{figure*}

\begin{figure*}
\includegraphics[width= 0.95\columnwidth]{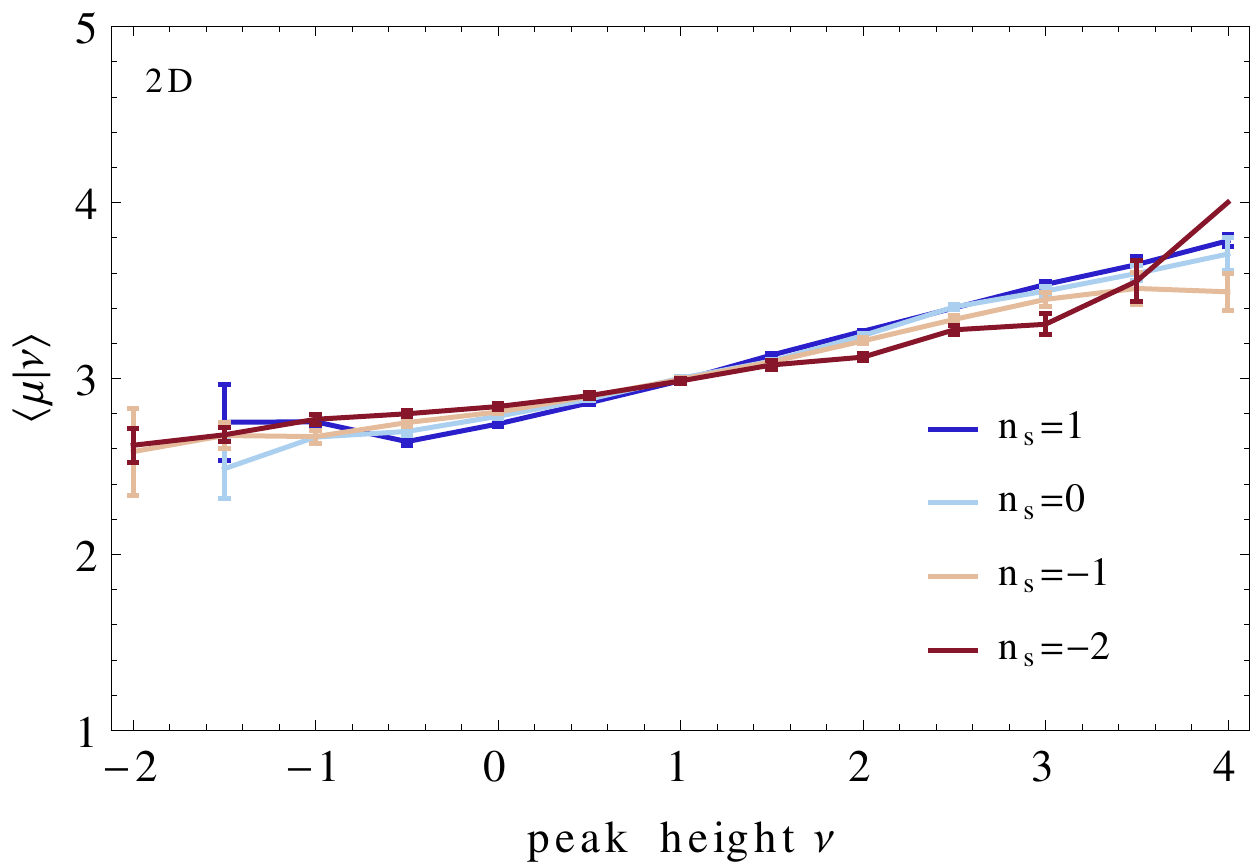} \hskip 1cm
\includegraphics[width= 0.95\columnwidth]{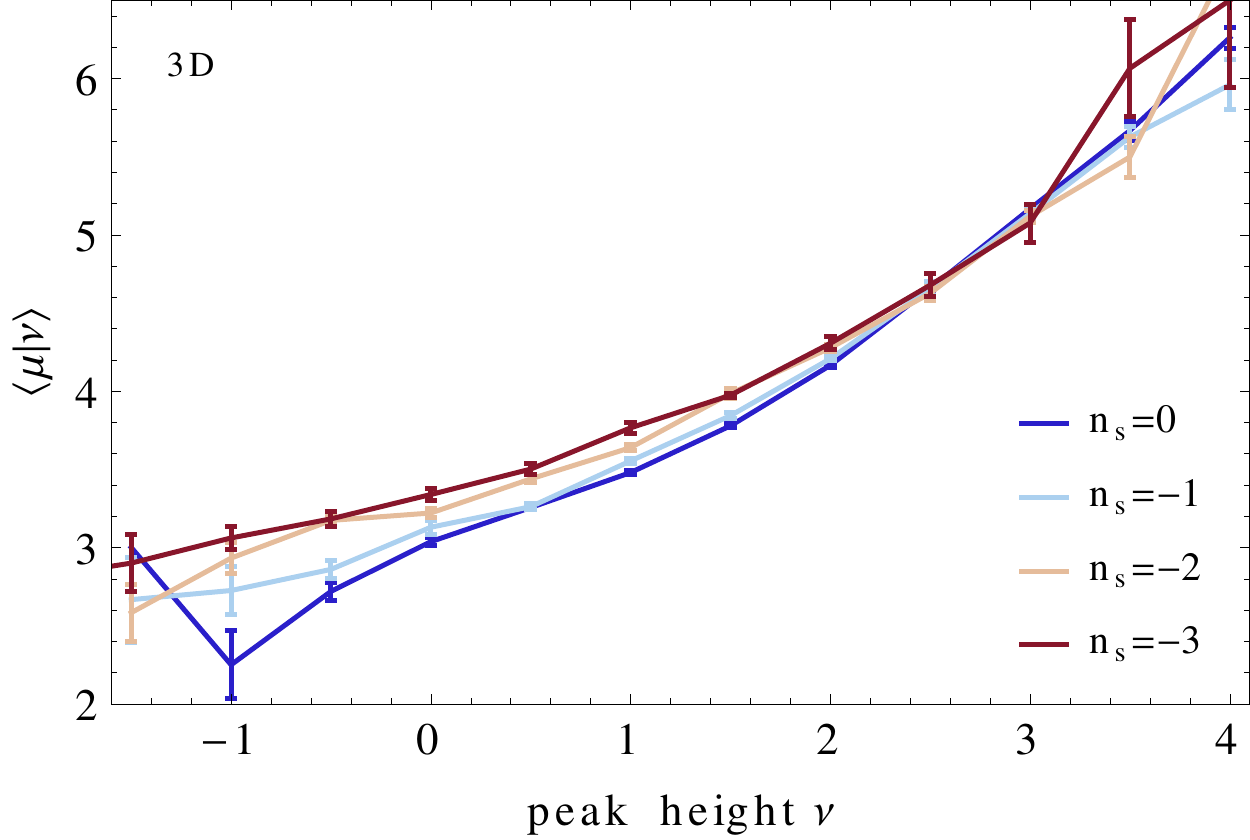} 
\caption{ Mean multiplicity as a function of the peak height 
from ten 2D ${2048}^{2}$GRF {\sl (left) } and ten 3D  $256^{3}$ GRF {\sl (right)} with spectral index as labelled.
}
\label{fig:GRF-mean-mu-given-nu}
\end{figure*}

In particular, the rareness (or height) of the peak changes the mean multiplicity. To investigate this effect, Figure~\ref{fig:GRF-2D-kappa-max} shows the joint distribution of peak's multiplicity and height. As expected, the larger the multiplicity, the higher the peak. For instance, 3D peaks of multiplicity $\mu=$2, 3, 4, 5 are expected to have increasing height on average: $\left\langle\nu|\mu\right\rangle\approx$ 1.07, 1.15, 1.38, 1.68 ($\pm 0.01$) in the case where $n_{s}=-2$ as displayed on that figure. As can be seen in Figure~\ref{fig:GRF-mean-mu-given-nu}, this relation is almost linear with slope about 0.2 almost independent from the spectral index and y-intercept varying from $\sim 0.3$ for $n_{s}=-3$ to $\sim 0.8$ for $n_{s}=0$.
 
\cite{pogo09}  identified the point process allowing us to build a proxy for the  number of bifurcation points which were found to lie in 
the vicinity of where the skeleton splits. 
It relied on the degenerate condition of equal eigenvalues of the Hessian (i.e. local isotropy), so that the next order in the Taylor expansion 
defines the direction of the splitting branches. 
Since  we expect that the third eigenvalue  remains distinct, the bifurcation will generically be co-planar.
Appendix~\ref{sec:geometry} quantifies the geometry of the bifurcation as a function of the relative strength of 
components of the gradient and that of the third-order derivative of the field.

%%%%%%%%%%%%%%%%%%%%%%%%%%%
\subsection{Theory of the peak connectivity}\label{thconmax}
%%%%%%%%%%%%%%%%%%%%%%%%%%%

In this section, we develop a theoretical framework to predict GRF connectivity from first principles. This is a challenging question as connectivity is by nature a global property and is difficult to catch from purely local considerations. 
Yet we will show that the mean connectivity in the field can be easily 
predicted. We will also develop a simplified theoretical approach that
explains the increase of connectivity with increase of rare peaks height and
even predicts reasonably well quantitative aspects of this behaviour.

\subsubsection{Global analysis}\label{sec:global}
The theoretical estimate of the average number of connections of a
peak 
comes simply from counting
the total number of saddles and peaks in the field. 
Since every saddle point is connected to two and only two peaks and each peak to $\kappa$ saddles, the mean connectivity of a peak simply reads
\begin{equation}
\label{eq:kappa-th}
\left\langle \kappa\right\rangle=\frac{2\bar n_{\rm sad}}{\bar n_{\rm max}}\,,
\end{equation}
which for GRF, 
from equations~(\ref{eq:extremacounts-2D})-(\ref{eq:extremacounts-3D}),
is exactly 4 in 2D and $\approx$ 6.11 in 3D.
This means that in 2D, the connected skeleton on average is similar to a
cubic lattice, while this is not exactly the case in 3D and some (small) "crystallographic" defects should appear.
However, we have not detected the departure from 
$\left\langle \kappa \right\rangle =6$ having limited statistics
in our 3D measurements.
Note that the connectivity can also be predicted for GRF in higher dimensions, it can be shown to follow
 \begin{equation}
 \kappa= 2d+\left(\frac{2d-4}{7}\right)^{7/4},
 \end{equation}
at least for dimension $d$ between 2 and 11 (see Appendix~\ref{sec:NDcon} for a quick derivation of this relation).

Equation~(\ref{eq:kappa-th}) is general and in particular not restricted to
GRF. It can therefore be computed in the weakly non-linear regime using the
predictions for extrema counts given, for instance, in \cite{GPP2012} by means of a Gram-Charlier expansion.
In 3D, the result at first order in non-Gaussianity reads
\begin{equation}
\label{eq:kappa-GC}
\left\langle \kappa\right\rangle=\kappa^{\mathrm G}
\left(1+\sum_{i\geq 1}\kappa^{(i)}\sigma_{0}^{i}\right),
\end{equation}
with the first two terms given by
\begin{equation}
\kappa^{ \mathrm G}
=2 \times \frac{29 \sqrt{3} +18 \sqrt{2}}{29\sqrt{3}-18\sqrt{2}}\approx 6.11,
\end{equation}
and
\begin{equation}
\label{eq:kappa1NL}
\kappa^{(1)}=\frac{4\sqrt 3}{35\sqrt{\pi}\sigma_{0}}\left(8\left\langle J_{1}^{3}\right\rangle-10\left\langle J_{1} J_{2}\right\rangle-21\left\langle q^{2} J_{1}\right\rangle\right).
\end{equation}
Equation~(\ref{eq:kappa1NL}) is written in terms of the extended skewness parameters $\left\langle J_{1}^{3}\right\rangle/\sigma_{0}$, $\left\langle J_{1} J_{2}\right\rangle/\sigma_{0}$ and $\left\langle q^{2} J_{1}\right\rangle/\sigma_{0}$ where $\sigma_{1}^{2}=\left\langle|\nabla \rho|^{2}\right\rangle$ is the variance of the density gradient, $\sigma_2^2=\left\langle\left(\Delta \rho\right)^{2}\right\rangle$ the variance of the density Hessian,  $\sigma_{1}^{2}q^{2}=|\nabla \rho|^{2}$ the modulus square of the gradient, $\sigma_{2}J_{1}=\Tr \rho_{ij}$ the trace of the Hessian matrix and $\sigma_{2}J_{2}=3/2 \Tr( \rho_{ij}^{2})-1/2 (\Tr \rho_{ij})^{2}$. In the cosmological context,   these extended skewness parameters  are constant in time at tree order in perturbation theory -- similarly to  $S_{3}=\left\langle \rho^{3}\right\rangle/\sigma_{0}^{4}$ -- and depend on the slope of the underlying power spectrum\footnote{They correspond to isotropic moments of the underlying Bispectrum, see  \cite{GPP2012}.}. 
Note that Section~\ref{sec:DM} and in particular
Figure~\ref{fig:PT-3D} will compute Equation~(\ref{eq:kappa1NL}) in the cosmological context.
Going to second order in the variance then requires to compute extended kurtosis parameters appearing in the next order term of the Gram Charlier expansion described in \cite{GPP2012}.

Note that as usual, equation~\eqref{eq:kappa-GC} can be interpreted as a measure 
of the temporal and scale evolution  of the connectivity given that $\sigma$ is a measure of the amplitude of non-linearities and depends on both time and scale.
When conducting a dark energy experiment based on the cosmic evolution of the connectivity, the leading 
contribution will come from $\sigma_0=D(z)\sigma_{0}(z=0) $ in equation~\eqref{eq:kappa-GC}, 
where $\sigma_{0}^{2}(z=0)$ is the linear variance of the density field at redshift zero, $D(z)$ is the growth rate given by
\begin{equation}
D(z)=\frac{5\Omega_m H_0^2}{2} H(a)\int_0^a \frac{{\rm d} a'}{a'^3 H^3(a')}\,,
  \label{eq:cosmo}
\end{equation}
 where $\!H\!(a)\!\!=\!\!H_0 \sqrt{\Omega_m/{a^3}\!+\! \Omega_\Lambda\! \exp[3 \!\int_0^z ({1\!+\!w(z')})/({1\!+\!z'}) {\rm d} z' ]}$,
with $\Omega_m$, $\Omega_\Lambda$ and $H_0$, respectively, the dark matter and
dark energy  densities and the Hubble constant at redshift zero,
$a\equiv1/(1+z)$ the expansion factor and $w(z)=w_0+ w_1 /({1+z})$
the parameterized equation of state of dark energy.
Since $\kappa^{(1)}$ is a number which can be predicted from
cosmological perturbation theory, the cosmic variation of $\langle \kappa \rangle$ puts constraints on $(w_0,w_1)$.
In practice, one also needs to  control the redshift evolution of the tracer threshold (e.g. the 
luminosity cut), since the mass function of haloes also evolves with redshift.

In turn, in 2D we can also easily obtain the Gaussian 
\begin{equation}
\kappa^{ \mathrm G}
=4\,,
\end{equation}
and first non-Gaussian order contribution to the global connectivity
\begin{equation}
\kappa^{ (1)}=
\frac{2}{27}\sqrt{\frac{6}{\pi}}
\left(
5\frac{\left\langle J_{1}^{3}\right\rangle}{\sigma_{0}}-6\frac{\left\langle J_{1} J_{2}\right\rangle}{\sigma_{0}}-18\frac{\left\langle q^{2} J_{1}\right\rangle}{\sigma_{0}}
\right), \label{eq:k1-2D}
\end{equation}
where $\kappa^{ (1)}$ is again independent of $\sigma_0$ at tree order in perturbation theory.
In equation~\eqref{eq:k1-2D} the definition of $J_2$ is now changed to 
$\sigma^2_{2}J_{2}=2 \Tr( \rho_{ij}^{2})- (\Tr \rho_{ij})^{2}=(\lambda_{1}-\lambda_{2})^{2}$ while $J_{1}$ an $q^{2}$ have the same definition as in the 3D case.
Again, the first non-Gaussian correction scales like $\sigma_{0}$ and is therefore a direct tracer of the growth of structure.

\subsubsection{Connectivity as a function of peak height}
\label{sec:kappa-th}
A simple idea to estimate the change of connectivity with the peak height $\nu_c$ 
is to count the number of saddles in the volume 
around the peak through which the filamentary connectors
to the neighbouring peaks pass. The main issue is therefore to quantify the typical
volume where bridging saddles are. We suggest that this should be the volume
of the peak-patch where there is a suppressed probability for other peaks to 
be present.  Then, the average saddle count, conditional on the presence
of the peak of height $\nu_c$ at the centre of the volume,
and therefore the connectivity of the peak,  is
\begin{equation}
\label{eq:kappath}
\kappa  (\nu)=\bar n_{\rm sad}\int_{0}^{R_{\rm pp}}\textrm {d}^{D}r\,(1+\xi_{\rm pk\!-\!sad}(r,\nu))\,,
\end{equation}
where $\xi_{\rm pk\!-\!sad}(r,\nu)$ is the correlation between peaks of height $\nu$
and saddles of any height, and $\bar n_{\rm sad}$ is the mean number density
of saddle points.

\begin{figure*}
\includegraphics[width= 0.95\columnwidth]{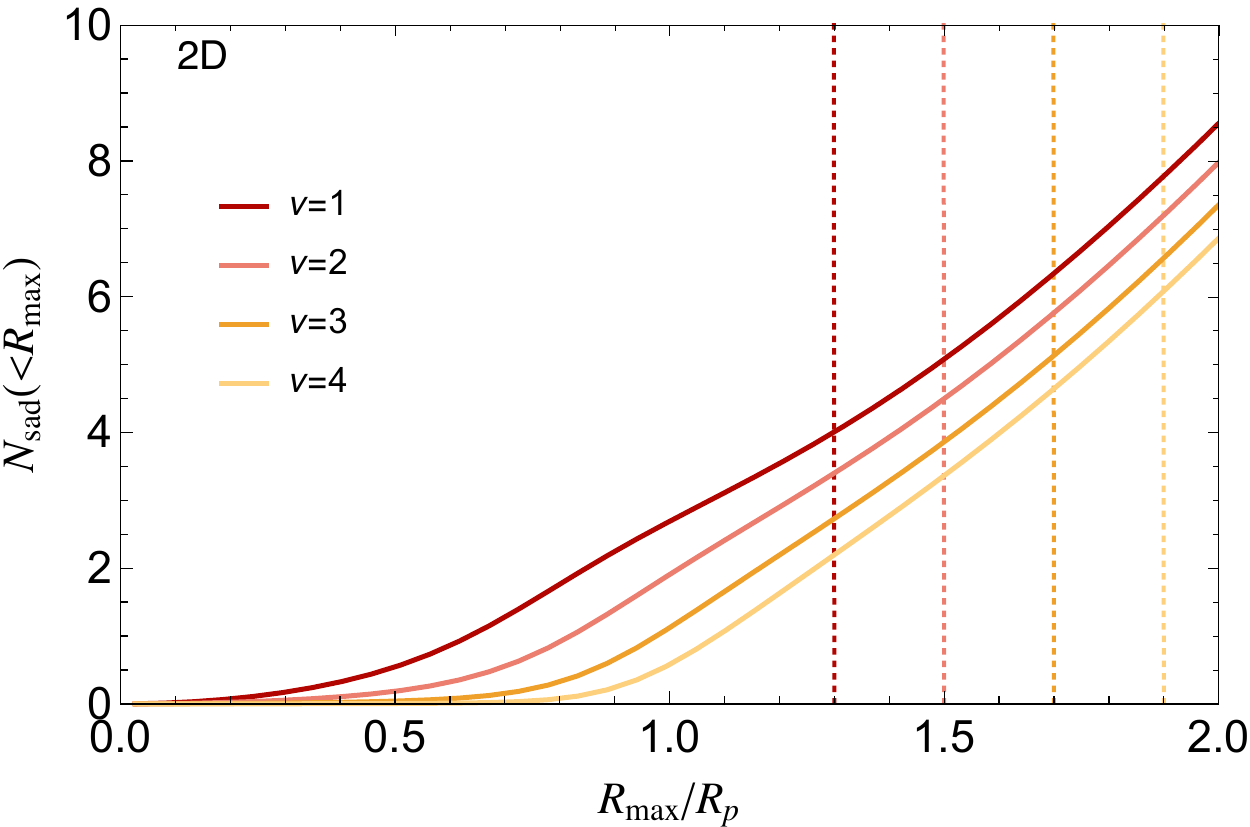}
\includegraphics[width= 0.95\columnwidth]{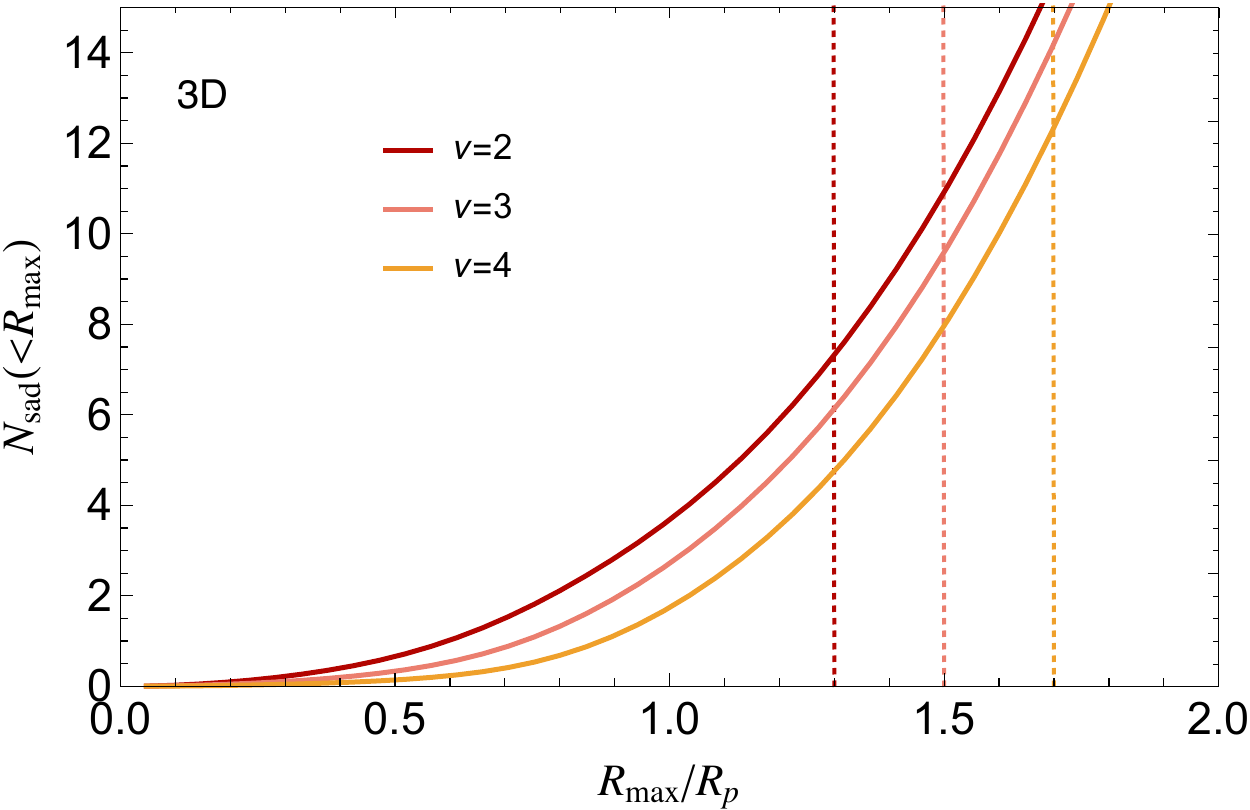}
\caption{Expected total number of saddles in spheres of increasing radius $r$ in 2D ($n_{s}=0$) and 3D ($n_{s}=-2$) GRFs around a central peak of height $\nu$ as labelled. The dotted lines represent the size of the corresponding peak patch $r=R_{\rm pp}$ according to $R_{\rm pp}=1.1 + \nu/5$ in 2D and $R_{\rm pp} = 0.9 + \nu/5$ in 3D, as described in Section~\ref{sec:corr_func}. }
\label{fig:Nsadtotal}
\end{figure*}

\begin{figure}
\includegraphics[width= 0.95\columnwidth]{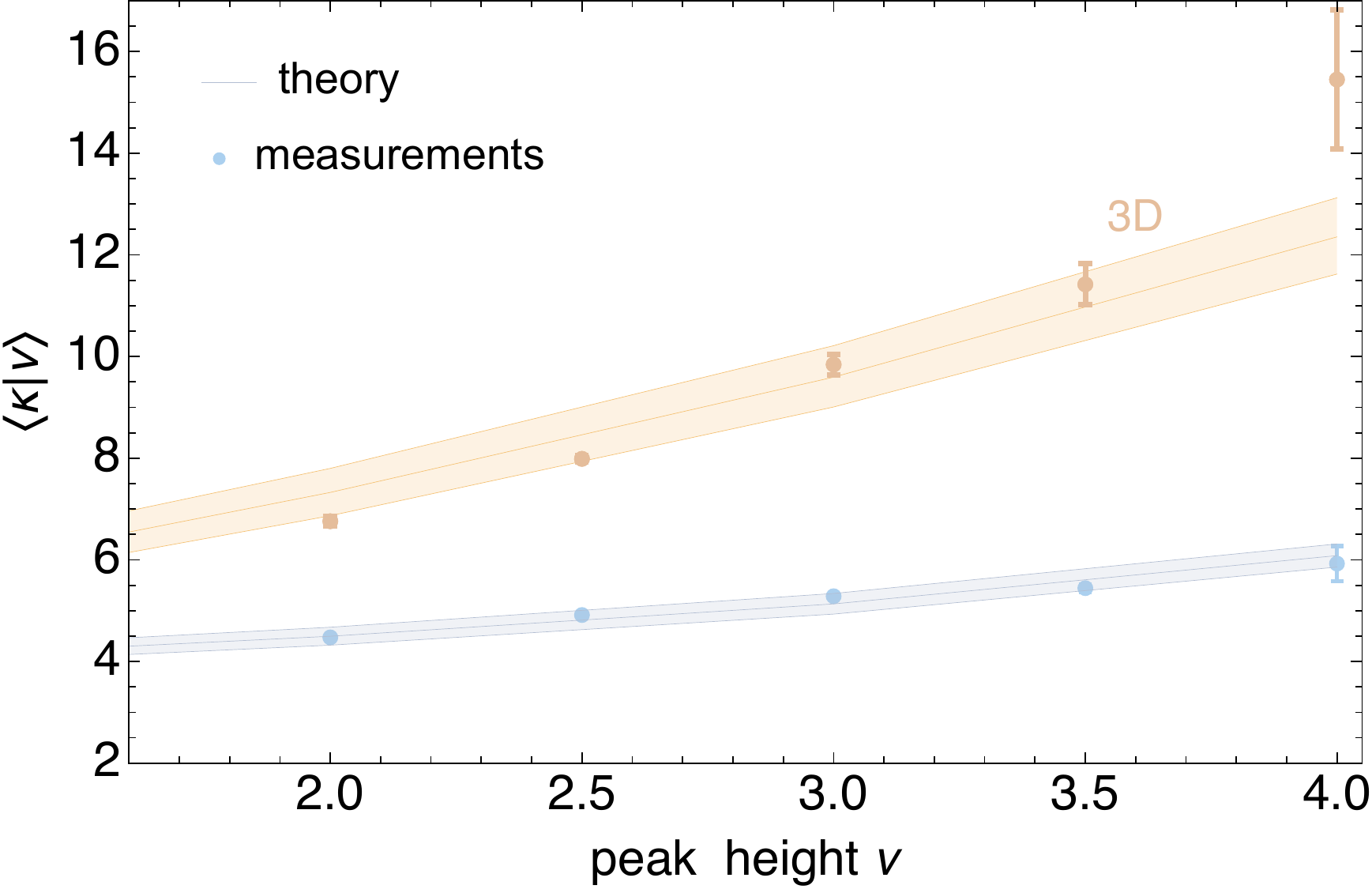}
\caption{Mean connectivity of peaks of height $\nu$ in 2D (blue) and 3D (orange) GRF.  
In 2D, we consider a spectral index $n_{s}=0$ while in 3D we choose for this figure $n_{s}=-2$. The error bars display the measured relation shown in Figure~\ref{fig:connect2D-nu} and Figure~\ref{fig:connect3D-nu}. The shaded areas represent the theoretical predictions given by Equation~(\ref{eq:kappath}) given that $R_{\rm pp}\approx 1.1 + \nu_c/5\pm 0.03$ in 2D and $R_{\rm pp}\approx 0.9 + \nu_c/5\pm 0.03$ in 3D as justified in Section~\ref{sec:kappa-th}. }
\label{fig:connectivity-th-2D}
\end{figure}

To understand what connectivity equation~(\ref{eq:kappath}) predicts
we refer back to Section~\ref{sec:corr_func}
for necessary ingredients, namely the peak-saddle correlation
function and the size of peak patch $R_{\rm pp}$.
The resulting dependence on $\nu_c$ is the result of the competition between
two effects: on the one hand the suppression of saddles near the peak as the height of this central peak
increases and on the other hand the increase of the peak-patch volume over which
saddles need to be counted. The first effect is practically contained
within $r < R_p$ radius, while the peak-patch radius $R_{\rm pp}$ increases linearly
with $\nu_c$ beyond $R_p$, including the outer region where the density of
saddles becomes equal to the average one.
Figure~\ref{fig:Nsadtotal} shows how the conditional saddle
counts accumulate with the radius.  The Figure demonstrates that with $R_{\rm pp}$
taken to be $R_{\rm pp}=1.1 + \nu_c/5$ in 2D and $R_{\rm pp} = 0.9 + \nu_c/5$ in
3D as discussed in Section~\ref{sec:corr_func}, 
the volume effect dominates the local (near peak) saddle suppression and 
on balance the theory predicts a growth of connectivity with peak height.
Perhaps surprisingly, our estimate for $R_{\rm pp}$ gives even quantitative
values for $\kappa(\nu_c)$ in the interval $\nu_c=1.5 -  4$  close to
the 2D and 3D results measured in Figures~\ref{fig:connect2D-nu} and 
\ref{fig:connect3D-nu}, as Figure~\ref{fig:connectivity-th-2D} attests. 
As a note of caution, one should not overestimate
the quantitative accuracy of this prediction given the precision of
our measurements and simplicity of the theoretical model.

\subsection{Local analysis of peak connectivity, multiplicity and bifurcations}

The complexity of predicting peak connectivity arises from the non-local
topological nature of the global skeleton. Remarkably, the study of the number of 
real physical overdense filaments emanating from a peak
lends itself more readily to the rigorous local analysis. A way to proceed is
to compute the total number of 2D maxima at the surface of the sphere centred
on the central peak and study this number as the function of the radius
of the sphere. 
This is an approximation to the number of filaments crossing the sphere assuming that for high peaks, filaments are sufficiently ``stiff''.

This approach gives a detailed picture of how many filaments leave
the immediate peak volume, where they bifurcate at larger radii (as the count
increases) and, coupled with an estimate for the size of peak-patch from
Section~\ref{sec:corr_func}, gives an alternative way to predict 
the global connectivity of the central peak.
Being a straightforward conditional maxima count, albeit on
2D spherical sections of the underlying 3D field, this analysis can be
formulated via the joint distribution of the field values and
its derivatives in two different locations: the central peak and the position on a sphere around it. 
We refer the reader to Appendix~\ref{app:2Dpks} for the 
technical details of extrema counts on a curved spherical surface, 
and only describe the results here in the main text.

\begin{figure*}
\includegraphics[width= 0.95\columnwidth]{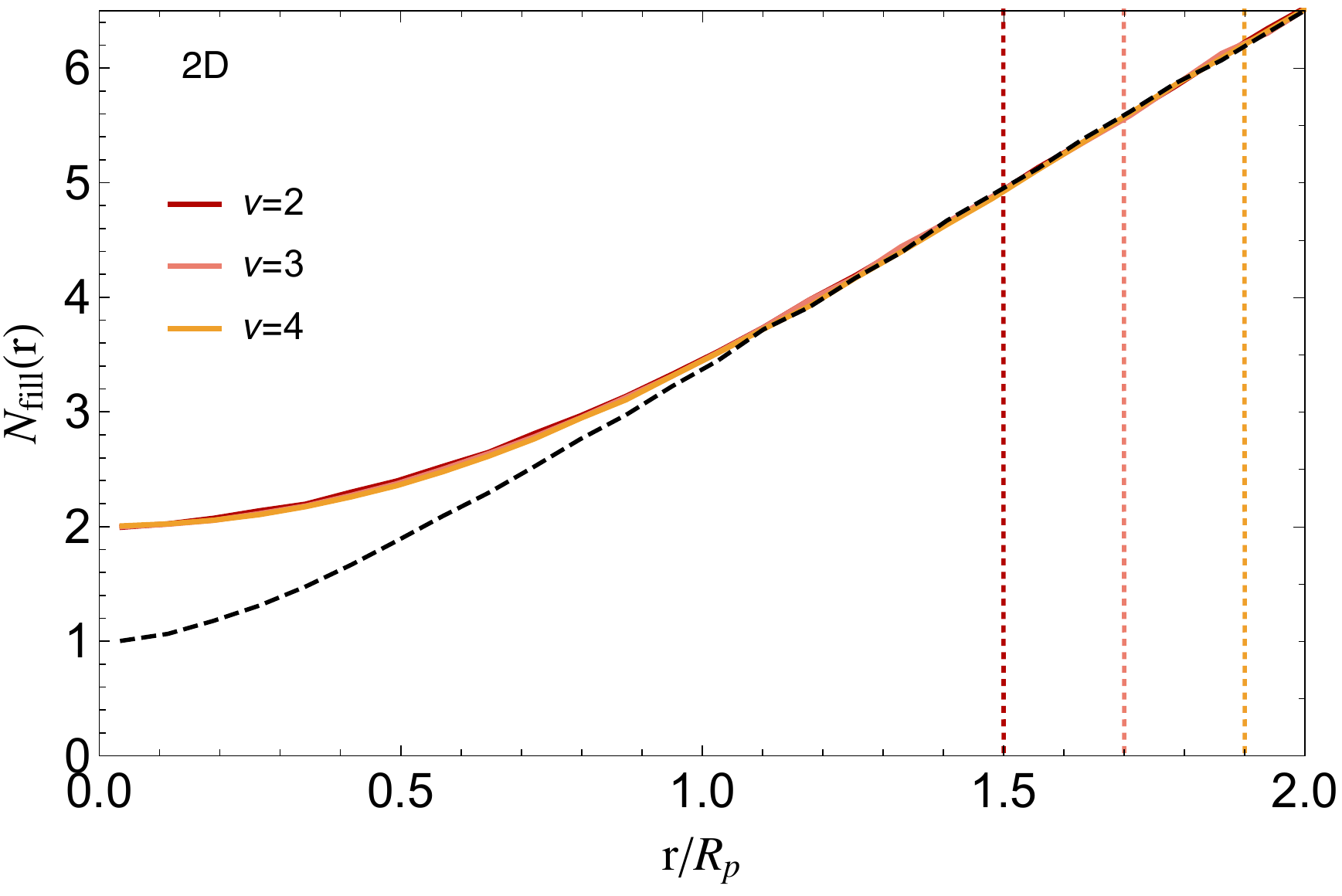}
\includegraphics[width= 0.95\columnwidth]{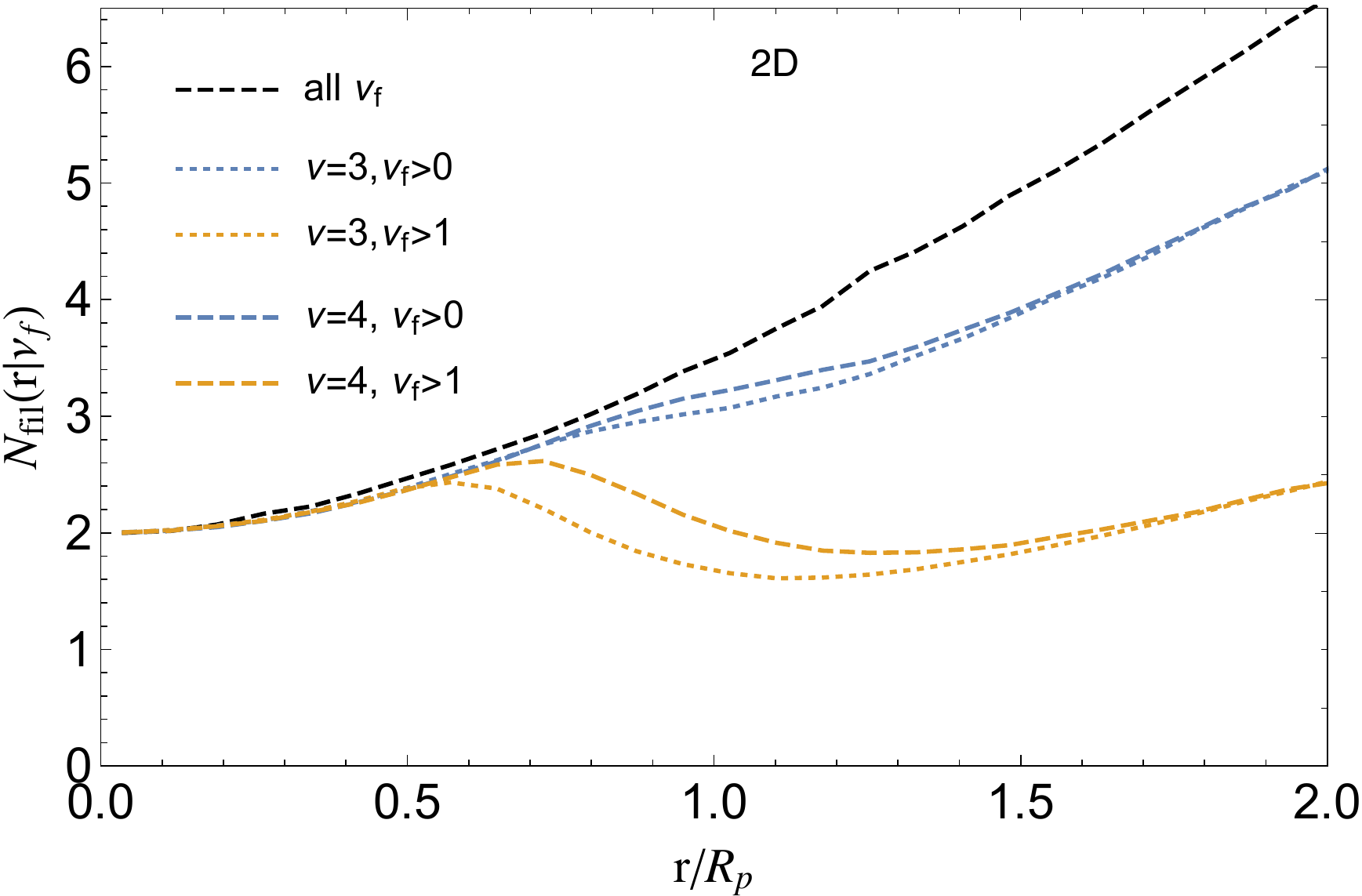}
\caption{{{\sl Left-hand panel:} Mean number of maxima on a circle of radius $r$
around a peak of height $\nu$ as a proxy for the number of filaments crossing that circle.
The prediction is for a Gaussian random field with $k^{0}$ power spectrum filtered with a Gaussian kernel. The dashed line represents the case where the central point is random (not necessarily a peak position).
Vertical dotted lines displays the peak patch size $R_{\rm pp}(\nu)$.}
{\sl Right-hand panel: }Same statistics as in the left-hand panel with restriction
that only filaments with density contrast $\nu_{f}$ exceeding 1(blue) or 2 (yellow) are counted.}
\label{fig:connectivity-2D-pred}
\end{figure*}

\begin{figure*}
\begin{center}
\includegraphics[width=0.95\columnwidth]{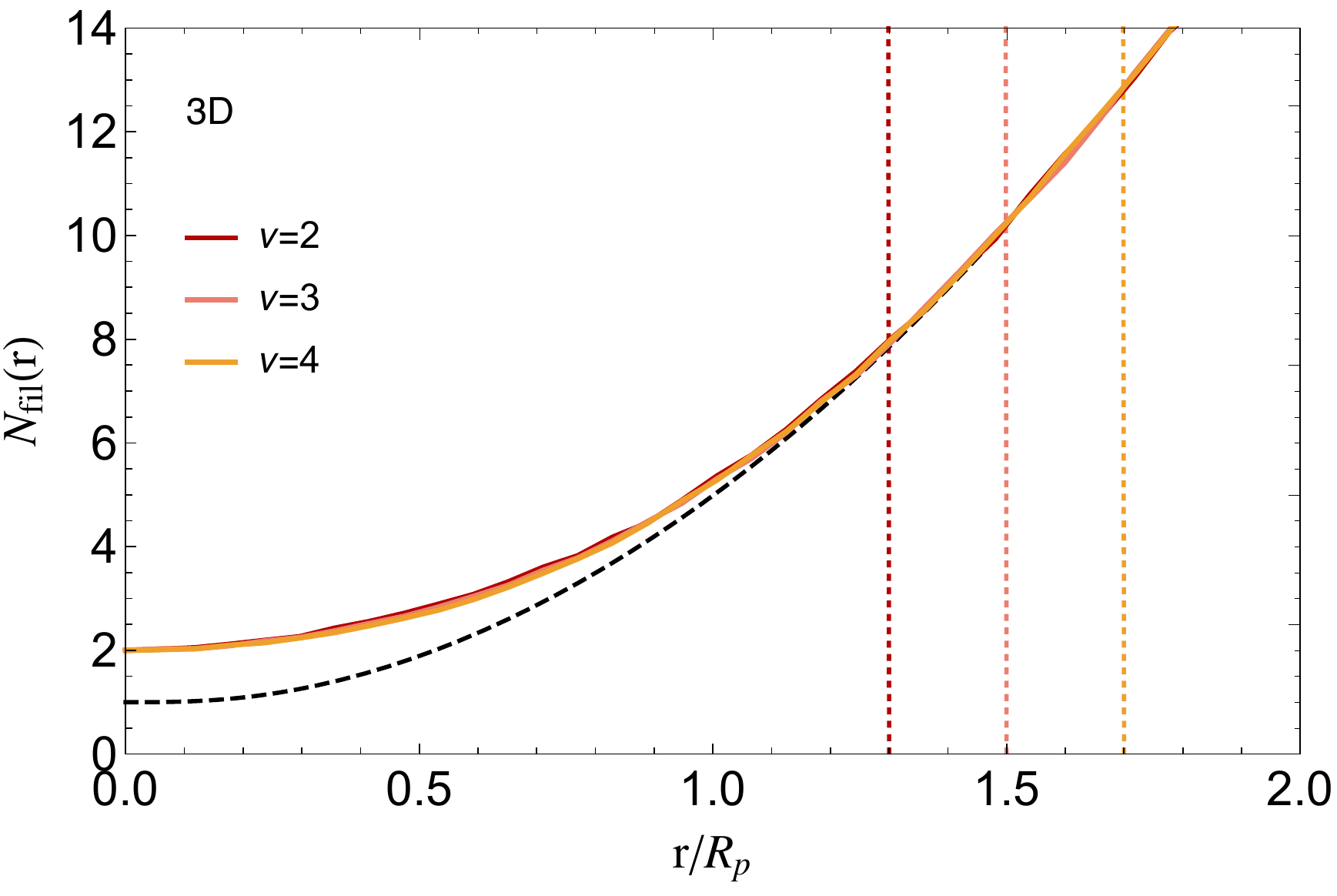} 
\includegraphics[width=0.95\columnwidth]{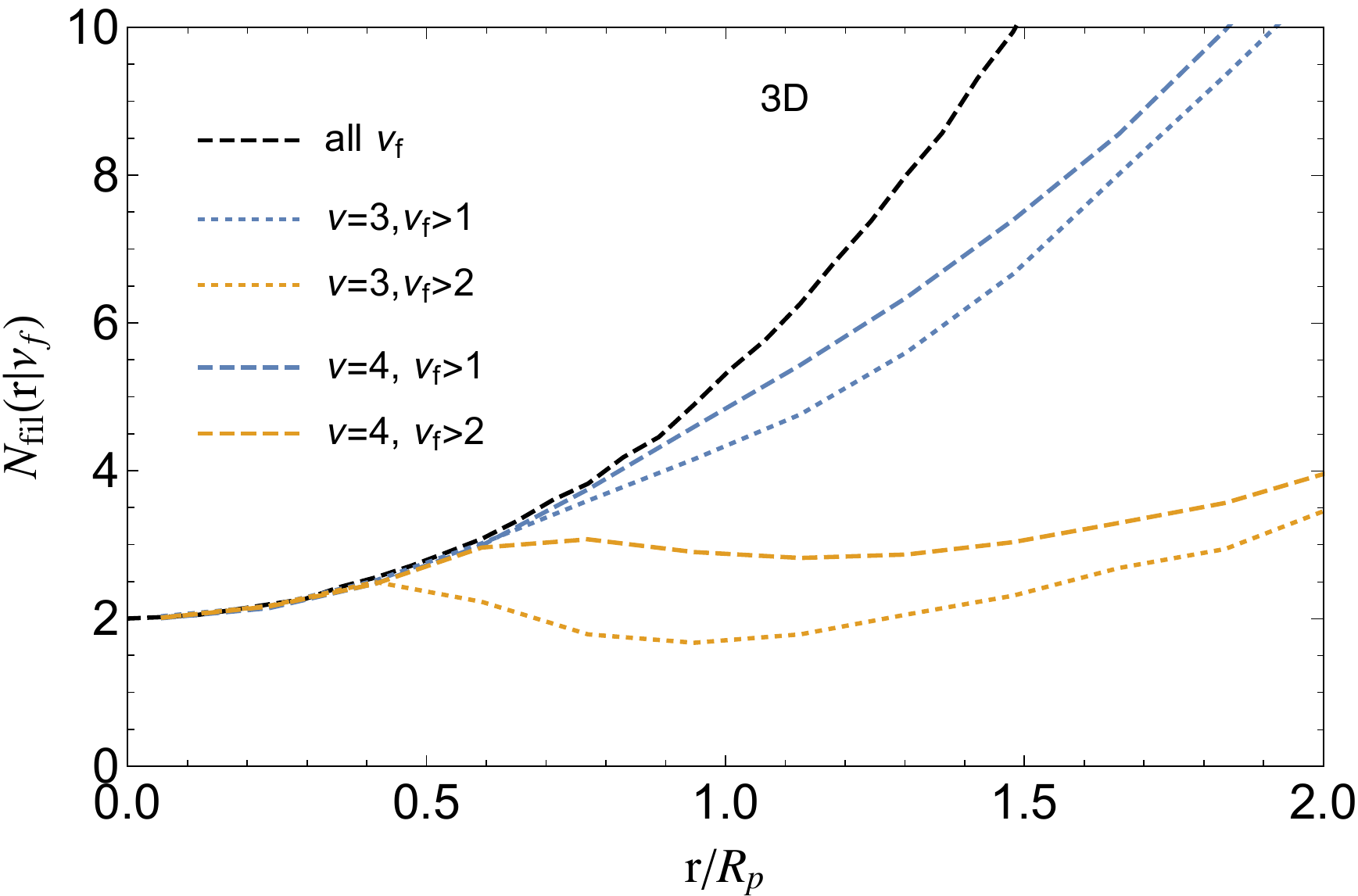} 
\caption{
{{\sl Left-hand panel: } Mean number of density maxima on a sphere of radius $r$
around a peak of height $\nu$, as a function of $r$
The prediction is for a Gaussian random field with $\Lambda$CDM power spectrum filtered with a Gaussian kernel on 5 Mpc$/h$.
The black dashed line represents the case where the central point is random (not necessarily a peak position). Vertical dotted lines displays the peak patch size $R_{\rm pp}(\nu)$.}
{\sl Right-hand panel: } Same statistics as in the left-hand panel with restriction
that only filaments with density contrast $\nu_{f}$ exceeding 1(blue) or 2 (yellow) are counted.
 }
\label{fig:connectivity-N2Dpeaks}
\end{center}
\end{figure*}

Figures~\ref{fig:connectivity-2D-pred} and
\ref{fig:connectivity-N2Dpeaks} demonstrate the main results for 2D and 3D
respectively.
The number of filaments crossing the sphere of radius $r$,
$N_\mathrm{fil}(r)$, is
shown in the left panels of Figures~\ref{fig:connectivity-2D-pred} and
\ref{fig:connectivity-N2Dpeaks}.
We find that the function $N_\mathrm{fil}(r)$ 
is essentially universal, independent of the central peak height and
the slope of the power spectrum.
At small distances, peaks have an ellipsoidal shape with two 
filaments (that is to say two branches of a same physical filament aligned
with the peak longest semi-axis),  $N_\mathrm{fil}(0) = 2$.
The expected number of filaments then grows with distance to $\sim 3$ 
and $\sim 4$ in 3D at the scale of the smoothing length. This is 
what corresponds to the mean multiplicity $\mu$ of the peaks,
since our definition of the bifurcations points does not identify them as such
within one smoothing radius from the peak.

At larger $r$, our algorithm detects bifurcations of the filaments that lead
to further increase of the number of filaments with radius.
At $r > R_p$ the effect of the central peak is negligible and the 2D maxima
count simply follows the area of a spherical surface, growing linearly with $r$
in 2D and quadratically as $r^2$ in 3D. The predicted connectivity $\kappa$
is obtained by evaluating $N_\textrm{fil}$ at the peak-patch radius $R_{\rm pp}$, 
$\langle\kappa|\nu_c\rangle = N_\textrm{fil}(R_{\rm pp})$.
The predicted connectivity does depend on the height because again $R_{\rm pp}$
is a growing function of the peak height.
For instance in 2D, using the typical peak-peak separation found in the
previous section $R_{\rm pp}=1.3, 1.5, 1.7, 1.9   R_p$ for $\nu=1,2,3,4$,
the peak connectivity is found to go from $\sim 4 $ to $6$,
in full agreement with our measurements in Gaussian random fields.
A similar agreement is found in 3D where 
the connectivity ranges from 7 and 13 for peak of heights between $2\sigma$ and $4\sigma$
above the mean. This shows the consistency of the local approach with
the other discussed measures of connectivity.

Both the method based on saddle counts in section \ref{sec:kappa-th} and the one relying on
maxima count on surrounding spheres developed here
readily allow us to add a constraint on the overdensity
of the filaments and therefore to study only physically prominent ones. 
The information obtained with the two techniques is complimentary. Saddle counts
reflect the change in the topological properties of the connected web. In this context,
eliminating filamentary connections below a given density 
threshold changes the connectivity $\kappa$ of the web.
Conversely, the approach of this section describes the local properties of filaments. It allows us to count them 
as a function of  distance from
the peak and track the filaments that are dense near the peak even
if their density will drop below ``detectability" further away.
So it tracks the peak branches (peak mutliplicity), not necessarily the connecting segments (peak connectivity).

The right panels of Figures~\ref{fig:connectivity-2D-pred} and \ref{fig:connectivity-N2Dpeaks}, show the number of prominent filaments
(e.g. defined as $\nu_{f}>1$ in 2D and $\nu_{f}>2$ in 3D) leaving the peak
as function of the distance $r$. We consider two types of
peaks, the very rare ones,
$\nu_c=4$, and lower and more frequent ones, $\nu_c=3$. Focusing on the 3D
case, one sees that a high-contrast peak is typically surrounded
by three main filaments, while the patch of a less rare peak is
dominated by two branches of presumably one dense filament.
Notably, the most prominent filaments
do not experience bifurcations (hence the yellow line almost constant on Figures~\ref{fig:connectivity-2D-pred} and \ref{fig:connectivity-N2Dpeaks}), meaning that in any bifurcation only one child
follows the dense parent. This situation pervades far from the central peak and all the way to the
neighbouring peaks, therefore forming a low connectivity network. 
Of course, one expects the density in a filament to decrease as the distance
from the peak increases. If it drops below the threshold,
the filament is considered terminated, which is then evidenced by the drop of
$N_\textrm{fil}(r| \nu_f\! >\! \nu_{f*},\nu_c)$ with distance. Closer is
the limit on the filament's density $\nu_{f*}$ to the peak height $\nu_c$,
less probable it is for the filament to extend far out.
Our theory
therefore predicts the relation between 
the heights of the peaks
and the density in the filaments 
that allows to build a fully connected web,
i.e to percolate from peak to peak along the filamentary network.
The necessary criterium is to have 
$N_\textrm{fil}(r| \nu_f\! >\!\nu_{f*},\nu_c \!>\! \nu_{c*}) \!>\! 2$ for
all $r$.
We'll explore the applications of this criterium in future works.

%=====================
\section{Cosmic connection} \label{sec:cosmic}
%=====================
\begin{figure*}
\includegraphics[width=1\columnwidth]{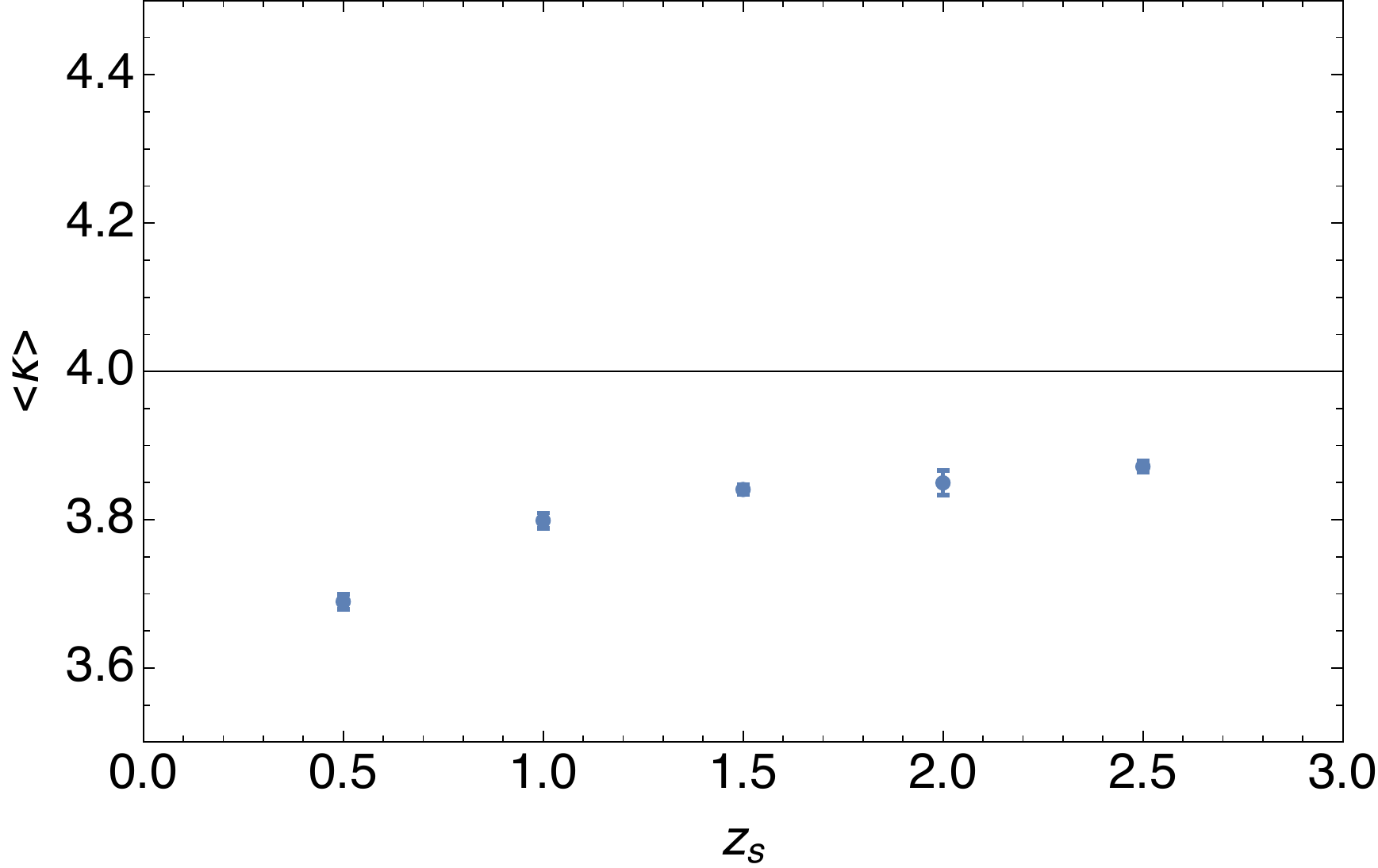}
\includegraphics[width=1\columnwidth]{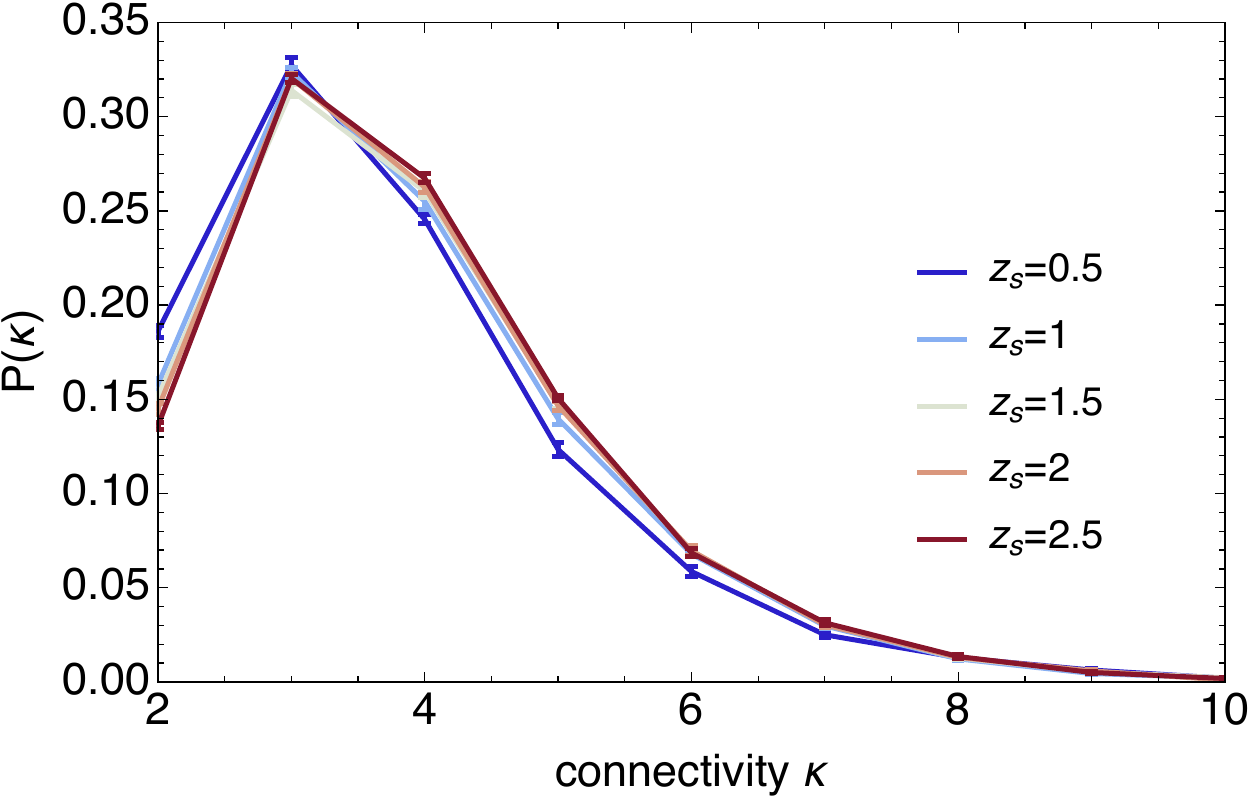}
\caption{Connectivity of convergence maps for source redshifts between 0.5 and 2.5 as labelled. {\sl The left panel} shows only the mean connectivity while {\sl the right panel} displays the full distribution function.
} \label{fig:connectivity-lensing}
\end{figure*}

Let us now turn to the connectivity of the evolved cosmic web as measured in dark matter simulations,
first in 2D convergence maps, then in 3D for dark matter and dark haloes respectively.

\subsection{Connectivity of cosmic convergence maps} \label{sec:conv}
%%%%%%%%%%%%%%%%%%%%%%%%%%%

We use convergence maps for four galaxy source redshifts  $z_s=0.5,$ 1, 1.5, 2 and 2.5 from the MassiveNuS suite of $\Lambda$CDM simulations\footnote{Those maps are publicly available at \url{columbialensing.org}} and corresponding to the case without massive neutrinos \citep{Liu17}. 
The cosmological parameters are set to $h=0.7$, $n_s=0.97$, $\Omega_b=0.046$, $\Omega_m=0.3$ and $\sigma_8=0.8523$ concordant with current observations. These maps cover 12.25 deg$^2$ with 0.1 arcmin resolution and were obtained from Gadget-2 runs using the LensTools Python code \citep{2016A&C....17...73P}.  We refer the reader to \cite{Liu17} for further details on the simulations and pipeline to generate the convergence maps.

For the sake of simplicity, we select 10 of the 1000 realisations of the MassiveNuS suite and smooth the convergence fields with a Gaussian kernel on 8 pixels corresponding to a FWHM of 1.9 arcmin. We then measure the connectivity of each two dimensional peak with {\tt DISPERSE}. 
The persistence threshold is chosen so that the number of peaks found by {\tt DISPERSE} is the same as the thoroughly tested {\tt map2ext} code \citep{Colombi2000,pogo11} which is described in Appendix~\ref{sec:persistence-cuts}.

Figure~\ref{fig:connectivity-lensing} shows the measured mean connectivity as a function of redshift and the corresponding full distribution on the left and right panels respectively. As expected higher galaxy source redshifts are the closest to the Gaussian prediction. At lower redshifts, the field becomes more non-Gaussian and peaks are less connected. A similar picture also appears for the three dimensional matter distribution as will be shown in the next section.

%%%%%%%%%%%%%%%%%%%%%%%%%%%
%\subsection{Dark matter cosmic connectivity} \label{sec:DM}
%%%%%%%%%%%%%%%%%%%%%%%%%%%
\subsection{Connectivity of the 3D matter distribution} \label{sec:DM}
%%%%%%%%%%%%%%%%%%%%%%%%%%%
Let us now focus on the connectivity of the three dimensional distribution of matter in the Universe by means of analytical arguments first and numerical simulations when non-linearities become important.

\begin{figure}
\includegraphics[width= 0.95\columnwidth]{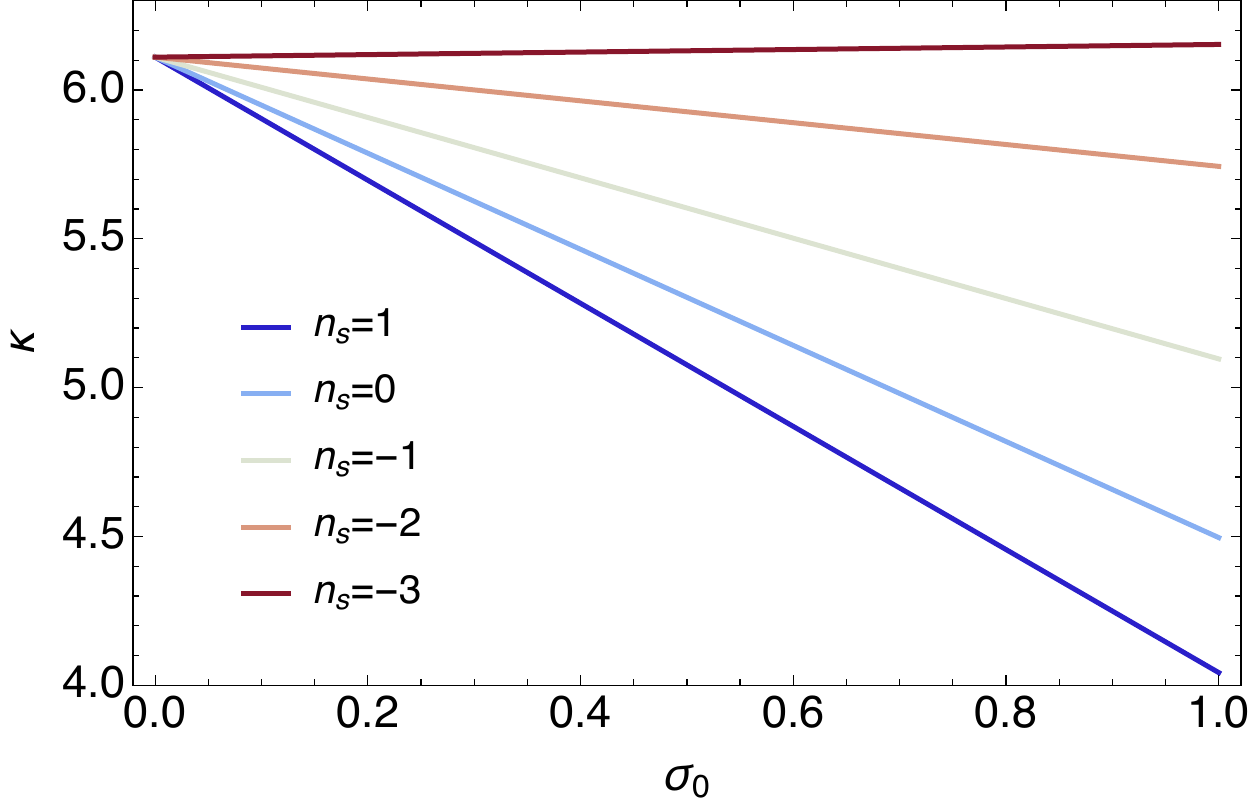} 
\caption{Evolution of the mean number of connectors in 3D as a function of the variance $\sigma_{0}$ -- i.e across cosmic time -- for different spectral indices as labeled. The early evolution of the 3D connectivity is rather slow. We expect the subsequent highly non-linear evolution to modify this result for higher $\sigma_{0}$. }
\label{fig:PT-3D}
\end{figure}

Figure~\ref{fig:PT-3D} shows the expected evolution of the connectivity of the three dimensional distribution of matter in the Universe at first order in perturbation theory.
To obtain this result, equation~\eqref{eq:kappa-GC} is used with the extended skewness parameters analytically computed at tree order in cosmological perturbation theory, for a Gaussian smoothing and different power law power spectra. Those leading order skewness-like terms involve hypergeometric functions \citep{GPP2012}.

Eventually, the combination of interest $\tilde S_{3}=8\left\langle J_{1}^{3}\right\rangle-21\left\langle q^{2}J_{1}\right\rangle+10\left\langle J_{1}J_{2}\right\rangle$ reads
\begin{equation}
\tilde S_{3}=8\gamma(s_{5}^{5}{\cal F}_{5}^{5}+s_{7}^{5}{\cal F}_{5}^{7}+s_{9}^{9}{\cal F}_{9}^{9}+s_{11}^{9}{\cal F}_{9}^{11}),
\end{equation}
with the Gauss hypergeometric functions
\begin{eqnarray}
&&\hskip -.7cm{\cal F}_{5}^{5}=\pFq{2}{1}{\frac {n_{s}+5} 2 }{\frac {n_{s}+5} 2}{\frac 3 2}{\frac 1 4},\nonumber\\
&&\hskip -.7cm{\cal F}_{5}^{7}=\pFq{2}{1}{\frac {n_{s}+5} 2 }{\frac {n_{s}+7} 2}{\frac 3 2}{\frac 1 4},\nonumber\\
&&\hskip -.7cm{\cal F}_{9}^{9}=\pFq{2}{1}{\frac {n_{s}+9} 2 }{\frac {n_{s}+9} 2}{\frac 9 2}{\frac 1 4},\nonumber\\
&&\hskip -.7cm{\cal F}_{9}^{11}=\pFq{2}{1}{\frac {n_{s}+9} 2 }{\frac {n_{s}+11} 2}{\frac {11} 2}{\frac 1 4},\nonumber
\end{eqnarray}
and coefficients
\begin{eqnarray}
&&\hskip -.7cm s_{5}^{5}\!\!=\!\!\frac{\!261212 n_{s}^4\!\!+\!\!2882763 n_{s}^3\!\!+\!\!6981958 n_{s}^2\!\!-\!\!35748720 n_{s}\!\!-\!\!197568000\!}{25725 n_{s}^2 (n_{s}\!+\!2)^2}\!,\nonumber\\
&&\hskip -.7cm s_{7}^{5}\!\!=\!\!\frac{\!51861600\!\!+\!\!6744654 n_{s}\!\!-\!\!1875041 n_{s}^2\!\!-\!\!633302 n_{s}^3\!\!-\!\!51932 n_{s}^4\!}{10290 n_{s}^2 (n_{s}\!+\!2)^2},\nonumber\\
&&\hskip -.7cm s_{9}^{9}\!=\!-\frac{2392 (n_{s}\!+\!5) (n_{s}\!+\!7)^2}{3087000},\nonumber\\
&&\hskip -.7cm s_{11}^{9}\!=\!\frac{1276 (n_{s}\!+\!5) (n_{s}\!+\!7)^2 (n_{s}\!+\!9)}{21609000}.\nonumber
\end{eqnarray}

\begin{figure}
\includegraphics[width= 1.\columnwidth]{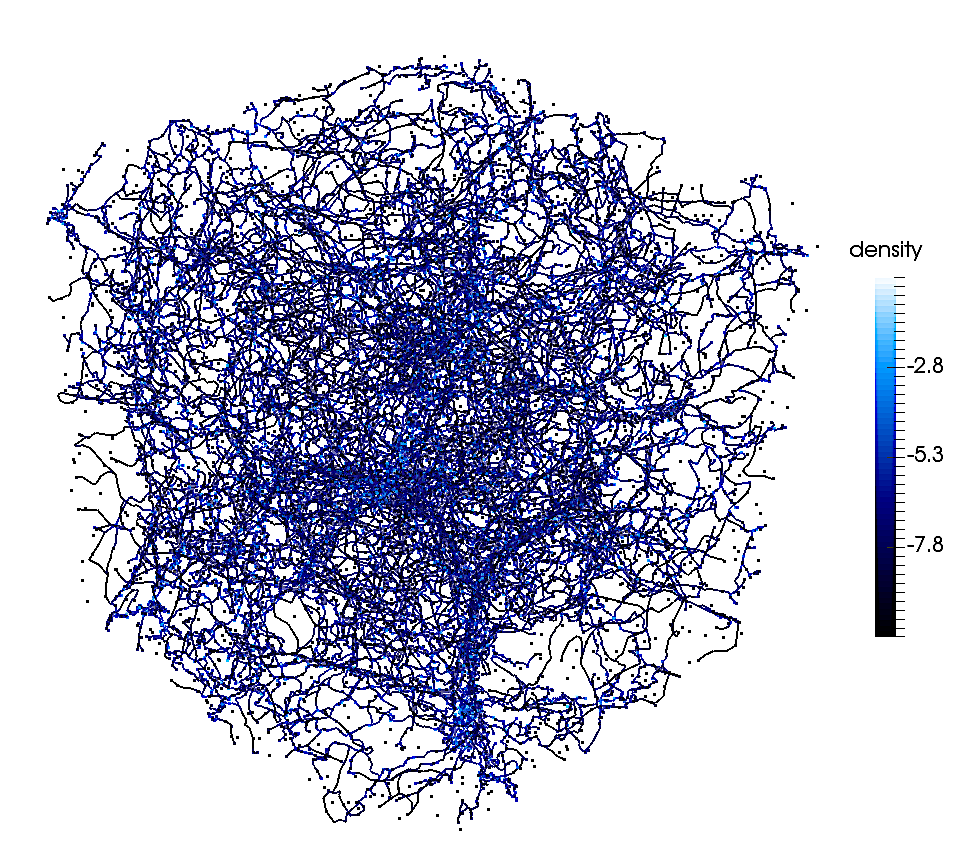} 
\caption{
The skeleton computed by {\tt DISPERSE} on  dark matter particles in one of the 
 $\Lambda$CDM simulations used in this paper using one particle in 40 at redshift zero. 
 }
\label{fig:skel3D}
\end{figure}

This result only holds for low values of the variance that is to say early times or large scales. At later times and smaller scales, this Gram Charlier expansion is proved to converge very slowly breaking down for $\sigma_{0}\lesssim1$ and numerical simulations are needed in order to investigate accurately the time-evolution of cosmic connectivity. To do so, we use 18 $\Lambda$CDM and 4 CDM simulations of a 50 Mpc$/h$  periodic box with $256^{3}$ particles evolved until redshift 0 with {\tt Gadget}
using as a cosmology  $\Omega_{\rm M}=0.3$ $\Omega_\Lambda=0.7$ $\sigma_8=0.9$. As an illustration, Figure~\ref{fig:skel3D} displays the skeleton of one of the  $\Lambda$CDM run.
From these persistent skeletons, we then measured the connectivity at various time slices and for a constant comoving smoothing $R\approx0.8$Mpc$/h$ corresponding to 4 pixels. 
Once again, the persistence cut is chosen at each time step to match the number of peaks found by the {\tt map2ext} code. 

\begin{figure*}
\includegraphics[width=1\columnwidth]{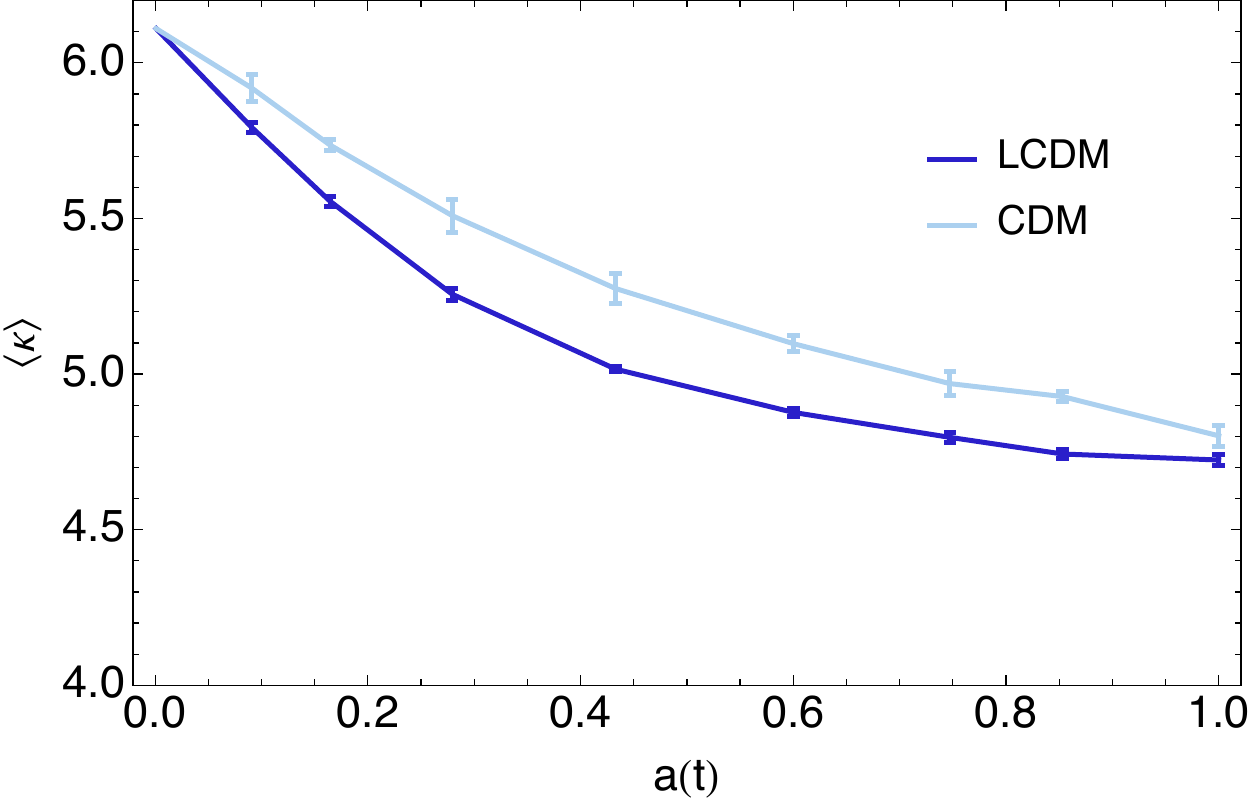}
\includegraphics[width=1\columnwidth]{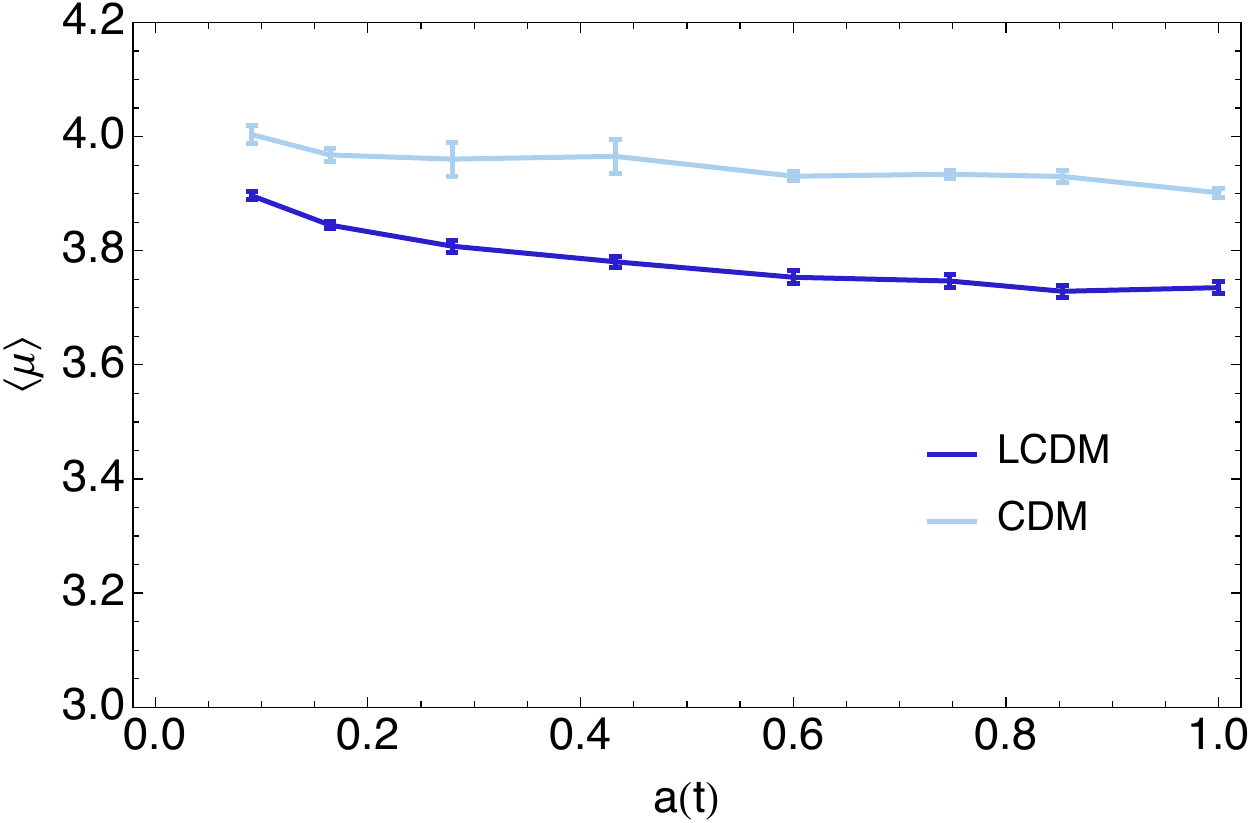}
\caption{Mean connectivity {\sl (left-hand panel)} and multiplicity {\sl (right-hand panel)} of the skeleton as a function of the expansion factor for $\Lambda$CDM and CDM simulations as labelled.
As expected, the CDM simulation is essentially  featureless, whereas the 
$\Lambda$CDM connectivity changes slope when the dark energy expansion kicks in.
} \label{fig:LCDM-CDM-compare}
\end{figure*}

Figure~\ref{fig:LCDM-CDM-compare} shows how connectivity and multiplicity evolve in these two types of simulations. As expected, connectivity decreases as voids grow and filaments merge with cosmic time. The presence of dark energy in the form of a cosmological constant changes the overall shape of this curve which is almost linear in the CDM case while in a $\Lambda$CDM Universe it is clearly convex which is expected since dark energy tends to slow down the growth of cosmic structure and therefore change the slope at late times. On the right-hand panel, multiplicity is shown to vary much less with time. We observe an almost constant shift between the CDM and $\Lambda$CDM  Universes (with a slight decay in the $\Lambda$CDM case which might again be a manifestation of dark energy changing the slope). 
 
Hence the disconnection of filaments with cosmic time is a direct consequence of gravitational clustering, and can in fact also be  understood ab initio in the framework of excursion set theory -- which links the time evolution of cosmic structures to a random walk driven by cosmic variance/smoothing.  With smoothing of the initial GRF as a proxy for its upcoming time evolution, one can identify a special smoothing scales corresponding to the coalescence of  wall-type saddle and filament-type saddle points, which topologically correspond to the disappearance of a tunnel, or equivalently to two filaments merging into one\footnote{this is the analogue of 
maxima and filament-type saddle points coalescence  identified by \cite{hanami} as  slopping saddles tracing merger events.}. From the point of view of the local peak,
 its connectivity  decreases by one when the two filaments merge.
 The relative change in the  disconnection of filaments with cosmology  reflects the same process,
  but is also  in part driven by the role played by dark energy which will stretch and disconnect the filamentary structure 
through the increased expansion of  voids via the cosmological constant.
 
\begin{figure}
\includegraphics[width= 0.95\columnwidth]{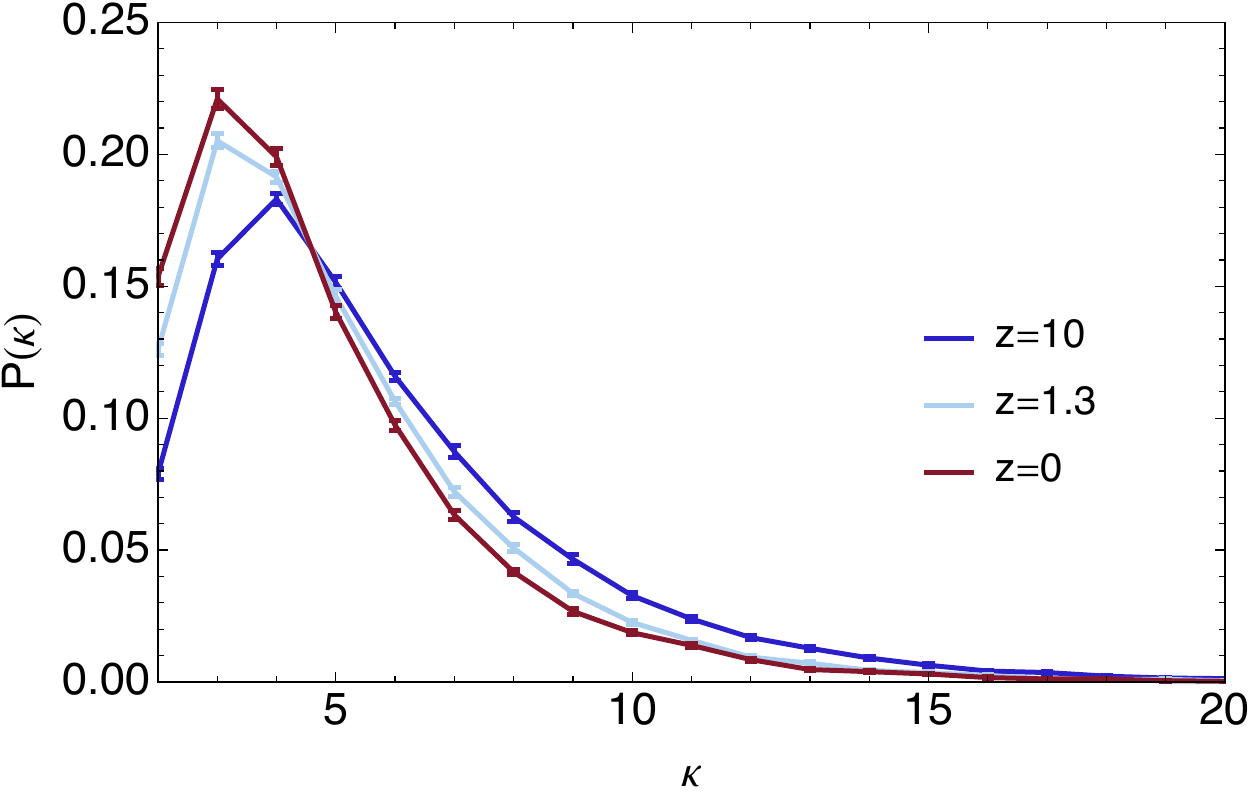}
\caption{PDF of cosmic connectivity at various redshifts as labelled is the $\Lambda$CDM simulations smoothed on a constant comoving length $R\approx0.8$Mpc$/h$.  }
\label{fig:connectivityLCDM}
\end{figure}

Figure~\ref{fig:connectivityLCDM} then shows the full PDF of the global connectivity in the $\Lambda$CDM simulations at various redshifts. At high redshift, the field is nearly Gaussian and the measured connectivity has a statistics close to the Gaussian prediction shown in Figure~\ref{fig:connect3D} for a spectral index about $n_{s}\approx -2$ as expected for a $\Lambda$CDM spectrum at one megaparsec scale. Towards lower redshifts, the PDF gets more and more skewed and concentrated, the most likely connectivity is shifted towards lower values (3 instead of 4) and the tail is suppressed. The global connectivity therefore decreases with cosmic time and highly connected nodes become rarer and rarer due to filaments merging.

Given those findings, we anticipate that measuring the redshift-evolution of cosmic connectivity could prove to be an interesting probe of dark energy.

%%%%%%%%%%%%%%%%%%%%%%%%%%%%
\subsection{Connectivity of dark matter haloes}
%%%%%%%%%%%%%%%%%%%%%%%%%%%%

We make use of the $43$ million dark matter haloes detected at redshift zero in the Horizon 4$\pi$ N-body simulation \citep{Teyssier2009}. 
This simulation contains $4096^3$ DM particles distributed in a 2 $h^{-1}$Gpc periodic box and is characterized by the following $\Lambda$CDM cosmology: $\Omega_{\rm m}=0.24 $, $\Omega_{\Lambda}=0.76$, $n=0.958$, $H_0=73 $ km$\cdot s^{-1} \cdot $Mpc$^{-1}$ and $\sigma _8=0.77$ within one standard deviation of WMAP3 results \citep{Spergel2003}. 
The initial conditions were evolved non-linearly down to redshift zero using the adaptive mesh refinement code RAMSES \citep{teyssier2002}, on a $4096^3$ grid. 
The Friend-of-Friend Algorithm \citep{Huchra1982} was then used  over $18^3$ overlapping subsets of the simulation with a linking length of  0.2 times the mean interparticle distance  to define dark matter haloes. 
In the present work, we only consider the 43 million haloes with more than 100 particles (the particle mass being $7.7\times 10^{9}M_{\odot}$). The mass dynamical range of this simulation spans about 5 decades.

\begin{figure*}
\includegraphics[width=1\columnwidth]{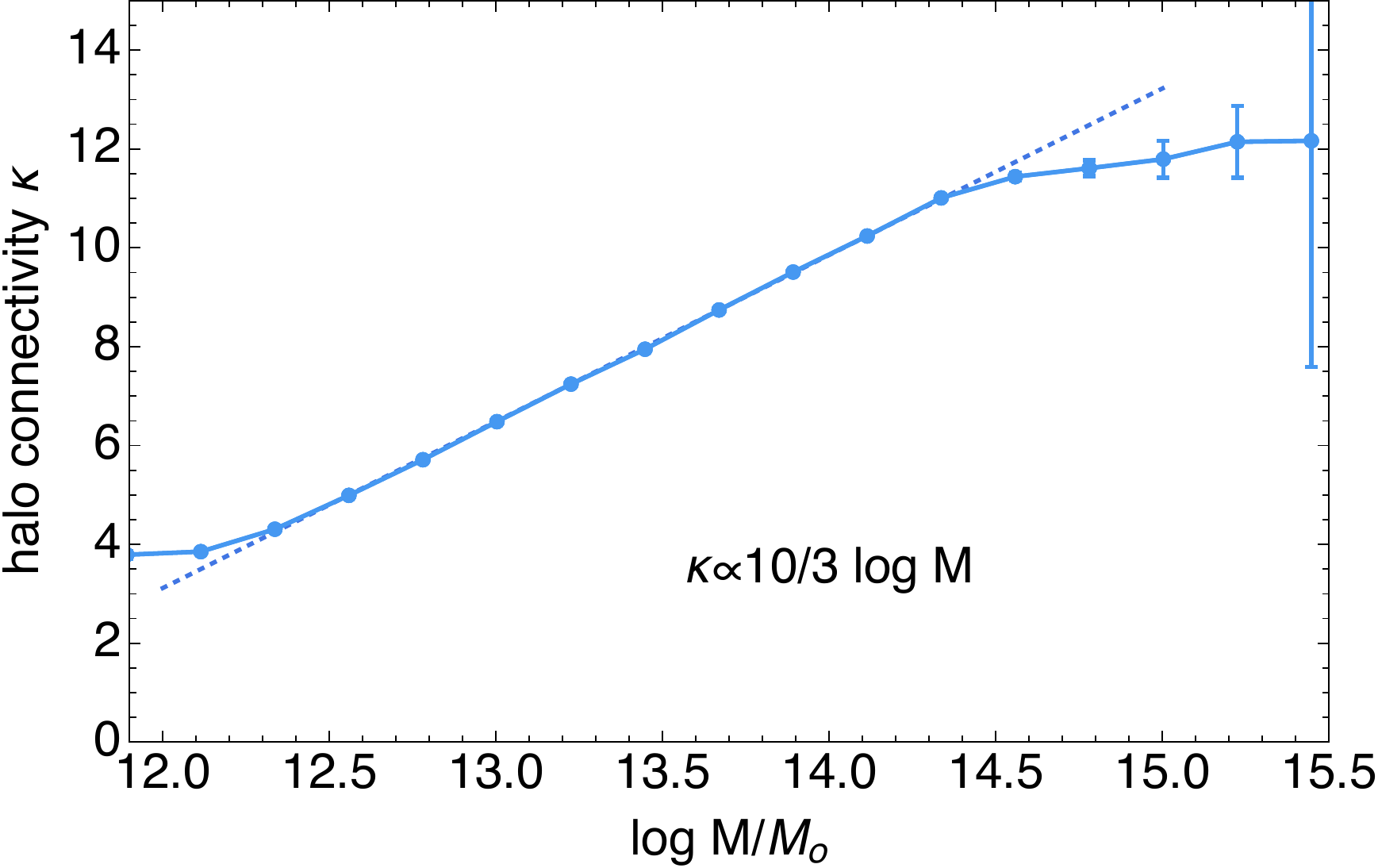}
\includegraphics[width=1\columnwidth]{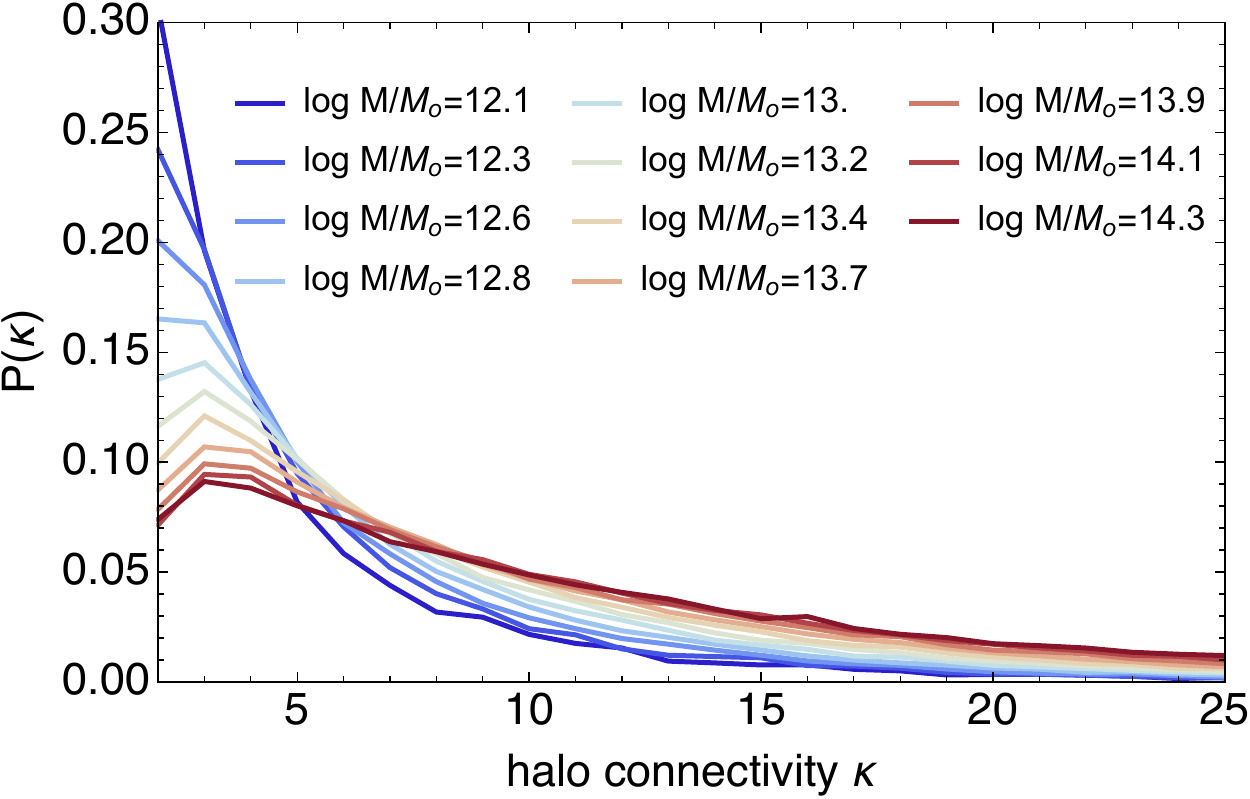}
\caption{ {\sl Left panel:} Mean connectivity of dark haloes at redshift zero  as a function of log mass as labelled for  about one million nodes of the 
cosmic web. The mean cosmic connectivity here is found to be well approximated by the simple linear relation $\kappa(M)\approx10 /3 \log (M/ 10^{11}M_{\odot})$.  {\sl Right panel}: the corresponding PDF $P(\kappa)=P(\kappa|M)$.
} \label{fig:connectivity-hz4pi}
\end{figure*}

The dark halo catalogue was split into 250 sub regions for which the skeleton was computed with  {\tt DISPERSE} using a persistence level of three.
For each identified node, the mass of  most massive dark halo in the vicinity was assigned to this node (out of five nearest neighbour; the 
answer shown not to depend strongly with this choice).
The connectivity was then sampled as a function of mass and its PDF$(\kappa,M)$ and mean value are shown on Figure~\ref{fig:connectivity-hz4pi}.
As expected the connectivity at low mass is lower than for GRF, but grows significantly with the mass of the cluster.
Its growth is well fitted by  $\kappa(M)\approx10 /3 \log (M/ 10^{11}M_{\odot})$.

Small peaks that typically host isolated galaxies tend to be fed by a small number of filaments. Typically those objects are  accreting matter from two main filaments which can bring coherent angular momentum from large scale flows. They will naturally be able to form disc-like galaxies with a spin aligned with galactic filaments.
On the other hand, rare peaks are multiply connected so that galaxies in this cluster-like environment can hardly acquire angular momentum consistently \citep{2012MNRAS.423.3616D}. The more isotropic accretion will favour the formation of elliptical galaxies in dense environments dominated by tidal forces. 
This relationship between number of filaments and spin was already highlighted in \cite{pichonetal11}. We have also shown that three main filaments are expected to dominate, a result which was empirically found in numerical simulation \citep{2015MNRAS.452..784P,2017MNRAS.464.4666G}. 

Using {\tt DISPERSE},  one could  also compute the skeleton of sub halo or (virtual or observed) galaxy catalogues, 
hence their connectivity  as a function halo mass or colour and morphology. The result will depend on the chosen persistence 
threshold, which in turn will depend on the sparsity of the catalogue (hence its limiting mass or magnitude).
This will be the topic of future work.

%=====================
\section{Discussion \& Conclusions}\label{sec:conclusion}
%=====================
This paper  studied the  connectivity  of cosmic random fields in two and three dimensions, quantifying preliminary results presented in  \cite{2010AIPC.1241.1108P}. Starting with generic Gaussian random fields, peaks were shown to be connected on average to 4 other peaks in 2D and $\sim 6.1$ in 3D (and 8.3, 10.7 and 13.2 in 4,5 and 6 D) independently of the shape of the power spectrum. The full statistics of the number of connectors was precisely characterised using  {\tt DISPERSE}: the distribution is skewed, picking around 3 in 2D (resp. 4 in 3D) with a strong tail that can extend up to quite large values $~10$ in 2D (respectively 20 in 3D). The PDF of the number of connectors was shown to depend on the power spectrum and more importantly on the height of the peak. Rare peaks (with high density contrast) tend to be connected to numerous neighbours, the rarest ones having more than 7 (15 in 3D) connections on average.
Interestingly, the overall number of connected neighbours does not correspond to the local number of filaments sticking out of a peak. Locally, the geometry of a peak is ellipsoidal with two filaments following the axis of minimal curvature. Further away those filaments split and bifurcation points appear. Typically, some of those bifurcations occur so close to the central peak that they are not detectable in a map with finite resolution.
 The corresponding multiplicity of a peak is given by its connectivity minus the measured number of bifurcations. For Gaussian random fields, the mean multiplicity was shown to be 3 in 2D and 4 in 3D. Here the dependence with the power spectrum and the typical scatter is reduced. However, the dependence with peak height remains strong.

We have also developed a theoretical framework which  explains the connectivity properties measured with the persistent skeleton. This theory relies on the statistical properties of saddle points in the vicinity of peaks and allowed us to understand precisely the mean connectivity but also its dependence on peak height.
A precise count of the number of filaments crossing a sphere centred on a peak was used to investigate how bifurcation points appear from the peak to the boundary of its peak patch. These calculations allowed us to study the peak's multiplicity and quantify how many filaments dominate (typically we find 3 dominant filaments for high peaks in 3D, 2 for smaller ones).

Finally, the subsequent evolution of  connectivity across cosmic time was also investigated. We developed predictions based on perturbation theory to probe the first stages of structure formation and relied on numerical simulations to confirm the predictions and extend them to a more non-linear regime. As expected, the evolution of cosmic connectivity  depends primarily  on the growth factor   \citep[and on generalized cumulants, see ][]{GPP2012} and therefore on cosmology.
Also as expected, the non-linear evolution reduces the number of connections as filaments merge. 
Compared to alternative probes of dark energy, the connectivity (or the multiplicity) may prove to be a robust 
estimator given that it is a topological property that can be measured locally.

As an astrophysical application we focussed on  galaxy lensing convergence maps and dark halo catalogues. As expected, the connectivity of kappa maps also decreases towards low redshifts, although the evolution is milder than for the three dimensional fields as the lensing kernel tends to smooth the signal along the line-of-sight
\citep[see also][for an investigation of the conditional connectivity around clusters using strong lensing]{2017A&A...605A..27G,2017A&A...605A..80C}. 
In the near future, it should be possible to compute the 3D  connectivity of  HI density maps 
reconstructed from QSO absorption fluxes from the  PSF, WEAVE surveys, or  intensity maps from CHIME, MeerKAT, ASKAP, MWA or HERA, and at some later stage
the  E-ELT  and  SKA resp. 
We also computed  the connectivity of dark haloes and found that it scales logarithmic  with mass with a scaling going like 10/3. 

Cosmic connectivity is not only of interest in the context of cosmology:
on smaller scales it is also paramount to understand galactic assembly.  Indeed 
 dark matter  filaments  have a baryonic continuation within dark haloes, 
 which connect closely the cosmic environment to the galaxies within dark haloes.
 Beyond the number of connected filaments to a given galaxy, the mass load, geometry and 
 torques advected along filaments is also of interest. In this context, one should investigate in more details the small scale connectivity of galaxies  and dark haloes, feeding them with cold gas with stratified angular momentum which 
allows them to re-form stellar discs \citep{pichonetal11}.
 As a topological quantity it would be of interest to quantify the (expected)
robustness of the multiplicity  and connectivity w.r.t. redshift space distortions,
shot noise, photometric errors etc. 
As shown in Sec.~\ref{sec:global}  non-linear gravitational coupling decreases the connectivity of dark haloes.  
This could in principle be quantified by predicting the
 rate of coalescence of wall-saddle  and filament-saddle critical  points,  which 
 reflect filaments coalescence. 
 
Beyond astrophysics, the present theory of connectivity  could prove to be of importance in the context of percolation theory. For instance, the percolation threshold can be explained in terms of the statistical properties of the connectivity of the relevant nodes (keeping track of heights of peaks).
At a more abstract level, the present work on GRF could be of interest to generically connect their properties to graph theory, since the skeleton provides a mapping from the density field to sets of connected vertices \citep[see, e.g.][]{2016arXiv160403236C}. 
This is left for future work.

\subsection*{Acknowledgements}
{ \sl
We thank St\'ephane Colombi, Yohan Dubois, Julien Devriendt, Simon Prunet and Thierry Sousbie for numerous discussions 
about this project over the last 10 years!
We thank our collaborators and the staff at the CCRT for producing the {\tt horizon-4$\pi$}  \href{http://www.projet-horizon.fr/article323.html}{simulation},
D.~Munro for freely
distributing his Yorick programming language and opengl interface
(available at {\em\tt \url{http://yorick.sourceforge.net/}}),
and the community of \href{mathematica.stackexchange.com}{mathematica.stackexchange} for help.
Many thanks to St\'ephane Rouberol for customising the Horizon cluster, hosted by the Institut d'Astrophysique de Paris, for our purposes, and Eric Pharabod for Figure~6.
This work is partially supported by the Spin(e) grant \href{http://www.cosmicorigin.org}{ANR-13-BS05-0005} of the French \textit{Agence Nationale de la Recherche}. 
Special thanks also go to Jia Liu who kindly pointed us to the publicly available Columbia lensing group lensing maps. These maps were obtained with support from NSF grant AST-1210877 and NSF XSEDE allocation AST 140041. We thank New Mexico State University (USA) and Instituto de Astrofisica de Andalucia SCIC (Spain) for hosting the Skies and Universes site for cosmological simulation products.
SC and CP thank Lena for hosting multiple visits, DP thanks the ILP for a senior visiting fellowship, 
 while
CP and DP thank CITA for hospitality during the completion of the project.
}

\bibliographystyle{mnras}
\bibliography{methbib}

\appendix
\section{Extrema correlations}
\label{sec:appA}
Peak theory originates from the so-called
Kac-Rice formula 
\cite{Kac1943,Rice1945} and was first used in a cosmological context in the pioneering work of \cite{BBKS}. Let us sketch here how the one and two point statistics of extrema can be obtained. First, for a random field $\rho$ (e.g the cosmic density field), we define 
the moments
\begin{align}
{\sigma_0}^2 &= \langle \rho^2 \rangle, 
& {\sigma_1}^2 &= \langle \left( \nabla \rho \right)^2 \rangle, 
& {\sigma_2}^2 &= \langle (\Delta \rho)^2 \rangle.
\end{align}
Two characteristic lengths can be built from those moment
$R_0 ={\sigma_0}/{\sigma_1}$ and $R_\star = {\sigma_1}/{\sigma_2}$, as well as the 
spectral parameter
\begin{equation} 
\gamma=\frac{{\sigma_1}^2}{\sigma_0 \sigma_2} =\frac{d \log \sigma(R)}{d \log R}.
\end{equation}
For simplicity, let us work with fields having unit variances:
\begin{align} 
x&=\frac{1}{\sigma_0} \rho, 
& x_i&=\frac{1}{\sigma_1} \nabla_i \rho, 
& x_{ij} &=\frac{1}{\sigma_2} \nabla_i \nabla_j \rho.
\end{align}
The one-point probability density (PDF) will be denoted  ${\cal P}(\boldsymbol{X})$ and the joint PDF
${\cal P}(\boldsymbol{X},\boldsymbol{Y})$, where $\boldsymbol{X}=\{x,x_{ij},x_i\}$ and $\boldsymbol{Y}=\{y,y_{ij},y_i\}$ of dimension $(d+1)(d+2)/2$ represent the normalized 
field and its (up to second) derivatives, 
at two prescribed comoving locations (${\boldsymbol r}_{x}$ and ${\boldsymbol r}_{y}$ separated by a distance 
$r=|{\boldsymbol r}_{x}-{\boldsymbol r}_{y}|$). In the supposingly Gaussian initial conditions, this 
joint PDF is a multivariate Normal distribution
\begin{equation}
{\cal N}(\boldsymbol{X},\boldsymbol{Y})= \frac{\exp\left[-\frac{1}{2}
\begin{pmatrix}
 \boldsymbol{X} \\
 \boldsymbol{Y}
\end{pmatrix}
^{\rm T}
 \cdot
  \mathbf{C}
 ^{-1}\cdot
\begin{pmatrix}
 \boldsymbol{X}\\
 \boldsymbol{Y}
\end{pmatrix}
\right]}{{\rm det}|\mathbf{C}|^{1/2} \left(2\pi\right)^{\rm (d+1)(d+2)/2 }} \,, 
\label{eq:defPDF}
\end{equation} 
where $\mathbf{C}_{0}\equiv \langle  \boldsymbol{X}\cdot \boldsymbol{X}^{\rm T} \rangle$ and  
$\mathbf{C}_{\gamma}\equiv \langle  \boldsymbol{X}\cdot \boldsymbol{Y}^{\rm T} \rangle$ are the diagonal
and off-diagonal components of the covariance matrix
\begin{equation}
\quad \mathbf{C}=
\begin{pmatrix}
\mathbf{C}_{0} &\mathbf{C}_\gamma \\
\mathbf{C}_\gamma^{\rm T}  &\mathbf{C}_{0}
\end{pmatrix}
\,.
\end{equation}
All these quantities solely depend on the separation $r$ because of homogeneity and isotropy. 
Equation~(\ref{eq:defPDF}) is sufficient to compute the expectation of any quantity involving the 
field, its first and second derivatives.
In particular, the two-point correlation $\xi_\text{crit}(r,\nu)$ of critical points 
at threshold $\nu$ separated by $r$ is given by
\begin{equation}
\label{eq:xicrit}
1+\xi_{\rm crit}(r,\nu)=
 \frac{
\big\langle \rho_{\rm crit}(\boldsymbol{X})  \rho_{\rm crit}(\boldsymbol{Y}) \big\rangle}
{\big\langle \rho_{\rm crit}(\boldsymbol{X}) \big\rangle^2}   \,,
\end{equation}
with the ``localized'' density of critical point 
\begin{equation}
\label{eq:ncrit}
\rho_{\rm crit}(\boldsymbol{X})=\frac{1}{R_{\star}^{\text d}}
|{\rm det}(x_{ij})|\delta_{\rm D}(x_i)\delta_{\rm D}(x-\nu) \,.
\end{equation}
This density is formally zero unless the condition for a critical point is satisfied. In particular,
\begin{eqnarray}
\big\langle \rho_{\rm crit}(\boldsymbol{X})  \big\rangle
&=& \left(\frac{\sigma_2}{\sigma_1}\right)^\text{d}\!\! \int\!{\rm d} \boldsymbol{X} \,
  {\rm det}(x_{ij})  \delta_{\rm D}(x_i)  \delta_{\rm D}(x-\nu)     {\cal P}(\boldsymbol{X})\nonumber\\
&\equiv &\partial_{\nu}{n}_\text{crit}\,,
\nonumber
\end{eqnarray}
which appears in the denominator of equation~(\ref{eq:xicrit}), is the average number density 
of critical points at threshold $\nu$ while
\begin{equation}
\big\langle \rho_{\rm crit}(\boldsymbol{X}) \rho_{\rm crit}(\boldsymbol{Y})\big\rangle 
=  \int\!{\rm d}\boldsymbol{X}\!\int\!{\rm d}\boldsymbol{Y} \, {\cal P}(\boldsymbol{X},\boldsymbol{Y}) \,
\rho_{\rm crit}(\boldsymbol{X})
\rho_{\rm crit}(\boldsymbol{Y})\nonumber
\end{equation}
is the cross-correlation.
If one wants to restrict the signature of the critical points ($---$ for peaks, $- - +$ for filament-type saddles, $- + +$ for wall-type saddles and $+++$ for minima),
an additional constraint on the sign of the second derivatives is required. Then, the peak two-point correlation function reads
\begin{equation}
\label{eq:xipeak}
1+\xi_{\rm pp}(r,\nu)= 
 \frac{\big\langle \rho_{\rm pk}(\boldsymbol{X}) \rho_{\rm pk}(\boldsymbol{Y})\big\rangle}
{\big\langle \rho_{\rm pk}(\boldsymbol{X})\big\rangle^2} \,.
\end{equation}
where the localized peak number density $\rho_{\rm pk}(\boldsymbol{X})$, 
\begin{equation}
\label{eq:npk}
\rho_{\rm pk}(\boldsymbol{X}) =
\frac{1}{R_{\star}^{\text d}}
|{\rm det}(x_{ij})|  \delta_{\rm D}(x_i)  \Theta_{\rm H}(-\lambda_{i})  \delta_{\rm D}(x-\nu) \,,
\end{equation}
implements the peak condition.
For $d>1$,  $\delta_{\rm d}(x_i)$ is understood as
$\prod_{i\le d} \delta_{\rm D}(x_i)$,
while  $ \Theta_{\rm H}(-\lambda_{i}) $  stands for $\prod_{l\le d} \Theta_{\rm H}(-\lambda_l)$, with 
$\{\lambda_l\}_l$ the eigenvalues of the Hessian.
Because of these restrictions on the signature, the integral typically is not analytical.
In dimension $\rm d$, we define the conditional probability that $x_{ij}$ and $y_{ij}$ 
satisfy the PDF, subject to the condition that $x_i=y_i=0$ and $x=y=\nu$ and 
rely on Monte-Carlo 
methods in {\small MATHEMATICA} in order to evaluate numerically equation~(\ref{eq:xipeak}). Namely, we draw random 
numbers of dimension 
$ d(d+1)$ from the conditional probability that $x_{ij}$ and $y_{ij}$ satisfy 
the PDF, subject to the condition that $x_i=0$ and $x=y=\nu$. 
For each draw $^{(k)}$ if $\lambda_l(x^{(k)}_{ij})<0 $ and  $\lambda_l(y^{(k)}_{ij})<0 $ 
($l\le d$) we keep the sample and evaluate $ {\rm det}(x^{(k)}_{ij})   {\rm det}(y^{(k)}_{ij}) $ 
and otherwise we drop it; eventually, 
\begin{equation}
\left\langle \rho_{\rm pk}(\!\boldsymbol{X}\!)\rho_{\rm pk}(\!\boldsymbol{Y}\!)\right\rangle
\! \approx\!\! \frac{ \!{\cal P}(\!x\!=\!y\!=\!\nu,x_{i}\!=\!y_{i}\!=\!0\!)\!}{N}
\!\sum_{k\in {\cal S}} \! \!  {\rm det}(x^{\!(k)}_{ij})  {\rm det}(y^{\!(k)}_{ij}),
\nonumber
\end{equation}
where $N$ is the total number of draws, and $\cal S$ is the subset of the indices of draws 
satisfying the constraints on the eigenvalues.
The same procedure can be applied to evaluate the denominator
$\left\langle \rho_{\rm pk}(\boldsymbol{X}) \right\rangle\equiv \bnpk(\nu)$. 
Equation~(\ref{eq:xipeak}) then yields an estimation of  $\xi_{\rm pp}(r,\nu)$. This algorithm is embarrassingly 
parallel and can be easily generalized, for instance, to the computation of the correlation 
function $\xpk(r,>\nu)$ of peaks {\sl above} a given threshold in density, the correlation between peaks and saddle points and to any dimension $d$. 
In practice it is fairly efficient as the draw is customized to the shape of the underlying 
Gaussian PDF.
Obviously, if correlation functions above a given threshold are considered, the required number of draws 
is larger and increases with the value of the threshold (as the event $x>\nu$ becomes rarer).

\section{Filament crossings}
\label{app:2Dpks}
We propose here to describe how to compute peak counts on the surface of a sphere around a central peak.
The case of curved manifold for one-point statistics was addressed in \cite{2016JCAP...04..058M,2016IAUS..308...61P}. Here we develop a formalism to deal with two-point statistics when spherical sections of higher dimensional spaces are considered.

\subsection{One-point statistics on the sphere}
Following \cite{Codis2013}, we  use cartesian coordinates with indices 1 to 3. Here 3 will refer to the direction of the separation $\mathbf r$ between the central peak and the current point. The unit variance field under consideration is again denoted $x$ and its unit variance first and second derivatives $x_{i}$ and $x_{ij}$ for $i,j$ between 1 and 3. We want to find the function $B_{fil}(x(0),x_{i}(0),x_{ij}(0),x(r),x_{i}(r),x_{ij}(r))$ such that the mean number of 2D peaks at a distance $r$ from a central peak reads $\left\langle B_{fil}\right\rangle $ where all variables follow a known Gaussian distribution.  To do so, we need to express the gradient, trace and determinant of the Hessian at the surface of the sphere in terms of the 3D field variables.

The surface gradient can be defined as
\begin{equation}
\nabla_{\perp}x=\nabla x-\frac{\partial x}{\partial r}\,.
\end{equation}
This operator acts over the unit sphere and is trivially independent of the frame orientation.
In our frame, the surface gradient has coordinates $(x_{1},x_{2},0)$.

The surface Laplacian (often called the Beltrami-Laplace operator) can also be defined in a frame-orientation independent way
\begin{equation}
r^{2}\Delta_{\perp}x=r^{2}\Delta x-r\frac{\partial^{2}(r x)}{\partial r^{2}}\,,
\end{equation}
which in our frame reads $I_{1}\equiv\Delta_{\perp}x=x_{11}+x_{22}-2x_{3}/r$.

The only remaining quantity to compute is now the determinant on the sphere. It can be shown that the Hessian matrix of covariant derivatives reads
\begin{equation}
H_{\rm 2D}=\left(
\begin{array}{cc}
x_{11}-x_{3}/r& x_{12}\\
x_{12}&x_{22}-x_{3}/r
\end{array}
\right)\,,
\end{equation}
so that the determinant simply reads $I_{2}\equiv x_{11}x_{22}-x_{12}^{2}-x_{3}/r(x_{11}+x_{22})+x_{3}^{2}/r^{2}$.
This surface determinant $I_{2}$ has therefore mean $\left \langle I_{2}\right\rangle=\sigma_{1}^{2}/(3r^{2})$ and the variance of the surface Laplacian $I_{1}$ has a curvature-correction given by $\left\langle I_{1}^{2}\right\rangle=8/15\sigma_{2}^{2}+4/(3r^{2})\sigma_{1}^{2}$.

In particular, the genus (i.e signed critical points) can be computed $\chi=4\pi r^{2}\times \sigma_{1}^{2}/(3r^{2})\times 3/(2\pi\sigma_{1}^{2})=2$ as expected.
 Note that the factor $3/(2\pi\sigma_{1}^{2})$ comes from the zero gradient condition which reads $P(x_{1}=0,x_{2}=0)=(1/\sqrt{2\pi\sigma_{1}^{2}/3})^{2}$ as the variance of the gradient along each direction is $\sigma_{1}^{2}/3$.

We also recover that $J_{2}=I_{1}^{2}-4I_{2}=(x_{11}-x_{22})^{2}+4x_{12}^{2}$ is independent of the curvature.

\subsection{Two-point statistics}

We now consider the vector containing the field and its first and second derivatives at the origin and on a sphere at a distance $r$, $X=\{y,y_{i},y_{ij},x,x_{i},x_{ij}\}$. 
The genus on the sphere given a peak constraint at the center can then be computed as
\begin{equation}
\chi|_{\rm pk}=\frac{\left\langle I_{2}\delta_{D}(x_{i}^{\rm 2D})\times \det y_{ij} \delta_{\rm D}(y_{i})\delta_{\rm D}(y-\nu) {\cal B}(\lambda_{3}<0)\right\rangle}{\left\langle  \det y_{ij} \delta_{\rm D}(y_{i})\delta_{\rm D}(y-\nu) {\cal B}(\lambda_{3}<0)\right\rangle}\nonumber\,,
\end{equation} 
where $\lambda_{1}<\lambda_{2}<\lambda_{3}$ are the eigenvalues of the Hessian matrix $y_{ij}$  $\delta_{\rm D}(y_{i})=\delta_{\rm D}(y_{1})\delta_{\rm D}(y_{2})\delta_{\rm D}(y_{3})$ and $\delta_{D}(x_{i}^{\rm 2D})=\delta_{\rm D}(x_{1})\delta_{\rm D}(x_{2})$,
and is found to be exactly 2 as expected.

As a proxy for the number of skeleton branches around a peak, we can now compute the mean number of 2D peaks on the surface of the sphere around a central peak. For that purpose, a condition on the sign of the eigenvalues has to be implemented and leads to
\begin{equation}
N_{\rm fil}=\frac{\left\langle I_{2}\delta_{D}(x_{i}^{\rm 2D}){\cal B}(\mu_{2}\!<\!0)\det y_{ij} \delta_{\rm D}(y_{i})\delta_{\rm D}(y\!-\!\nu) {\cal B}(\lambda_{3}\!<\!0)\right\rangle}{\left\langle  \det y_{ij} \delta_{\rm D}(y_{i})\delta_{\rm D}(y\!-\!\nu) {\cal B}(\lambda_{3}\!<\!0)\right\rangle}\nonumber
\end{equation} 
where $\mu_{1}<\mu_{2}$ are the eigenvalues of the Hessian matrix $H_{\rm 2D}$.
As displayed in Fig.~\ref{fig:connectivity-N2Dpeaks}, the number of peaks is two in the zero separation limit (while it would be one without zero gradient constraint at the center of the sphere) as the local geometry is ellipsoidal and it does not change with the height of that central peak. As separation grows, bifurcation points occurs and the number of filaments increases. 

\section{Persistence cuts} \label{sec:persistence-cuts}
\begin{figure*}
\begin{center}
\includegraphics[width= 0.98\columnwidth]{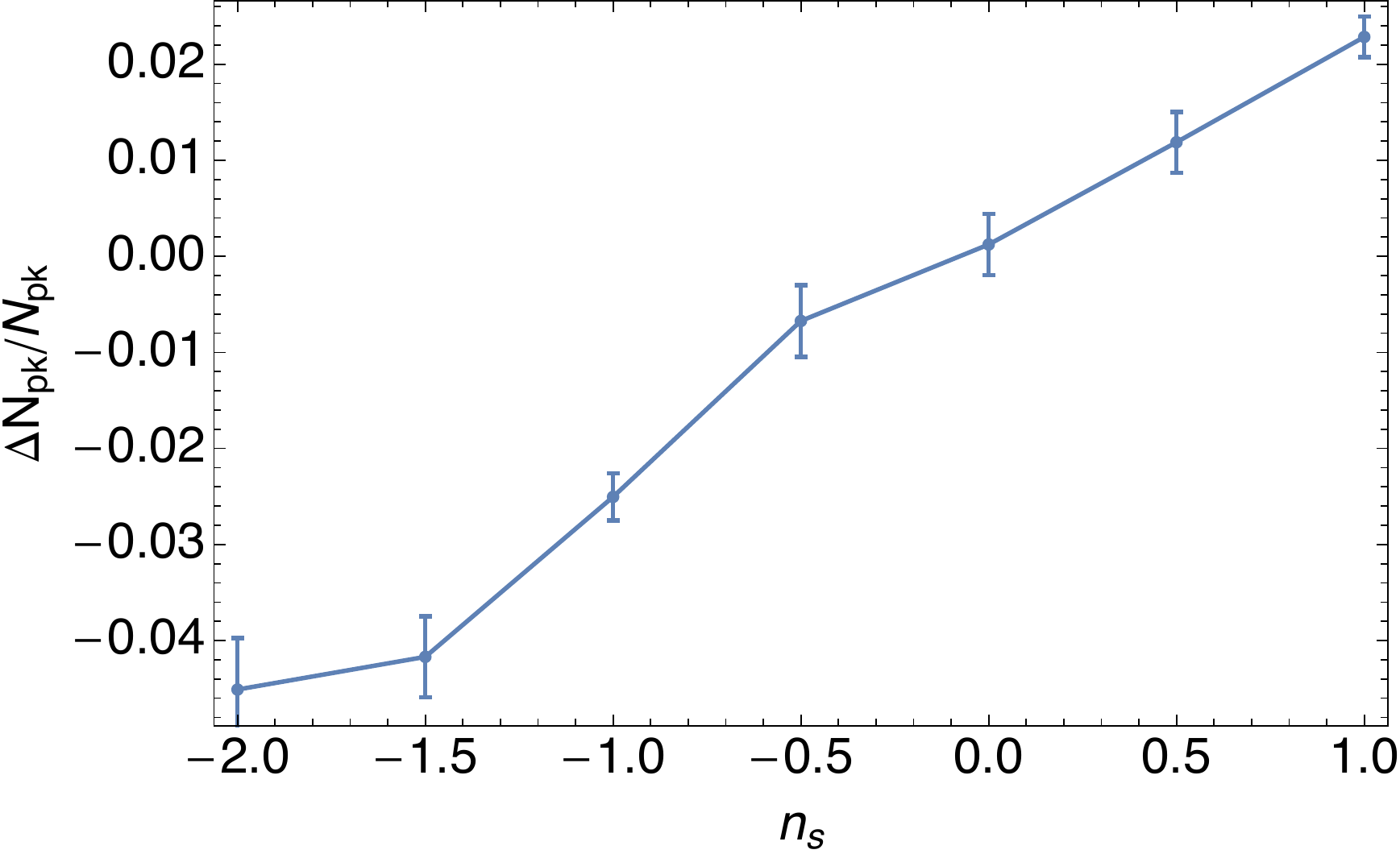} \hskip 1cm
\includegraphics[width= 0.91\columnwidth]{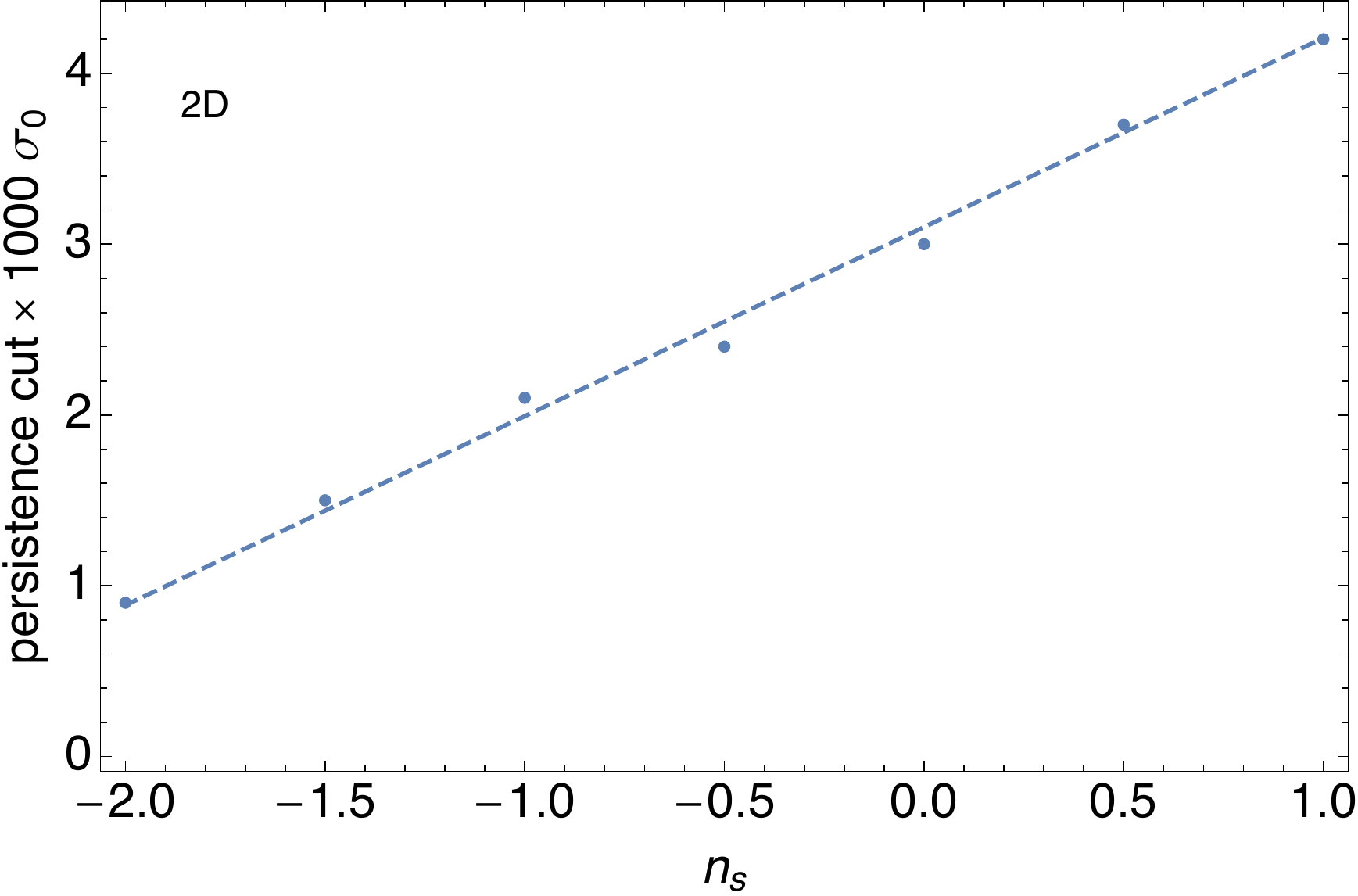}
\caption{
Left-hand panel: error on the number of 2D peaks measured compared to expected as a function of the spectral index when a fixed cut of $p_{\rm min}= 0.003 \; \sigma_0$
is chosen for the persistence. Right-hand panel: cut in persistence (in units of $\sigma_{0}/1000$) for different values of the spectral index $n_s$ chosen so that the number of peaks is not different from the expected value by more than one per cent. This mapping is well fitted by the linear relation $p_{\rm min}=0.0031(1+n_{s}/2.8)\sigma_{0}$.}
\label{fig:cut}
\end{center}
\end{figure*}

To get the persistent skeleton of our two and three dimensional Gaussian random field realisations, we use the code {\tt DISPERSE}. In addition to its ability to work with  sampled data sets while assuming nothing about its geometry or homogeneity, this code can select structures on the basis of their significance via the notion of persistence ratio,  a measure of the strength of the topological connection between individual critical points. 
This persistence threshold is expressed in terms of the typical RMS of the noise. 
To set this threshold, in practice we use a persistence cut $p_{\rm min}$ so that 
 the number of peaks in the maps matches the expected number of peaks (which is known exactly for Gaussian random fields) with better than 0.5\% accuracy.
 This cut has to depend on the shape of the power spectrum. Indeed, if a constant persistence cut is chosen, the left panel of Figure~\ref{fig:cut} shows that the error on the total number of peaks in the map is significantly non-zero and is a strong function of the spectral index. Instead, we therefore select a cut different for each spectral index.
In 2D, adopted persistence cuts are given in Table~\ref{tab:persistencecuts} and display on the right panel of Figure~\ref{fig:cut}. The relation seems to be almost linear with the spectral index and well fitted by $p_{\rm min}=0.0031(1+n_{s}/2.8)\sigma_{0}$.

The same procedure is followed in 3D. In that case, the  persistence cuts are found to be $p_{\rm min}=\{4.5,6,9,12\}\sigma_{0}/1000$ for $n_{s}=\{-3,-2,-1,0\}$. 

For non-Gaussian random fields, the analytical prediction is not known in general. Instead we compare the measurements of {\tt DISPERSE} \citep{Sousbie2011b} and subsequently adapt the persistence cut) to measurements from the 
{\tt map2ext} code \citep{Colombi2000,pogo11}. Here, for every pixel a segment of quadratic surface is fit in the tangent plane based on the field values at the pixel of origin and its neighbours. The position of the extremum of this quadratic surface, its height and its Hessian are computed. The extremum is counted into the tally of the type determined by its Hessian (two negative eigenvalues for peaks) if its position falls within the original pixel. 
Several additional checks are performed to preclude registering extrema in the neighbouring pixels and minimize missing extrema due to jumps in the fit parameters as region shifts to the next pixel. This procedure performs with better than 1\% accuracy when the map is smoothed with a Gaussian filter whose full width at half maximum exceeds 6 pixels. 

\begin{table}
\centering
\begin{tabular}{|c|c|c|c|c|c|c|c|}
$n_{s}$&$1$ &$0.5$ &$0.$ &$-0.5$ &$-1$ &$-1.5$&  $-2$\\\hline
$p_{\rm min}$&
4.2 &
3.7&
3 &
2.4 &
2.1&
1.5&
0.9\\
\end{tabular}
\caption{Cut in persistence $p_{\rm min}$ (in units of $\sigma_{0}/1000$) for different values of the spectral index $n_s$ chosen so that the number of peaks is not different from the expected value by more than 10. This mapping is well fitted by the linear relation $p_{\rm min}=0.0031(1+n_{s}/2.8)\sigma_{0}$.  }
\label{tab:persistencecuts}
\end{table} 

%%%%%%%%%%%%%%%%%%%%%%%%%%%
\section{Geometry of connection}\label{sec:geometry} 
%%%%%%%%%%%%%%%%%%%%%%%%%%%

Let us briefly investigate the geometry of the field in the vicinity of bifurcation points.
As defined by \cite{pogo09}
 bifurcation points will locally correspond to two eigenvectors of the Hessian having the 
same eigenvalue. 

\subsection{Bifurcation in 2D}
%%%%%%%%%%%%%%%%%%%%%%%%%%%

Let us expand the field, $x$, in the Hessian eigenframe to third order in the vicinity of $x_2=0, x_{11}=x_{22}$ along the  direction 
$\delta \mathbf{r}=\epsilon (\cos\theta,\sin \theta)$; its variation, $\Delta x$ will obey (having set $\epsilon$ to one)
\[
6 \Delta x=6 x_1 \cos (\theta )+3 x_{22}+x_{111} \cos ^3(\theta )+3 \sin (\theta
   ) x_{112} \cos ^2(\theta )+
   \]
   \[\qquad\,\,\,\,\,\,\,\,
   3 \sin
   ^2(\theta ) x_{122} \cos (\theta )+\sin
   ^3(\theta ) x_{222} +{o}(x_{ijk}) \,.
\]
Calling $y=\cos^2\theta$, the extrema of $\Delta x $ with respect to $\theta$ on the $\varepsilon$-circle satisfy the algebraic equation
\be
y\left(\frac{ \left(y \left(3
   x_{112}-x_{222}\right)-2 x_{112}+x_{222}\right)^2}{\left(2
   x_1+y \left(x_{111}-3
   x_{122}\right)+x_{122}\right)^2}+1\right)=1\,,
\ee
{  or geometrically }
\be
\quad \left(\frac{1}{2}
\delta \mathbf{r} \cdot \mathbf{\nabla \nabla \nabla \rho } \cdot \delta \mathbf{r} + \nabla \rho \right)
\times   \delta \mathbf{r} =\mathbf{0}  \,,\label{eq:defy}
   \ee
which is a cubic in $y$; hence at most six solutions are possible for $\theta$ which correspond to 3 maxima and 3 minima.
Note that if the gradient dominates equation~(\ref{eq:defy}) relatively to the third derivative tensor, the solutions  $\pm \delta \mathbf{r} $ should be globally in the direction of  $\nabla \rho $ (the dipole), 
whereas if it is subdominant (e.g. for the first bifurcation near the peak),  $\delta \mathbf{r} $ should follow the ``eigen directions'' of $\nabla \nabla \nabla \rho $ (the octupole), possibly  weakly skewed by the local gradient.
The inspection of equation~(\ref{eq:defy}) shows indeed that the stronger the gradient $x_1$, the weaker the deviation of the bifurcation from the gradient's direction ($y=1$). 
If the octupole has special symmetries ($x_{111}=3x_{122} $ and $x_{222}=3 x_{112}$) then only one solution in $y$ remains: $y_0= \left({x_{112}^2}/{\left(2
   x_1+x_{122}\right)^2}+1\right)^{-1}$. 
   Note also that if $x_{222}= x_{112}=0$  only $\cos \theta =\pm 1$  is a solution: this corresponds to a straight bifurcation since the corresponding curvature is 
   also zero.

\subsection{Bifurcation in 3D}

Let us expand the field, $x$ to third order in the vicinity of $x_2=x_3=0$, $x_{11}=x_{22}$ along the infinitesimal direction 
$\delta \mathbf{r}=\epsilon (\cos\theta \cos \phi,\cos\theta \sin \phi,\sin \theta)$.
The extrema of $\Delta x$ on the infinitesimal sphere of radius $\epsilon$ obey
\[
\frac{\partial \Delta x}{\partial \delta \mathbf r} \times   \delta \mathbf{r} =
\left(\frac{1}{2}
\delta \mathbf{r} \cdot \mathbf{\nabla \nabla \nabla \rho } \cdot \delta \mathbf{r} + \nabla \nabla \rho \cdot \delta \mathbf{r}  + \nabla \rho \right)
\times   \delta \mathbf{r} =\mathbf{0} \,.
\] 
Again the critical directions correspond to a mixture between the  eigenvector of the anisotropic part of the Hessian, the gradient, and the 
``eigenvectors'' of $\nabla \nabla \nabla \rho $, depending on the relative strength of the three components.
Unless the  anisotropic part of the Hessian is null the bifurcation will be co-planar, and can be investigated with the 2D formalism presented above.

\section{Nd connectivity} \label{sec:NDcon}
%%%%%%%%%%%%
In order to compute the connectivity in arbitrary dimensions it is best to move to the 
eigenframe of the hessian and rely on the joint statistics of its eigenvalues.
From \cite{pogo09} the probability of measuring the set of d (ordered) eigenvalues of the d dimensional  Hessian $\{\lambda_{i}\}$  and 
density $\nu$ obeys 
\begin{equation}
\prod_{i\le { d}}  d \lambda_i 
 \prod_{i<j} (\lambda_j-\lambda_i)
  \exp\left( -\frac{1}{2}
Q_{\gamma}(\nu,\{\lambda_{i}\})
\right)\,,  \label{eq:defQprob}
\end{equation}
where $Q_{\gamma}$ is a quadratic form in $\lambda_{i}$ and $\nu$ given by
\begin{equation}
Q_{\gamma}(\nu,\{\lambda_{i}\})=\nu^2+ \frac{\left(\sum_{i}\lambda_{i}+\gamma \nu\right)^2}{ (1-\gamma^2)}+
{\cal Q}_{d}(\{\lambda_{i}\})\,, \label{eq:Qgam}
 \end{equation}
with 
\begin{equation}
{\cal Q}_{d}(\{\lambda_{i}\})= (d+2)\left[\frac{1}{2}(d-1) \sum_{i} \lambda^{2}_{i}- \sum_{i\neq j} \lambda_{i}\lambda_{j}\right].
\end{equation}
It now follows that the extrema number counts in dimension d read: 
\begin{equation}
{\partial_\nu {n}^{{d}}}=
\left\langle   \Theta_{\rm H}(-\lambda_{i})   \left| \prod  \lambda_i \right| \right\rangle\,, \label{eq:NDdiff}
\end{equation}
where this expectation is computed using equation~\eqref{eq:defQprob}.
From equation~\eqref{eq:Qgam}, integration over $\nu$ yields the marginal 
probability of  $\{\lambda_{i}\}$:
\begin{equation}
\hskip -0.1cm \prod_{i\le {d}}\!  d \lambda_i \!
 \prod_{i<j} (\lambda_j\!-\!\lambda_i)
  \exp\left(\! -\frac{1}{2}
{\cal Q}_{d}(\{\lambda_{i}\})-\!\frac{1}{2} \left(\sum_i \lambda_i \right)^2
\right)\,. \label{eq:defQprob} 
\end{equation}
Finally, the $d$ dimensional  connectivity is simply given by the ratio
\begin{equation}
\kappa = \frac{2 \left\langle   \Theta_{\rm H}(-\{\lambda_{i}\}_{i<d})  \Theta_{\rm H}(\lambda_{d})   \left| \prod  \lambda_i \right| \right\rangle}
{\left\langle   \Theta_{\rm H}(-\{\lambda_{i}]\}_{i\leq d})   \left| \prod  \lambda_i \right| \right\rangle}\,,
\label{eq:defkapND}
\end{equation}
where this expectation is now computed using equation~\eqref{eq:defQprob}.
Implementing equation~\eqref{eq:defkapND} in dimension up to 10 yields
\begin{table}
\centering
\begin{tabular}{|c|c|c|c|c|c|c|c|c|c|c|c|}
$d$       & $2$ &$3$ &$4$ &$5$ &$6$&$7$&$8$&$9$&$10$&$11$ \\\hline
\!\!$\kappa$\!\!&\!$4$ \!&\!\!\!$6.11$\!\!\!&\!\!\!$8.35$\!\!\!&\!\!\!$10.73$\!\!\!&\!\!\!$13.23$\!\!\!&\!\!\!$15.85$\!\!\!&\!\!\!$18.7$\!\!\!&\!\!\!$21.4$\!\!\!&\!\!\!$24.4$\!\!\!&\!\!\!$27.4$\!\!\!\\
\end{tabular}
\caption{$\kappa_{\rm G}$ as a function of dimension $d$ for GRFs.
The scaling is well fitted by $\kappa= 2d+\left((2d-4)/7\right)^{7/4}$.}
\label{tab:kappaDim}
\end{table} 
 Table~\ref{tab:kappaDim} which gives a few values of $\kappa$ as a function of dimension. 
 The global connectivity of a Gaussian random field is shown to slightly depart from a cubic lattice in dimensions greater than 3. These defects are shown to affect the connectivity even more so as the dimension increases with a scaling proportional to the power $7/4$ of the dimension, so that a good fit to the global connectivity is given by
 \begin{equation}
 \kappa= 2d+\left(\frac{2d-4}{7}\right)^{7/4}.
 \end{equation}
We checked that integration over the field variables, $\mathbf{X}=(x,x_{ij},x_i)$ also yield the same numbers.

\label{lastpage}

\end{document}